\documentclass[12pt]{iopart}

\usepackage[dvips]{graphicx}
\usepackage{cite}
\usepackage[perpage]{footmisc}

\newcommand{\dd}     {\mathrm{d}}
\newcommand{\epem}   {\ensuremath{\mathrm{e^+e^-}}}
\newcommand{\mpmm}   {\ensuremath{\mu^+\mu^-}}
\newcommand{\qbar }  {\ensuremath{\mathrm{\overline{q}}}}
\newcommand{\qqbar}  {\ensuremath{\mathrm{q\qbar}}}
\newcommand{\qqbarg} {\ensuremath{\mathrm{q\qbar g}}}
\newcommand{\bbbar}  {\ensuremath{\mathrm{b\overline{b}}}}
\newcommand{\ccbar}  {\ensuremath{\mathrm{c\overline{c}}}}
\newcommand{\as}     {\ensuremath{\alpha_{\mathrm{S}}}}
\newcommand{\asq}    {\ensuremath{\as(Q)}}
\newcommand{\ash}    {\ensuremath{\hat{\alpha}_{\mathrm{S}}}}
\newcommand{\asb}    {\ensuremath{\bar{\alpha}_{\mathrm{S}}}}
\newcommand{\ashq}   {\ensuremath{\hat{\alpha}_{\mathrm{S}}(Q)}}
\newcommand{\asmu}   {\ensuremath{\as(\mu)}}
\newcommand{\asnp}   {\ensuremath{\as^{\mathrm{NP}}}}
\newcommand{\oaaa}   {\ensuremath{\mathcal{O}(\as^3)}}
\newcommand{\oaa}    {\ensuremath{\mathcal{O}(\as^2)}}
\newcommand{\oa}     {\ensuremath{\mathcal{O}(\as)}}
\newcommand{\oan}    {\ensuremath{\mathcal{O}(\as^n)}}
\newcommand{\bt}     {\ensuremath{B_T}}
\newcommand{\bw}     {\ensuremath{B_W}}
\newcommand{\bn}     {\ensuremath{B_N}}
\newcommand{\mh}     {\ensuremath{M_H}}
\newcommand{\ml}     {\ensuremath{M_L}}
\newcommand{\dmh}    {\ensuremath{\Delta M_H}}
\newcommand{\thr}    {\ensuremath{1-T}}
\newcommand{\tmaj}   {\ensuremath{T_{\mathrm{maj.}}}}
\newcommand{\tmin}   {\ensuremath{T_{\mathrm{min.}}}}
\newcommand{\cp}     {\ensuremath{C}}
\newcommand{\dpar}   {\ensuremath{D}}
\newcommand{\bcf}    {\ensuremath{\overline{C}_F}}
\newcommand{\bca}    {\ensuremath{\overline{C}_A}}
\newcommand{\cf}     {\ensuremath{C_F}}
\newcommand{\ca}     {\ensuremath{C_A}}
\newcommand{\tf}     {\ensuremath{T_F}}
\newcommand{\nc}     {\ensuremath{N_C}}
\newcommand{\nf}     {\ensuremath{n_f}}
\newcommand{\anull}  {\ensuremath{\alpha_0}}

\newcommand{\anulltwo} {\ensuremath{\anull(2~\mathrm{GeV})}}
\newcommand{\bm}[1]  {\mbox{\boldmath\ensuremath{#1}}}
\newcommand{\order}[1]  {\mbox{\ensuremath{{\cal O}(#1)}}}
\newcommand{\mui}    {\ensuremath{\mu_{\mathrm{I}}}}
\newcommand{\mur}    {\ensuremath{\mu_{\mathrm{R}}}}
\newcommand{\muq}    {\ensuremath{\mu_{\mathrm{q}}}}

\newcommand{\roots}  {\ensuremath{\sqrt{s}}}
\newcommand{\rootsp} {\ensuremath{\sqrt{s'}}}
\newcommand{\znull}  {\ensuremath{\mathrm{Z^0}}}
\newcommand{\wpm}    {\ensuremath{\mathrm{W^{\pm}}}}
\newcommand{\mz}     {\ensuremath{m_{\znull}}}
\newcommand{\mtau}   {\ensuremath{m_{\tau}}}
\newcommand{\mhiggs} {\ensuremath{m_{\mathrm{H}}}}
\newcommand{\gz}     {\ensuremath{\Gamma_{\znull}}}
\newcommand{\zgstar} {\ensuremath{(\znull/\gamma)^*}}
\newcommand{\asmz}   {\ensuremath{\as(\mz)}}
\newcommand{\asmt}   {\ensuremath{\as(\mtau)}}
\newcommand{\ycut}   {\ensuremath{y_{\mathrm{cut}}}}
\newcommand{\mb}     {\ensuremath{m_{\mathrm{b}}}}
\newcommand{\mbmz}   {\ensuremath{\mb(\mz)}}
\newcommand{\ddel}   {\ensuremath{\mathrm{d}}}
\newcommand{\xmu}    {\ensuremath{x_{\mu}}}
\newcommand{\msbar}  {\ensuremath{\overline{\mathrm{MS}}}}
\newcommand{\lmsbar} {\ensuremath{\Lambda_{\msbar}}}
\newcommand{\lmqcd}  {\ensuremath{\Lambda_{\mathrm{QCD}}}}
\newcommand{\sigtot} {\ensuremath{\sigma_{\mathrm{tot}}}}
\newcommand{\sigt}   {\ensuremath{\sigma_{\mathrm{T}}}}
\newcommand{\sigl}   {\ensuremath{\sigma_{\mathrm{L}}}}
\newcommand{\signull}{\ensuremath{\sigma_{\mathrm{0}}}}
\newcommand{\ymax}   {\ensuremath{y_{\mathrm{max}}}}
\newcommand{\ytwothree} {\ensuremath{y_{23}}}
\newcommand{\ytwothreed} {\ensuremath{\ytwothree^{\mathrm{D}}}}
\newcommand{\nch}    {\ensuremath{\momone{n_{\mathrm{ch.}}}}}
\newcommand{\rguds}  {\ensuremath{R_{\mathrm{g/uds}}}}
\newcommand{\rgb}    {\ensuremath{R_{\mathrm{g/b}}}}
\newcommand{\rgq}    {\ensuremath{R_{\mathrm{g/q}}}}
\newcommand{\degr}   {\ensuremath{^{\mathrm{o}}}}
\newcommand{\kshort} {\ensuremath{\mathrm{K_S^0}}}
\newcommand{\kplus}  {\ensuremath{\mathrm{K^+}}}
\newcommand{\pizero} {\ensuremath{\pi^0}}
\newcommand{\stat}   {\ensuremath{\mathrm{(stat.)}}}
\newcommand{\syst}   {\ensuremath{\mathrm{(syst.)}}}
\newcommand{\expt}   {\ensuremath{\mathrm{(exp.)}}}
\newcommand{\had}    {\ensuremath{\mathrm{(had.)}}}
\newcommand{\theo}   {\ensuremath{\mathrm{(theo.)}}}
\newcommand{\nonp}   {\ensuremath{\mathrm{(non-pert.)}}}
\newcommand{\evol}   {\ensuremath{\mathrm{(evol.)}}}
\newcommand{\chisq}  {\ensuremath{\chi^2}}
\newcommand{\chisqd} {\ensuremath{\chi^2/\mathrm{d.o.f.}}}
\newcommand{\qbarjet}{\ensuremath{\overline{Q}_{\mathrm{jet}}}}
\newcommand{\qjet}   {\ensuremath{Q_{\mathrm{jet}}}}
\newcommand{\ejet}   {\ensuremath{E_{\mathrm{jet}}}}
\newcommand{\gcc}    {\ensuremath{g_{\ccbar}}}
\newcommand{\gbb}    {\ensuremath{g_{\bbbar}}}
\newcommand{\dstar}  {\ensuremath{\mathrm{D}^{*}}}
\newcommand{\evis}   {\ensuremath{E_{\mathrm{vis}}}}
\newcommand{\chibz}  {\ensuremath{\chi_{\mathrm{BZ}}}}
\newcommand{\momone}[1] {\mbox{\ensuremath{\langle#1\rangle}}}
\newcommand{\momn}[2] {\mbox{\ensuremath{\langle#1^{#2}\rangle}}}
\newcommand{\lnr}    {\ensuremath{\ln(R)}}
\newcommand{\bthree} {\ensuremath{\mathrm{B}_3}}
\newcommand{\rthree} {\ensuremath{\mathrm{R}_3}}
\newcommand{\rthreeb}{\ensuremath{\mathrm{R^b}_3}}
\newcommand{\rthreel}{\ensuremath{\mathrm{R^l}_3}}
\newcommand{\kperp}  {\ensuremath{k_{\mathrm{t}}}}
\newcommand{\mbmb}   {\ensuremath{\mb(\mb)}}
\newcommand{\fpt}    {\ensuremath{F_{\mathrm{PT}}}}
\newcommand{\cy}     {\ensuremath{c_{\mathrm{y}}}}
\newcommand{\rz}     {\ensuremath{R_{\mathrm{Z}}}}
\newcommand{\rhad}   {\ensuremath{R_{\epem}}}
\newcommand{\rtau}   {\ensuremath{R_{\tau}}}
\newcommand{\gammaz} {\ensuremath{\Gamma_{\mathrm{Z}}}}
\newcommand{\gammah} {\ensuremath{\Gamma_{\mathrm{h}}}}
\newcommand{\gammainv} {\ensuremath{\Gamma_{\mathrm{inv}}}}
\newcommand{\gammal} {\ensuremath{\Gamma_{\ell}}}
\newcommand{\sigmah} {\ensuremath{\sigma_{\mathrm{h}}}}
\newcommand{\sigmal} {\ensuremath{\sigma_{\ell}}}
\newcommand{\rfour}  {\ensuremath{\mathrm{R}_4}}
\newcommand{\rfourd} {\ensuremath{\mathrm{R_4(D)}}}
\newcommand{\rfourc} {\ensuremath{\mathrm{R_4(C)}}}
\newcommand{\kaph}   {\ensuremath{\kappa_{\mathrm{H}}}}
\newcommand{\xe}     {\ensuremath{x_{\mathrm{E}}}}
\newcommand{\nchgg}  {\ensuremath{N^{\mathrm{ch.}}_{\mathrm{gg}}}}

\newcommand{\ksi}    {\ensuremath{\xi}}
\newcommand{\ksistar}{\ensuremath{\xi^*}}
\newcommand{\delbl}  {\ensuremath{\delta_{\mathrm{bl}}}}

\begin{document}


\title{ Tests of Quantum Chromo Dynamics at \bm{\epem} Colliders }

\author{ Stefan Kluth }

\address{ Max-Planck-Institut f\"ur Physik, F\"ohringer Ring 6, D-80805
Munich, Germany }

\begin{abstract}

The current status of tests of the theory of strong interactions,
Quantum Chromo Dynamics (QCD), with data from hadron production in
\epem\ annihilation experiments is reviewed.  The LEP experiments ALEPH,
DELPHI, L3 and OPAL have published many analyses with data recorded on
the \znull\ resonance at $\roots=91.2$~GeV and above up to
$\roots>200$~GeV.  There are also results from SLD at
$\roots=91.2$~GeV and from reanalysis of data recorded by the JADE
experiment at $14\le\roots\le 44$~GeV.  The results of studies of jet
and event shape observables, of particle production and of quark gluon
jet differences are compared with predictions by perturbative QCD
calculations.  Determinations of the strong coupling constant \asmz\
from jet and event shape observables, scaling violation and
fragmentation functions, inclusive observables from \znull\ decays,
hadronic $\tau$ decays and hadron production in low energy \epem\
annihilation are discussed.  Updates of the measurements are performed
where new data or improved calculations have become available.  The
best value of \asmz\ is obtained from an average of measurements using
inclusive observables calculated in NNLO QCD:
\begin{displaymath}
  \asmz = 0.1211 \pm 0.0021
\end{displaymath}
where the error is dominated by theoretical systematic uncertainties.
The other measurements of \asmz\ are in good agreement with this
value.  Finally, investigations of the gauge structure of QCD are
summarised and the best values for the colour factors are determined:
\begin{eqnarray} \nonumber
  \ca & = & 2.89 \pm 0.21 \\ \nonumber
  \cf & = & 1.30 \pm 0.09 \nonumber
\end{eqnarray}
with errors dominated by systematic uncertainties and in good
agreement with the expectation from QCD with the SU(3) gauge symmetry.

\end{abstract}


\maketitle

\tableofcontents

\newpage

\section{Introduction}
\label{sec_intro}

The interactions between the constituents of matter are successfully
described by the four forces: the weak, electromagnetic and strong
forces and the gravitational force.  The weak and the strong
interactions occur at small atomic to subatomic distances, the
electromagnetic interaction is observed at subatomic to macroscopic
distances while effects of Gravitation only play a role at macroscopic
distances.

The strong interaction, the main focus of this review, is responsible
for the existence of all composite elementary particles (hadrons) by
providing the binding force between the constituents and also for most
of the short lived hadron decays.  Furthermore, the binding of protons
and neutrons in nuclei may be explained in analogy to chemical binding
of molecules based on the strong interaction of the proton and neutron
constituents.

The constituents of hadrons are known as partons or quarks.  The known
spectrum of hadrons can be explained as composites of five (out of a
total of six) quark flavours, where each flavour occurs in three
variants distinguished by the so-called colour.  In
table~\ref{tab_quarks} the basic properties of the six known quarks
are shown.

\begin{table}[htb!]
\caption[ bla ]{ Basic properties of the six quarks in the standard model }
\label{tab_quarks}
\begin{indented}\item[]
\begin{tabular}{ccccc} 
\hline
Generation & 1st & 2nd & 3rd & Charge \\ 
\hline
up-type    & u (up)  & c (charm) & t (top) & +2/3 \\
down-type  & d (down) & s (strange) & b (bottom) & $-$1/3 \\
\hline
\end{tabular}
\end{indented}
\end{table}

A dynamic theory of strong interactions at the constituent level,
Quantum Chromo Dynamics (QCD), is constructed as a renormalised field
theory in close analogy to Quantum Electro Dynamics (QED), the quantum
field theory of the electromagnetic interaction, see
e.g.~\cite{fritzsch73,gross73a,gross73b,politzer73,ellis96,muta87}.
The quarks are viewed as carriers of three different strong charges,
referred to as colours.  Interactions between colour charged quarks
are mediated by gluons, in analogy to the exchange of photons in QED.
The theory is constructed to remain invariant under exchange of the
three colour charges, i.e.\ local gauge transformations in the colour
space with the SU(3) symmetry.  From the requirement of local gauge
invariance under SU(3) the special properties of the gluons are
derived: there are eight gluons each carrying a colour charge and
anti-charge and thus the gluons can directly interact with each other.
The gluon-gluon interactions of QCD are not present in QED; QCD is
referred to as a non-abelian gauge theory while QED is an example of
an abelian gauge theory.  For massless quarks the strong coupling
constant \as\ is the only free parameter of the theory.

Qualitatively, some important properties of QCD already follow from
the possibility of direct gluon-gluon interactions.  In QED, the scale
dependence of the coupling constant $\alpha$, the so-called running,
may be seen as a consequence of screening of the bare electric charge
due to vacuum polarisation.  At small momentum scales, the exchanged
photon resolves only a large volume around the bare charge and thus is
exposed to stronger charge screening.  At large momentum scales a
smaller volume is resolved and less charge screening occurs leading to
a rising value of the coupling constant.  In QCD the effect has the
opposite direction, because a cloud of virtual colour charged gluons
around a bare colour charge effectively spreads the colour charge over
a volume around the quark carrying the bare colour charge.  Thus at
small momentum scales exchanged gluons interact with an apparently
stronger colour charge while at large momentum scales a
correspondingly smaller fraction of the colour charge is resolved.  In
the limit of infinite momentum scales the resolved charge would
vanish, leading to the prediction of asymptotic freedom for quarks.
At very small momentum scales the resolved colour charge becomes so
large that quarks are tightly bound into hadrons; this explains the
observation that quarks are confined into hadrons and cannot be
observed as free particles in an experiment.

The process of electron positron annihilation into hadrons,
$\epem\rightarrow\mathrm{hadrons}$, is an ideal laboratory for studies
of strong interaction phenomena at the parton, i.e. quark and gluon
level, for several reasons:

\begin{itemize}

\item
There is no interference between the initial and the
final state with the consequence that interpretation of the hadronic
final state in terms of QCD processes is simplified.

\item
The four-momentum of the initial state is, in the absence of
significant QED initial state radiation (ISR), fully transferred to
the final state.  In most experiments the electron and positron beams
have equal energies such that the centre of mass system (cms) of the
final state coincides with the laboratory frame.

\item
The experimental conditions in \epem\ annihilation are usually rather
clean, because in past and existing colliders there were no multiple
interactions in a given bunch crossing and backgrounds coming directly
from the accelerator like lost beam particles or synchrotron
radiation were kept at a low level in the experiments.

\item
The presently available \epem\ annihilation data allow detailed
studies of QCD in a large range of cms energies covering more than an
order of magnitude from 12~GeV to 209~GeV.  This unique collection of
data forms in particular the basis of tests of the energy scale
dependence of QCD predictions.

\end{itemize}

\begin{figure}[htb!]
\includegraphics[width=1.0\textwidth, bb=0 70 567 385]{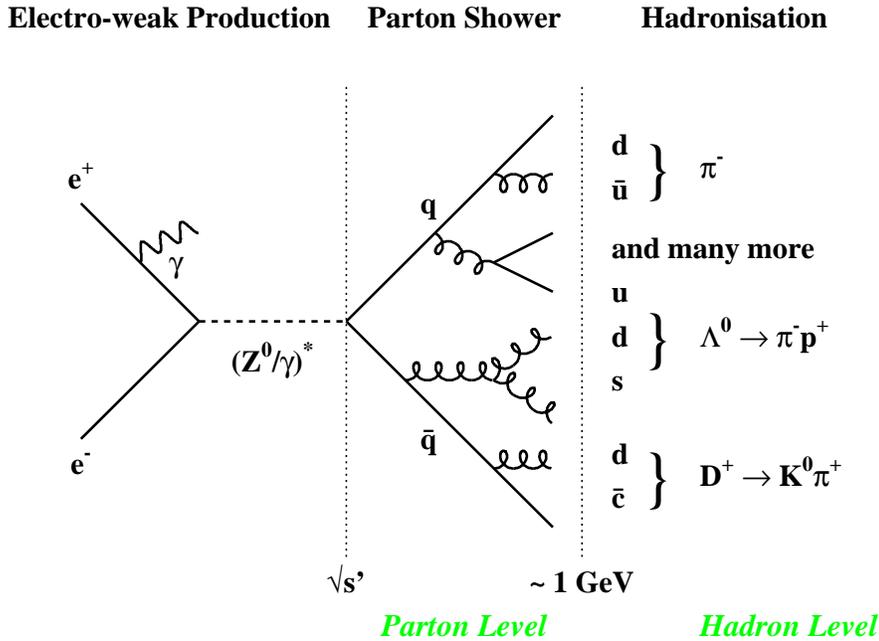}
\caption[ bla ]{ Schematic view of a
$\epem\rightarrow\mathrm{hadrons}$ event; \rootsp\ denotes the
invariant mass of the hadronic system, usually close the nominal cms
energy \roots\ of the experiment. }
\label{fig_qcdshower}
\end{figure}

Figure~\ref{fig_qcdshower} presents an overview of our current
understanding of the process $\epem\rightarrow\mathrm{hadrons}$.  The
incoming electron and electron annihilate into an intermediate vector
boson, shown as \zgstar, possibly after initial state radiation.  The
quark pair produced in the vector boson decay will start to radiate
gluons which in turn radiate gluons or turn into quark pairs
themselves.  In this way a so-called parton shower develops until all
parton interactions happen at low energy scales of about 1~GeV where
confinement and thus hadron formation sets in.  The parton shower
occupies a volume with radius $\sim 10^{-1}$~fm, according to the
uncertainty principle.

This review will begin in section~\ref{sec_basics} with a brief
introduction into fundamentals of QCD predictions and
section~\ref{sec_hist} gives a short overview of milestones in the
history of QCD tests in \epem\ annihilation together with a summary of
the \epem\ colliders and experiments which produced the currently
available data.  The following section~\ref{sec_jetsshapes} discusses
studies based on jet reconstruction algorithms and event shape
observables.  The subject of section~\ref{sec_incl} is inclusive
particle production and section~\ref{sec_inclobs} presents results
from inclusive observables.  A summary of determinations of the strong
coupling \as\ is given in section~\ref{sec_as} while the properties of
jets originating from quarks or gluons are discussed in
section~\ref{sec_qg} Studies of the gauge structure of QCD are
reviewed in section~\ref{sec_gauge}, followed by conclusions and
outlook in section~\ref{sec_conc}.

\section{Basics of QCD and hadron production}
\label{sec_basics}

\begin{figure}[htb!]
\includegraphics[width=1.0\textwidth, bb=0 90 567 210]{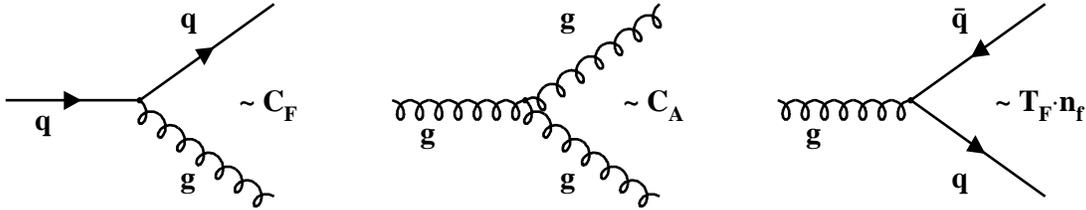}
\caption[ bla ]{ Feynman diagrams of the three fundamental QCD
processes. }
\label{fig_qcdprocs}
\end{figure}

We collect here the important predictions of QCD which can be studied
with data from hadron production in \epem\ annihilation.  In QCD with
the SU(3) gauge symmetry the following fundamental processes are
possible~\cite{muta87,ellis96}:

\begin{description}

\item[Gluon radiation from quarks]
This process is the analog of photon bremsstrahlung and occurs with
relative strength $\cf=4/3$.

\item[Quark pair production]
This process is the analog of e.g. \epem\ pair production from a
photon and has a relative strength $\tf=1/2$.  However, it is possible
to create quark pairs of all kinematically accessible quark flavours
\nf\ and thus the contribution of this process is effectively $\tf\nf$.

\item[Triple gluon vertex]
This process is unique to QCD because the gluons are themselves colour
charged, the relative strength is $\ca=3$.  

\end{description}
There is in addition the quartic gluon vertex which is at least
\order{\as^3} in \epem\ annihilation and known to be a negligible
contribution to radiative corrections to lower order
processes~\cite{nagy98a}.  Figure~\ref{fig_qcdprocs} shows the Feynman
diagrams of the three fundamental processes.

The overall strength of the strong coupling is given by a constant of
nature, the strong coupling constant \as.  This constant is the only
free parameter of the theory when quarks are treated as massless.
Therefore many studies of QCD can be formulated as measurements of the
strong coupling constant and comparisons of the results for \as\ of
different analyses give insight into the consistency of the theory.

\subsection{Everything runs}
\label{sec_qcdrun}

\subsubsection{Renormalisation scale dependence}
\label{sec_rgitheo}

The predictions of QCD are finite, because the theory has been
renormalised in order to remove divergent terms.  The renormalisation
introduces a dependence of QCD predictions on the energy scale $\mu$,
at which the renormalisation is performed.  The renormalisation scale
$\mu$ is arbitrary and physical results should be independent of
$\mu$.  Several prescriptions for renormalisation exist, the
so-called renormalisation schemes, and in this report we use the
widely adopted \msbar\ renormalisation scheme~\cite{muta87,ellis96}
throughout.

A QCD prediction of an observable quantity $R$ measured in \epem\
annihilation, e.g.\ a differential cross section, can be written as: 
\begin{equation}
  R = \sum_{n=0}^{\infty} R_n \as^n\;\;,
\label{equ_r}
\end{equation}
where the $R_n$ are the coefficients in $n$th order of the
perturbation series.  The quantity $R$ depends on \as\ and the ratio
$\xmu=Q/\mu$ of the physical scale $Q$ of the process and the
renormalisation scale $\mu$, for reasons of consistency of the
dimension of $R$.  In the limit $n\rightarrow\infty$ this dependence
on the renormalisation scale is expected to vanish; however, for
truncated perturbative calculations a residual dependence will remain.
The requirement of renormalisation scale independence is formally
expressed as the renormalisation group equation (RGE), also known as
Callan-Symanzik equation~\cite{callan70,symanzik70}, see
e.g.~\cite{ellis96}:
\begin{equation}
  \mu^2 \frac{\ddel}{\ddel\mu^2}R\left(\xmu^2,\as \right) =
  \mu^2 \left( \frac{\partial}{\partial\mu^2} + 
    \frac{\partial\as}{\partial\mu^2}\frac{\partial}{\partial\as}
  \right) R = 0\;\;.
\label{equ_rge}
\end{equation}
The renormalisation scale dependence of the coefficients $R_n$ will be
compensated by a renormalisation scale dependence of the coupling
constant $\as(\mu)$:
\begin{equation}
  \mu^2\frac{\partial\as}{\partial\mu^2} = 
  \beta(\as(\mu^2)) = 
  -\beta_0\as^2(\mu^2)-\beta_1\as^3(\mu^2)-\beta_2\as^4(\mu^2)
  +\order{\as^5}\;\;.
\label{equ_beta}
\end{equation}

Equation~(\ref{equ_beta}) introduces the $\beta$-function of QCD, with
coefficients~\cite{ellis96}
\begin{eqnarray}
\label{equ_betacoeffs}
  \beta_0 & = & \frac{11\ca-4\tf\nf}{12\pi}\;\;, \\ \nonumber
  \beta_1 & = & \frac{17\ca^2-5\ca\nf-3\cf\nf}{24\pi^2}
    \;\;\mathrm{and} \\ \nonumber
  \beta_2 & = & \frac{2857\ca^3+(54\cf^2-615\cf\ca-1415\ca^2)\nf+
                      (66\cf+79\ca)\nf^2}{3456\pi^3}\;\;.
\end{eqnarray}
The coefficient $\beta_2$ depends on the renormalisation scheme and is
shown here for the \msbar\ scheme.  The variable \nf\ specifies the
number of active quark flavours which are considered in the
calculation and thus from kinematics it depends on the energy scale of
the process.  At low energies below the thresholds for heavy quark
production direct contributions as well as contributions to radiative
corrections are suppressed resulting in $\nf=3$.  When with increasing
energy scale a heavy quark production threshold is crossed \nf\ must
be increased by one unit, see section~\ref{sec_heavythresh} below.

Combining equations~(\ref{equ_r}), (\ref{equ_rge}) and
(\ref{equ_beta}) results in
\begin{eqnarray}
\label{equ_rscale}
0 & = & \mu^2\frac{\partial R_0}{\partial\mu^2}+
  \as(\mu^2)\mu^2\frac{\partial R_1}{\partial\mu^2}+
  \as^2(\mu^2)\left(\mu^2\frac{\partial R_2}{\partial\mu^2}-
                    R_1\beta_0\right)+ \\ \nonumber
&&\as^3(\mu^2)\left(\mu^2\frac{\partial R_3}{\partial\mu^2}-
        (R_1\beta_1+2R_2\beta_0) \right) + \ldots \;\;. 
\end{eqnarray}
A solution for equation~(\ref{equ_rscale}) can be found by demanding
that the coefficients of $\as^n$ vanish for all orders $n$.  After
integration from a specific choice of renormalisation scale $\mur^2$ 
to another scale $Q^2$ the coefficients can be written as
\begin{eqnarray}
\label{equ_rcoeff}
  R_0 & = & const.\;\;, \\ \nonumber
  R_1 & = & const.\;\;, \\ \nonumber
  R_2(\xmu^2) & = & R_2(1)+\beta_0R_1\ln\xmu^2\;\;\mathrm{and} \\ \nonumber
  R_3(\xmu^2) & = & R_3(1)+(2R_2(1)\beta_0+R_1\beta_1)\ln\xmu^2+
                           R_1\beta_0^2\ln^2\xmu^2
\end{eqnarray}
with $\xmu=Q/\mur$.  The $R_i(1)$ are the coefficients evaluated at
the renormalisation scale \mur.  We note that all coefficients $R_n,
n\geq2$ depend explicitly on the renormalisation scale parameter \xmu.
However, the coefficients $R_n(\xmu)$ only depend on \xmu\ via terms
containing $\ln\xmu^2$.  This observation has the important
consequence that QCD predictions for physical observables at any scale
given by \xmu\ can be computed in terms of constant coefficients
derived using the renormalisation scale given at $\xmu=1$.  The scale
dependence of the prediction then enters only through the scale
dependent strong coupling constant $\as(Q^2)$.

The choice of the renormalisation scheme (RS) is arbitrary and usually
dictated by convenience to perform a calculation.  It is also evident
that a result of a QCD prediction for a physical observable like a
cross section should in principle not depend on the choice of RS.  In
practice with perturbative predictions truncated at e.g.\ 2nd or 3rd
order there are significant dependences on the RS which may be
interpreted as indications of the size of missing higher order terms.

This problem and attempts to provide solutions have already been
discussed more than 20 years ago when the first next-to-leading (NLO)
QCD calculations became
available~\cite{grunberg80,grunberg84,stevenson81,brodsky83,dhar84}.
Three different ways to choose an ``optimal'' renormalisation scheme
or scale for sufficiently inclusive observables depending on a single
energy scale were devised, see e.g.~\cite{chyla95} for a review.
Examples for such observables are mean values (or higher moments) of
event shape distributions or $R_{\epem}$.  It is important to note
that at NLO a variation of the renormalisation scale is equivalent to
a variation of the RS yielding the same results~\cite{stevenson82}.
The three methods for choosing an optimal renormalisation scheme are
the following:

\begin{description}

\item[Principle of minimum sensitivity (PMS)] The principle of
minimum sensitivity is based on the observation that at a given order
all possible RSs can be labelled by their renormalisation point (scale)
\xmu\ and the coefficients of their
$\beta$-functions~\cite{stevenson81}.  Now one may find the RS for
which a QCD prediction $R(\as)$ for a physical observable has minimal
sensitivity w.r.t.\ choosing a RS.  This corresponds to a stationary
point of the function $R(\as,l_i)$ with the RS labels $l_i$.  At NLO
finding a stationary point of $R(\as,\xmu)$ is sufficient, since at
this order the coefficients of the $\beta$-function are universal.

\item[Method of effective charges (ECH)] Using the freedom to choose the
RS the perturbative expansion is rearranged such that higher order
terms vanish.  The running coupling absorbs all scale dependent
effects and becomes a process dependent so-called effective charge
which satisfies a generalised
$\beta$-function~\cite{grunberg80,grunberg84}.  This method is also
referred to as ``fastest apparent convergence'' or FAC.  At NLO a
convenient way to find the FAC scheme is to find the value of \xmu\
for which the complete \oaa\ term vanishes.

\item[Brodsky-Lepage-Mackenzie method (BLM)] The BLM method proposes
to adjust the renormalisation point (scale) \xmu\ for a given RS such
that the dependence of the NLO term on the number of quark flavours
\nf\ is cancelled~\cite{brodsky83,brodsky95}.  This prescription 
implies that vacuum polarisation corrections due to fermion pairs are
absorbed by the running coupling \asmu.

\end{description}

\subsubsection{The running strong coupling constant}
\label{sec_asrun}

The scale dependence or running of the strong coupling constant can be
derived by integrating equation~(\ref{equ_beta}) to obtain an
expression relating values of \as\ at two different scales $Q$ and
$\mur$.  Since the choice of renormalisation scale is arbitrary the
expressions can be used to relate values of \as\ at any two scales.
Solutions differ according to how many orders (often referred to as
the number of loops) in perturbation theory have been considered in
the calculation of the $\beta$-function.  The first term corresponds
to one order (loop), the second to two orders (loops), etc.  The
solutions valid for one and two loops are:
\begin{equation}
  \as(Q^2)= \frac{\as(\mur^2)}{1+\as(\mur^2)\beta_0\ln\xmu^2}\;\;\mathrm{and}
\label{equ_asrunone}
\end{equation}
\begin{equation}
  \beta_0\ln\xmu^2 = \frac{1}{\as(Q^2)}-\frac{1}{\as(\mur^2)}+
                     \frac{\beta_1}{\beta_0}
                     \ln\left(\frac{\as(Q^2)}{\as(\mur^2)}\cdot
         \frac{\beta_0+\beta_1\as(\mur^2)}{\beta_0+\beta_1\as(Q^2)}\right)\;\;.
\label{equ_asruntwo}
\end{equation}
The 2-loop solution shown in equation~(\ref{equ_asruntwo}) may be solved
numerically for $\as(Q^2)$.  The solution for a 3-loop $\beta$-function 
is also found directly by integrating equation~(\ref{equ_beta}):
\begin{equation}
  \beta_0\ln\xmu^2 = F(\as(Q)) - F(\as(\mur))\;\;.
\label{equ_asrunthree}
\end{equation}
The function $F(\as) = \beta_0 \int
-1/(\beta_0\as^2+\beta_1\as^3+\beta_2\as^4)\, \ddel\as$ is for $\nf\leq5$,
i.e.\ $\beta_2>0$, given by\footnote{Integration courtesy of
\texttt{integrals.wolfram.com}.}
\begin{eqnarray}
\label{equ_fas}
 F(\as) & = & \frac{1}{\as} 
 - \frac{\beta_1^2-2\beta_0\beta_2}{\beta_0\sqrt{-\beta_1^2+4\beta_0\beta_2}}
   \arctan\left(\frac{\beta_1+2\beta_2\as}{\sqrt{-\beta_1^2+4\beta_0\beta_2}}\right) \\ \nonumber
 & & + \frac{\beta_1\log(\as)}{\beta_0} 
 - \frac{\beta_1\log(\beta_0+\beta_1\as+\beta_2\as^2)}{2\beta_0}\;\;. 
\end{eqnarray}
As before equation~(\ref{equ_asrunthree}) may be solved numerically
for \asq. For the case $\nf\geq6$ we have $\beta_2<0$ and the second
term of the RHS of equation~(\ref{equ_fas}) is rewritten to contain
only real terms with
$z=(\beta_1+2\beta_2\as)/\sqrt{\beta_1^2-4\beta_0\beta_2}$:
\begin{equation}
 - \frac{\beta_1^2-2\beta_0\beta_2}{\beta_0\sqrt{\beta_1^2-4\beta_0\beta_2}}
   \frac{1}{2}\log\frac{1+z}{1-z} \;\;.
\end{equation}
The case $\nf=6$ becomes relevant when an evolution of \as\ to energy
scales above the threshold for t quark production is performed.

The perturbative running of the strong coupling constant breaks down
at sufficiently small scales $Q$ such that e.g.\ in LO
$\as(\mur^2)\beta_0\ln\xmu^2=-1$.  At such low scales values of the
strong coupling are $\as>1$ and thus perturbative expansions in \as\
will fail to converge.  The value of $Q=\lmqcd$ where this breakdown
of perturbative QCD occurs is known as the Landau pole.  Since typical
light hadron masses are \order{100}~MeV one might expect that \lmqcd\
has similar values.  Using the LO equation~(\ref{equ_asrunone}) this
yields $\asmz\approx 0.1$ and $\as(1~\mathrm{GeV}^2)\approx0.2$ which
sets a rough lower limit for the applicability of perturbative QCD.

\subsubsection{Running quark masses}
\label{sec_runmq}

QCD predictions discussed so far are valid for massless quarks.
However, it is well known that quarks have masses ranging from
\order{1}~MeV for the light u and d quarks to about 174~GeV for the t
quark~\cite{pdg04}.  Mass effects are incorporated into the theory via
mass terms in the Lagrangian which are subject to renormalisation like
the coupling constant.  The requirement of independence of QCD
predictions from the renormalisation scale introduces in the presence
of a quark mass a new term into the RGE equation~(\ref{equ_rge}):
\begin{equation}
  \mu^2 \left( \frac{\partial}{\partial\mu^2} + 
      \frac{\partial\as}{\partial\mu^2}\frac{\partial}{\partial\as} +
      \frac{\partial m}{\partial\mu^2}\frac{\partial}{\partial m}
  \right)R = 0
\label{equ_rgem}
\end{equation}
with
\begin{equation}
  \mu^2 \frac{\partial m}{\partial\mu^2} = -\gamma_m(\mu^2)m(\mu^2)
\label{equ_anomdim}
\end{equation}
where $\gamma_m(\mu^2)$ is the so-called mass anomalous dimension.
The mass anomalous dimension $\gamma_m$ has been calculated as a power
series in the strong coupling constant,
$\gamma_m(\as)=\gamma_0\as+\gamma_1\as^2+\order{\as^3}$, with
coefficients $\gamma_0=1/\pi$ and $\gamma_1=(303-10\nf)/(72\pi^2)$ in
the \msbar\ renormalisation scheme~\cite{vermaseren97}. 

Integrating equation~(\ref{equ_anomdim}) and substituting $\as(\mur^2)$
for $\mur^2$ gives an equation for the running quark mass:
\begin{equation}
  m(Q^2)= m(\mur^2)\exp\left( \int_{\as(\mur^2)}^{\as(Q^2)}
  \frac{\gamma_m(\as)}{\beta(\as)} \ddel\as \right)\;\;.
\label{equ_runmq}
\end{equation}
Solutions to equation~(\ref{equ_runmq}) in the \msbar\ renormalisation
scheme are e.g.\ given in~\cite{vermaseren97}.  

\subsubsection{Heavy quark thresholds}
\label{sec_heavythresh}

The evolution of the strong coupling constant,
equation~(\ref{equ_beta}), and the quark masses,
equation~(\ref{equ_runmq}), depends on the number of active quark
flavours \nf\ through the coefficients of the QCD $\beta$-function and
mass anomalous dimension, respectively.  When an evolution of e.g.\
$\as(\mu^2)$ across the excitation threshold \muq\ for production of
heavy quark pairs is attempted an explicit treatment of the changing
number of flavours, $\nf\rightarrow\nf+1$, is necessary.

One requires that the theory at scales below the heavy quark threshold
\muq\ with \nf\ active quark flavours is consistent with the theory at
scales above \muq\ with $\nf+1$ active quark
flavours~\cite{chetyrkin97b}.  This results in matching conditions for
the strong coupling constant at the heavy quark threshold; these are
$\as(\nf)=\as(\nf+1)$ in LO (leading order) and NLO.  At the next
order the matching involves a discontinuity of the value of the strong
coupling constant at the heavy quark threshold
\muq:
\begin{equation}
  \as(\nf,\muq) = \as(\nf+1,\muq) + \as^3(\nf+1,\muq)c_2\;\;,
\label{equ_asmatch}
\end{equation}
where the value of the coefficient $c_2$ depends on the definition of
the heavy quark mass.  For the running \msbar\ mass, i.e.\
$\muq=m_{\mathrm{q},\msbar}(m_{\mathrm{q},\msbar})$, one has
$c_2=11/(72\pi^2)$ while for the pole mass $M_{\mathrm{q}}$ with
$\muq= M_{\mathrm{q}}$ one has $c_2=-7/(24\pi^2)$~\cite{chetyrkin97b}.

In this report we use 3-loop running of the strong coupling constant
as explained in section~\ref{sec_asrun} together with matching at
heavy quark thresholds using equation~(\ref{equ_asmatch}) and \msbar\
masses $m_{c,\msbar}(m_{c,\msbar})=1.25$~GeV and
$m_{b,\msbar}(m_{b,\msbar})=4.25$~GeV~\cite{pdg04} when a value of
\as\ is evolved from an energy scale $Q_1$ to another scale $Q_2$.

\subsection{Perturbative QCD}
\label{sec_pqcd}

Predictions of perturbative QCD are possible for observables which
fulfil the Sterman-Weinberg criteria of infrared and collinear
safety~\cite{sterman77}:  an observable is infrared and collinear safe
when its value is not affected by the emission of low-momentum
partons or by the replacement of a parton by collinear partons with
the same total 4-momentum.  For such observables predictions as power
series in the strong coupling \as\ may be performed.  

The observables typically used in studies of \epem\ annihilation to
hadrons may be classified as follows:
\begin{description}

\item[exclusive]
Exclusive observables derived from jet clustering algorithms or based
on event shape definitions (see section~\ref{sec_jetsshapes} for
details) classify the hadronic final states according to their
topology.  Measurements of differential cross sections for such
observables may be compared with corresponding QCD predictions.

\item[semi-inclusive]
With semi-inclusive observables all hadronic final states are 
considered and average properties of the produced hadrons such
as particle multiplicity or momentum spectra are studied.

\item[inclusive]
The inclusive observables are simply based on counting of hadronic
final states.  Examples are the hadronic widths of the \znull\ boson
or the $\tau$ lepton or the total cross section for hadron production
in \epem\ annihilation.

\end{description}

Most exclusive observables used in hadron production from \epem\
annihilation are sensitive to the non-collinear emission of a single
energetic (or hard) gluon.  The corresponding final state is expected
to consist of three separated bundles of particles referred to as jets
and originating from the quarks and gluon produced in the \epem\
interaction.  These observables will be referred to as 3-jet
observables.  Some observables are defined such that they are only
sensitive to 4-jet final states, i.e.\ those involving one gluon
emission and one of the fundamental QCD processes listed above, and
will be referred to 4-jet observables.

\subsubsection{Fixed order predictions}

The basic prediction in NLO for the normalised
differential distribution of a 3-jet observable $y$ is given by a
power series in $\ashq=\asq/(2\pi)$:
\begin{equation}
  \frac{1}{\sigtot}\frac{\ddel \sigma}{\ddel y} =
  \frac{\ddel A}{\ddel y}\ashq +
  \left( \left(2\pi\beta_0\ln(\xmu^2)-2\right)\frac{\ddel A}{\ddel y} +
         \frac{\ddel B}{\ddel y} \right)\ash^2(Q)\;\;.
\label{equ_3jetnlo}
\end{equation}
The energy scale $Q$ at which the strong coupling is evaluated is
generally identified with the physical hard scale of the process such
as the cms energy of the \epem\ annihilation.  The coefficient
functions $\ddel A/\ddel y$ and $\ddel B/\ddel y$ correspond to the
coefficients $R_1$ and $R_2$ of equation~(\ref{equ_r}) and their
renormalisation scale dependence is given by
equation~(\ref{equ_rcoeff}) after replacing \as\ by \ash\ and
absorbing $(2\pi)^n$ in the coefficient $R_n$.  The normalisation to
the total hadronic cross section \sigtot\ has been accounted for by
the relation $\sigtot=\signull(1+2\ash)$~\cite{ert}.  The coefficient
functions $\ddel A/\ddel y$ and $\ddel B/\ddel y$ may be derived
numerically for any suitable observable~\cite{yellow1qcd,event2}.  The
derivation is done by Monte Carlo integration of the NLO QCD matrix
elements for the production of up to four partons~\cite{ert} over
contours in phase space given by the definition of the observable $y$.
In some cases the LO terms $\ddel A/\ddel y$ have been obtained
analytically, see e.g.~\cite{ellis96}.

A third order term $\sim \ddel C/\ddel y\ash^3$ can be added to
equation~(\ref{equ_3jetnlo}) once a numerical integration of the
corresponding next-to-next-to-leading order (NNLO) QCD matrix elements
is available to generate the NNLO coefficient function $\ddel C/\ddel
y$~\cite{gehrmann-deridder04,gehrmann-deridder05}.  The
renormalisation scale dependence will be expressed as in
equation~(\ref{equ_rcoeff}) while the normalisation to the total
hadronic cross section must now consider
$\sigtot=\signull(1+2\ash+5.6368\ash^2)$~\cite{gorishnii91}.  The
complete result for the NNLO term is
\begin{eqnarray}
\label{equ_nnlo}
&&\left( \frac{\ddel C}{\ddel y} + \left( 4\pi\beta_0 \frac{\ddel B}{\ddel y}
+ (2\pi)^2\beta_1\frac{\ddel A}{\ddel y} \right)\ln\xmu^2 
+\left(2\pi\beta_0\ln\xmu^2\right)^2\frac{\ddel A}{\ddel y} 
\right. \\ \nonumber
&& \left. -2\left(\frac{\ddel B}{\ddel y}+2\pi\beta_0\ln\xmu^2 
\frac{\ddel A}{\ddel y} \right)
-1.6368\frac{\ddel A}{\ddel y} \right)\ash^3(Q)\;\;.
\end{eqnarray}

For 4-jet observables NLO fixed order predictions are also
available~\cite{nagy98b,campbell99}.  Their form is derived from
equations~(\ref{equ_3jetnlo}) and (\ref{equ_nnlo}) by setting $\ddel
A/\ddel y=0$: 
\begin{equation}
  \frac{1}{\sigtot}\frac{\ddel \sigma}{\ddel y} =
  \frac{\ddel B}{\ddel y}\ash^2(Q) +
  \left( \left(4\pi\beta_0\ln(\xmu^2)-2\right)\frac{\ddel B}{\ddel y} +
         \frac{\ddel C}{\ddel y} \right)\ash^3(Q)\;\;.
\label{equ_4jetnlo}
\end{equation}

\subsubsection{NLLA predictions}

The NLO predictions described above work well in regions of phase
space where radiation of a single hard gluon dominates.  For
configurations with soft or collinear gluon radiation from the initial
quark-antiquark pair the Sterman-Weinberg criteria may not be
fulfilled anymore and consequently new divergent terms appear.  For
observables $y$ defined to vanish when no gluon radiation occurred,
i.e.\ in the 2-jet limit, the typical leading behaviour of the
cumulative cross section $R(y)=\int_0^y 1/\sigma\ddel\sigma/\ddel y' \,
\ddel y'$ is:
\begin{equation}
  R(y)\sim\as^n\ln^{2n}\frac{1}{y} = \as^nL^{2n}\;\;,
\end{equation}
for each order $n$ of the expansion in \as\ with
$\ln(1/y)=L$~\cite{nllathmh}.  For $y\rightarrow 0$ the simple NLO
prediction will be unreliable as $\as L^2 \ll 1$ will not hold
anymore.  A systematic analytic resummation of the leading logarithmic
terms $\sim L^{2n}$ and the next-to-leading logarithmic terms
$\sim L^{2n-1}$ to all orders in \as\ has been performed for a
special class of observables~\cite{nllathmh}.  Such observables {\em
exponentiate}, which means that $\ln R(y)\approx Lg_1(\as L)$ and
$g_1$ has a power series expansion in $\as L$~\cite{nllathmh}.  For
the cumulative cross section the following representation is possible:
\begin{eqnarray}
\label{equ_nlla}
  R(y) & = & C(\as)\Sigma(y,\as)+D(y,\as) \\ \nonumber
  C(\as) & = & 1 + \sum_{n=1}^{\infty} C_n\ash^n \\ \nonumber
  \ln\Sigma(y,\as) & = & \sum_{n=1}^{\infty}\sum_{m=1}^{n+1}
                         G_{nm}\ash^n L^m \\ \nonumber
  & = & Lg_1(\as L) + g_2(\as L) + \ldots 
\end{eqnarray}
where $\lim_{y\rightarrow 0} D(y,\as)=0$.  The function $g_1$ contains
the resummation of all leading terms $\sim\as^n L^{n+1}$ (LL) while
$g_2$ resums all next-to-leading terms $\sim\as^n L^n$ (NLL).  The
description of equation~(\ref{equ_nlla}) is expected to be valid in
the region $\as L<1$ which goes further into the 2-jet region at small
values of $y$ than the NLO prediction bounded by $\as L^2 \ll 1$.

A general numerical method for performing the resummation for a large
class of event shape observables has been presented in~\cite{banfi01}.
This makes resummation possible in cases where a purely analytical
calculation is impossible or too difficult.

\subsubsection{Matched NLLA and fixed order predictions}

The NLO and NLLA predictions described above can be combined to yield
a prediction which is valid in the 2- and 3-jet regions.  In order to
avoid double counting of terms $\sim\as$ and $\sim\as^2$ present in
both NLO and NLLA predictions one has to identify and remove such
terms from the NLLA prediction.  Table~\ref{tab_nlla} presents a
comparison of the fixed order with the NLLA prediction for the
quantity $\ln R(y)$.

\begin{table}[htb!]
\caption[ bla ]{ Comparison of NLLA and fixed order calculations.  The
dependence of the coefficient functions A, B, and C on $y$ has been
suppressed. }
\label{tab_nlla}
\begin{indented}\item[]
\begin{tabular}{ccccl}
\hline\hline
LL & NLL & subleading & non-log & fixed order \\
\hline
$G_{12}\ash L^2$   & $G_{11}\ash L$     &              & $\sim\as$
& $=A\ash$ \\
$G_{23}\ash^2 L^3$ & $G_{22}\ash^2 L^2$ & $G_{21}\ash^2 L$ & $\sim\as^2$
& $=(B-\frac{1}{2}A^2)\ash^2$ \\
$G_{34}\ash^3 L^4$ & $G_{33}\ash^3 L^3$ 
& $G_{32}\ash^3 L^2+G_{31}\ash^3L$ & $\sim\as^3$ 
& $=(C-AB+\frac{1}{3}A^3)\ash^3$ \\
 \vdots & \vdots & \vdots & \vdots & \vdots  \\ 
\hline
$Lg_1(\as L)$ & $g_2(\as L)$ & $+\ldots$ &  $+\ldots$ & $=\ln R(y)$ \\
\hline\hline
\end{tabular}
\end{indented}
\end{table}

The first two columns represent the resummation of leading and
next-to-leading logarithmic terms while the equalities in the last
column refer to summing each corresponding row.  The coefficient
functions $A(y)$, $B(y)$ and $C(y)$ correspond to the cumulative cross
sections, e.g.\ $A(y)=\int_0^y \ddel A/\ddel y'\ddel y'$ and analogously
for $B(y)$ and $C(y)$.  The normalisation of the coefficient functions
is such that $A(\ymax)=B(\ymax)=C(\ymax)=0$ where \ymax\ is the
maximum kinematically possible value of the observable.  This
normalisation generates all terms needed to account for the
normalisation of the prediction to the total hadronic cross section
\sigtot.  The $G_{nm}$ coefficients in the first two rows are
determined by expanding the functions $g_1$ and $g_2$ in terms of \as\
while subleading logarithmic terms can be determined
numerically~\cite{nllathmh,nllabtbw2,nllacp}.

The resulting matched prediction is for the cumulative cross section:
\begin{eqnarray}
\label{equ_lnr}
  \ln R(y) & = & A(y)\ash + (B(y)-\frac{1}{2}A^2(y))\ash^2 \\ \nonumber
  & & + Lg_1(\as L) + g_2(\as L) \\ \nonumber
  & & - (G_{12}L^2+G_{11}L)\ash - (G_{32}L^2+G_{22}L)\ash^2\;\;.
\end{eqnarray}
Renormalisation scale dependence and normalisation to the total
hadronic cross section can be inserted into equation~(\ref{equ_lnr})
by making the replacements $B(y)\rightarrow
B(y)+2A(y)(\pi\beta_0\ln\xmu^2-1)$ and $G_{22}\rightarrow
G_{22}+G_{12}2\pi\beta_0\ln\xmu^2$ as well as modifying $g_2$
according to~\cite{nllathmh}, equation~(8).  This form of matching is
referred to as \lnr-matching.  Other forms of matching are possible
with small differences in the treatment of higher order
terms~\cite{nllathmh,OPALPR075,dasgupta03}.  An important difference
between the \lnr-matching and other matching schemes is that in the
other schemes the coefficients $C_1$ and $C_2$ and in some cases
$G_{21}$ must be known explicitly.

In order to force the NLLA terms to vanish at the upper kinematic
limit \ymax\ of an observable $y$ the so-called {\em modified
matching} is used~\cite{nllathmh,dasgupta03}.  The modified matching
consists of replacing $L=\ln(1/y)$ by $L'=\ln(1+1/y-1/\ymax)$.  

A possible NNLO term in the fixed order prediction as discussed above
can be taken into account in the \lnr-matching scheme by adding
the following term to the right-hand-side (RHS) of
equation~(\ref{equ_lnr}):
\begin{equation}
\label{equ_lnrnnlo}
  \left( C(y)-A(y)B(y)+\frac{1}{3}A^3(y) -
         G_{34}L^4-G_{33}L^3 \right)\ash^3 \;\;.
\end{equation}
Renormalisation scale dependence and normalisation to the total
hadronic cross section can be considered by replacing
$C(y)=\int_0^y\ddel C/\ddel y'\ddel y'$ by the correspondingly
integrated equation~(\ref{equ_nnlo}) and by the replacement
$G_{33}\rightarrow G_{33}+G_{23}4\pi\beta_0\ln\xmu^2$.  For matching
schemes other than the $\ln(R)$-matching the coefficients $C_i, i=1,2,3$,
$G_{21}$ and in some cases $G_{3i}, i=2,1$ must be available.

\subsection{Inclusive observables}
\label{sec_incobs}

In the definition of inclusive observables no requirements on the
structure of the hadronic events are made, instead the cross
sections or branching ratios for the production of hadronic final
states are considered.  The objects of interest are QCD induced
corrections to electro-weak processes with hadronic final states.  

The total hadronic cross section in \epem\ annihilation \signull\
including $\gamma-\znull$ interference but without QCD corrections can
be written as follows (see e.g.~\cite{ellis96}):
\begin{equation}
\label{signull}
  \signull(s)= \frac{4\pi\alpha^2}{3s}\left[
  Q_f^2 - 2Q_fV_eV_f\chi_1(s) + (A_e^2+V_e^2)(A_f^2+V_f^2)\chi_2(s)
  \right]\;\;,
\end{equation}
where $\alpha$ is the QED coupling, $Q_f$ is the fermion charge and
the $A_f$ and $V_f$ are the axial and vector couplings of fermions to
the \znull.  These are given by $A_f=T_{f,3}$ and
$V_f=T_{f,3}-2Q_f\sin^2\Theta_W$ with $T_{f,3}=+1/2$ for neutral leptons
and up-type quarks and $T_{f,3}=-1/2$ for charged leptons and
down-type quarks with the electro-weak mixing angle $\Theta_W$.  
The functions $\chi_1=\kappa s(s-\mz^2)/((s-\mz^2)^2+\gz^2\mz^2)$ and
$\chi_2=\kappa^2 s^2/((s-\mz^2)^2+\gz^2\mz^2)$,
$\kappa=\sqrt{2}G_F\mz^2/(16\pi\alpha)$, take into account \znull\
exchange and its interference with photon exchange, respectively.  

The quantity $\rhad=\sigma(\epem\rightarrow\mathrm{hadrons})/
\sigma(\epem\rightarrow\mpmm)$ is
introduced to study the QCD corrections with suppressed electroweak
effects.  At values of the cms energy \roots\ far below the \znull\
resonance we have $\rhad=\nc\sum_q Q_q^2$ while for the hadronic
partial decay width of the \znull\ one gets $\rz=\nc\sum_q
(A_q^2+V_q^2)/(A_{\mu}^2+V_{\mu}^2)$ with the number of quark colours
$\nc=3$.

The QCD induced corrections to \rhad\ can be factorised such that one
gets for $\sqrt{s}<\mz$~\cite{surguladze91,gorishnii91,chetyrkin97a}:
\begin{equation}
  \rhad \rightarrow \rhad(1+2\ash+5.6368\ash^2-102.44\ash^3)\;\;.
\end{equation}

For the case of hadronic decays of $\tau$ leptons produced in \epem\
annihilation the quantity
$\rtau=\Gamma(\tau^-\rightarrow\mathrm{hadrons})/
\Gamma(\tau^-\rightarrow\nu_{\tau}\bar{\nu}_ee^-)$ is considered.  The
QCD corrections lead to the following expression for
\rtau~\cite{gorishnii91}:
\begin{equation}
  \rtau= 3(1+2\ash+20.809\ash^2+210.9\ash^3)\;\;.
\end{equation}
Section~\ref{sec_rtau} gives more details about the QCD description of
hadronic decays.

\subsection{Monte Carlo models}

Important tools for the analysis of hadronic \epem\ annihilation
events are event generation programs based on the Monte Carlo
method~\cite{james80}.  These programs simulate the production and
development of individual hadronic events according to probability
densities derived from the theory for the parton shower.  In addition
models are used to describe the transition from the parton to the
hadron state, the so-called hadronisation.  Brief reviews of the
various programs can be found in~\cite{yellow3qcd,yellowlep2qcdmc}
while reviews of the underlying theoretical methods are
e.g.\cite{webber86,ellis96}.  Figure~\ref{fig_qcdshower} gives an
overview of the various steps needed to simulate the production
of an hadronic event.

QCD calculations in the leading logarithmic approximation (LLA) form
the basis of the parton shower simulation.  The probability $P_{bc}$
in the LLA for a parton $a=\mathrm{q,g}$ to branch to two partons
$bc=\mathrm{qg,gg,qq}$ depends on the strong coupling \asq\ evaluated
at the energy scale $Q$ of the branching and the kinematics of the
branching process.  The differential DGLAP
equation~\cite{gribov72,altarelli77,dokshitzer77} specifies how
$P_{bc}$ depends on the energy scale of the branching through the
running of the strong coupling \as:
\begin{equation}
  \frac{\ddel P_{bc}}{\ddel t} = \cf\ash(Q^2)\int p_{bc}(z)dz
\label{equ_dglap}
\end{equation}
with $t=2\ln(Q/\lmqcd)$ and $z=E_b/E_a, 1-z=E_c/E_a$ where $E_i$ is
the energy of parton $i$.  Since the branching is understood to be a
part of a tree-like Feynman graph the partons $a$, $b$ and $c$ will be
virtual, i.e.\ off their mass shell.  The splitting functions
$p_{bc}(z)$ take into account the different kinematics of the three
possible branching processes $q\rightarrow qg, g\rightarrow gg$ and
$g\rightarrow\qqbar$ in the LLA.

The task of a Monte Carlo parton shower algorithm is to first choose
for a given parton at which value of the so-called evolution variable
$t$ the branching should take place and then to select a value of the
energy fraction $z$.  The branching probability $P_{bc}$
given by equation~(\ref{equ_dglap}) is used to write the probability
$P_{n.e.}(t_{max},t)$ that no emission takes place as the product of
constant no-emission probabilities valid in small $t$ intervals
$\Delta t$. These intervals range from the maximum possible $t$ value
$t_{max}$ given by the process which produced parton $a$ to the value
$t$ of interest:
\begin{eqnarray}
\label{equ_pne}
  \ln P_{n.e.}(t_{max},t) & = &
  \sum_{i=1}^N \ln(1-\frac{\ddel P_{bc}(t_i)}{\ddel t}\Delta t)
  \simeq \sum_{i=1}^N -\frac{\ddel P_{bc}(t_i)}{\ddel t}\Delta t \\ \nonumber
  & = &  -\int_t^{t_{max}}\frac{\ddel P_{bc}}{\ddel t'} {\ddel t'} \\ \nonumber
\end{eqnarray}
in the limit $\Delta t\rightarrow 0, N\rightarrow \infty$.
Equation~(\ref{equ_pne}) can be used to pick a value of $t$ from a
random number $R$ distributed evenly between 0 and 1 with
$R=P_{n.e.}(t_{max},t)$.  When the chosen value of $t$ falls below a
pre-defined value $t_{min}$ corresponding to a cut on $Q$ called $Q_0$
the parton cannot branch and is put on its
mass-shell.  This step will eventually stop the parton shower
evolution when all remaining partons cannot branch anymore. 

Once a value of $t$ is known the kinematics of the branching is
determined by choosing values of $z$ using the splitting functions
$p_{bc}(z)$.  The azimuthal angle of the branching must be chosen as
well to complete the configuration~\cite{ellis96}.  
The exact definition of the energy scale $Q$ of the branching has not
been given yet, since the existing parton shower algorithms use
different approaches, e.g. $Q^2=m_a^2$ or $Q^2=p_t^2$~\cite{webber86}.

The Monte Carlo parton shower simulation based on LLA QCD calculations
shown so far does not include interference between soft but acolinear
gluons produced in different branchings (colour
coherence)~\cite{webber86,yellow3qcd,ellis96}.  It turns out that soft
gluon interference can be treated in the LLA including subleading
corrections by the introduction of angular ordering where the opening
angles of successive branchings are required to decrease in addition
to the values of $t$ which decrease by construction.  The destructive
soft gluon interference is the QCD analogue to the Chudakov
effect~\cite{dokshitzer91}, where \epem\ pairs produced at high energy
are observed to generate less ionisation in the region where the two
tracks of the \epem\ pair are close together.  The intermediate
photons involved in the ionisation processes cannot resolve the two
individual charges of the nearby tracks thus leading to a suppressed
ionisation rate.  In the parton branching $a\rightarrow bc$ a
subsequent soft but acolinear gluon cannot resolve the daughters
$bc$ and thus it can be viewed as emitted by parton $a$.  This can be
implemented in the parton shower algorithm by the introduction of the
angular ordering requirement.  

The parton shower algorithm, in particular with the angular ordering
requirement, is not expected to describe effects of hard emissions
well, since it is based on the LLA valid for soft and collinear
emissions.  It is possible to correct the algorithms using the LO QCD
matrix element for the hard emission
process~\cite{bengtsson87,seymour95b,pythia62,herwig65}.  For
theoretical consistency it is necessary to apply the correction to
all branchings up to the hardest in the shower~\cite{seymour95b}.  An
alternative parton shower formulation based on the colour dipole
approach~\cite{gustafson88} does not need such corrections since it
uses the LO QCD matrix elements already at each branching.

Further corrections using NLO QCD matrix elements to obtain a better
description of multi-jet final states with hard and well separated
jets are actively
developed~\cite{andre98,lonnblad01a,catani01a,frixione02a,kurihara02}.

In the hadronisation phase hadrons are built from quarks and gluons
left after the end of the parton shower by algorithms based on
hadronisation models. In the hadronisation model it has to be
specified how much of the energy and momentum of a parton is taken
over by the hadron built from it.

The so-called string model~\cite{string,pythia62} is based on the
colour field line configuration between a separating quark and
antiquark. Naively, the colour field lines should look similar to
electric field lines but the self interaction of gluons causes a
contraction of the colour field lines into a narrow tube or {\em
string}. As the quark and the antiquark move apart energy is stored in
the stretching colour flux tube which can be expressed by a linear
term in the potential $V_{\mathrm{QCD}}(r)\sim\frac{a}{r}+b\cdot r$
between the quark and the antiquark. In this picture the colour flux
tube or string is stretched until the potential energy suffices to
create a new quark antiquark pair causing the string to break in
two. This process is repeated as long as the partons have enough
energy to support it. A gluon created in the parton shower will have
strings attached to it following the colour flow and stretching to
both the quark and the antiquark and will thus appear as a kink in the
string.

At the end of the string fragmentation quarks and antiquarks connected
by strings are treated as mesons.  Production of baryons is included
by the possibility to produce pairs of diquarks qq or
$\mathrm{\bar{q}\bar{q}}$ when a string breaks or by the popcorn
mechanism~\cite{popcorn} where the baryons are made up from
successively produced quarks or antiquarks.  The detailed behaviour of
the hadronisation is controlled by fragmentation functions and the
parameter $Q_0$ specifying the parton energy at which the parton
shower is terminated.  A typical value for $Q_0$ is about 1 GeV safely
above the Landau pole where perturbative QCD must fail.  The
fragmentation functions are probability densities $f(z)$ for the
fraction $z$ of energy and longitudinal momentum ($E+p_l$) the newly
created hadron $\mathrm{q_1\bar{q}_2}$ takes after a string connected
to the quark $\mathrm{q_1}$ broke to create a pair
$\mathrm{q_2\bar{q}_2}$.  Usually a fragmentation function
\begin{equation}
  f(z)\sim\frac{(1-z)^a}{z}e^{(bm_t^2/z)}
\label{equ_string}
\end{equation}
with $m_t^2=m^2+p_t^2$ and $a$ and $b$ free parameters is
used.  Alternatively a fragmentation function as proposed by Peterson
et al.~\cite{peterson83}
\begin{equation}
 f(z)\sim\left(z\left( 1-\frac{1}{z}-
  \frac{\epsilon_q}{1-z}\right)^2\right)^{-1}
\label{equ_peterson}
\end{equation}
may be used for the hadronisation of heavy quarks, i.e.\ b- and
c-quarks. The parameter $\epsilon_q$ is expected to scale like
$1/m_q^2$ but has to be adjusted for each heavy quark flavour
separately by comparison with experimental data.  The momentum
transverse to the string direction is given by a Gaussian distribution
with the free parameter $\sigma_Q$ to adjust the width.

The cluster hadronisation
model~\cite{gottschalk83,field83,webber84b,marchesini88,herwig65} is
based on the so-called preconfinement of colour in
QCD~\cite{amati79,azimov85a}, i.e.\ that partons in a shower build
clusters of colour singlets with masses ${\cal O}(Q_0)$.  The cluster
model assumes that hadronisation is a local process and does not need
fragmentation functions to describe hadronisation.  In the model after
the end of the parton shower gluons are split into
\qqbar\ pairs and quarks close in phase space are combined into colour
singlet objects called clusters~\cite{field83}.  The masses of the
clusters are ${\cal O}(Q_0)$ and are approximately independent of the
hard scale $Q$.  Clusters exceeding an upper limit on mass are further
split into lighter clusters.  The clusters are intermediate states in
the hadronisation algorithm allowed to decay isotropically into two
hadrons, or only one hadron when the cluster is too light.  Production
of mesons and baryons is handled by creating either a quark pair
\qqbar\ or by creating a pair of diquarks $\mathrm{qq\bar{q}\bar{q}}$.
The cluster decay products are then randomly identified with hadrons
fitting the quark contents with probabilities proportional to the spin
degeneracy of the hadrons and the available phase space for the decay,
i.e.\ the density of states.  Only kinematically allowed decays are
accepted and in the case of rejection the cluster decay algorithm
starts again for the cluster in question until an allowed decay is
found.

After the hadronisation is finished the unstable hadrons are allowed
to decay according to measured branching ratios and decay rates where
known. Especially for hadronic decays of hadrons containing heavy
quarks this information is not always available. In this case the
heavy quark is allowed to decay weakly into quarks and the hadrons are
created by the same hadronisation mechanisms as described
above~\cite{marchesini90,herwig65,pythia62}. 

The next subsections summarise some of the most popular programs
currently in use and briefly discuss the optimisation of the
description of the data by the models.

\subsubsection{PYTHIA}

The PYTHIA program~\cite{pythia62} implements a parton shower
algorithm essentially as described above with the invariant mass of
the parton (virtuality) as the ordering parameter and angular ordering
as additional constraint.  The parton shower is combined with the string
fragmentation model.

\subsubsection{HERWIG}

The HERWIG program~\cite{herwig65} provides a parton shower similar to
our description, but with the ordering variable replaced by the
opening angle of the branching.  This feature allows the inclusion of
angular ordering in a clean way.  Correlations between azimuthal
angles of subsequent parton branchings due to gluon polarisation
effects are also taken into account.  The parton shower is terminated
by setting minimal values for the invariant masses of quarks and the
gluon.  The hadronisation follows the cluster model including
splitting of gluons into \qqbar\ pairs after the parton shower
stopped.  In HERWIG clusters containing quarks produced in the parton
shower decay such that the hadron carrying the quark flavour travels
in the direction of the quark.  Hadronisation of \bbbar\ events can be
controlled with additional parameters to take account of their special
properties.

\subsubsection{ARIADNE}

The ARIADNE program~\cite{ariadne3} has a parton shower algorithm
based on the colour dipole model~\cite{gustafson88} and with the
relative momentum $p_{\perp}^2$ of the branching as the ordering
variable.  The branchings are strictly ordered with
$p_{\perp,1}>p_{\perp,2}>p_{\perp,3}>\ldots$ such that the angular
ordering requirement is fulfilled.  The parton shower terminates when
all $p_{\perp}^2$ are below a minimum value and the events are further
processed by the string fragmentation model as in PYTHIA.  

\subsubsection{COJETS}

The parton shower in the COJETS Monte Carlo
program~\cite{cojets1,cojetstuning} is formulated using the LLA in a
way so that a shower for each of the initial quarks or antiquarks can
be evolved separately.  The first branching is constrained to the \oa\
QCD matrix element for hard gluon radiation.  There is no imposition
of angular ordering in the parton shower, thereby neglecting the soft
gluon interference effects.  A limit on the energy of partons
generated in the shower is used to stop the parton shower.  The
running strong coupling \as\ is implemented using the one-loop
expression.

The hadronisation algorithm in COJETS is based on the Field-Feynman
model~\cite{indepfragm}. In this model each parton left after the
parton shower has stopped is fragmented independently using an
iterative algorithm.  A quark $\mathrm{q_1}$ is paired with an
antiquark $\mathrm{\bar{q}_2}$ from a newly created \qqbar\ pair to
build a hadron.  The remaining quark $\mathrm{q_2}$ is then fragmented
by the same procedure.  After the fragmentation is finished some extra
particles are added to ensure flavour conservation. The sharing of
energy and momentum is specified by a fragmentation function with the
Field-Feynman parametrisation $f(z) = 1-a+3a(1-z)^2$ where $a$ is a
free parameter and $z$ is the ratio of energy and momentum from the
hadron and the quark $\mathrm{q_1}$.  The models for the parton shower
and hadronisation are more simple than those of the other programs;
the COJETS program is now often used to study the effects of using
more simple models.

\subsubsection{Tuning of Monte Carlo Models}

The free parameters of the models can be varied in order to achieve an
optimal description of the data.  The important free parameters to
influence the behaviour of the parton shower are the value of the
strong coupling constant \as\ and the parameter to control termination
of the shower.  

The string hadronisation is controlled by the
parameters of the fragmentation functions for light or heavy quarks,
see equations~(\ref{equ_string}) and~(\ref{equ_peterson}), and by the
parameter setting the width of the Gaussian distribution of transverse
momentum.  Additional parameters regulate the production of baryons
and strange hadrons.  The cluster hadronisation model has free parameters
to control splitting of heavy clusters into light clusters before these
clusters are allowed to decay into hadrons.

The LEP collaborations have studied the Monte Carlo models in detail
and derived sets of parameters which optimise description of their
data.  The data consist of event shape observables, jet production
rates, charged particle multiplicities and production rates of
identified particles measured with the large event samples produced on
the peak of the \znull\ resonance. The latest values of the model
tuning parameters are given for ALEPH in~\cite{aleph141}, for DELPHI
in~\cite{delphi141}, for L3 in~\cite{l3290}, for OPAL
in~\cite{OPALPR141,OPALPR379} and for SLD in~\cite{sldmctune}.

\subsection{Power corrections}

At energy scales $Q\approx\lmqcd$ perturbative evolution of \asq\ breaks
down completely due to the Landau pole at \lmqcd.  It is therefore
impossible to attempt pQCD calculations of soft processes at low
values of $Q$ without further assumptions.  In the preceding sections
Monte Carlo event generators were discussed which successfully model
hadronisation effects.  However, it is interesting to attempt to treat
hadronisation in a more direct way.

\begin{figure}[!htb]
\begin{center}
\includegraphics[width=0.5\textwidth]{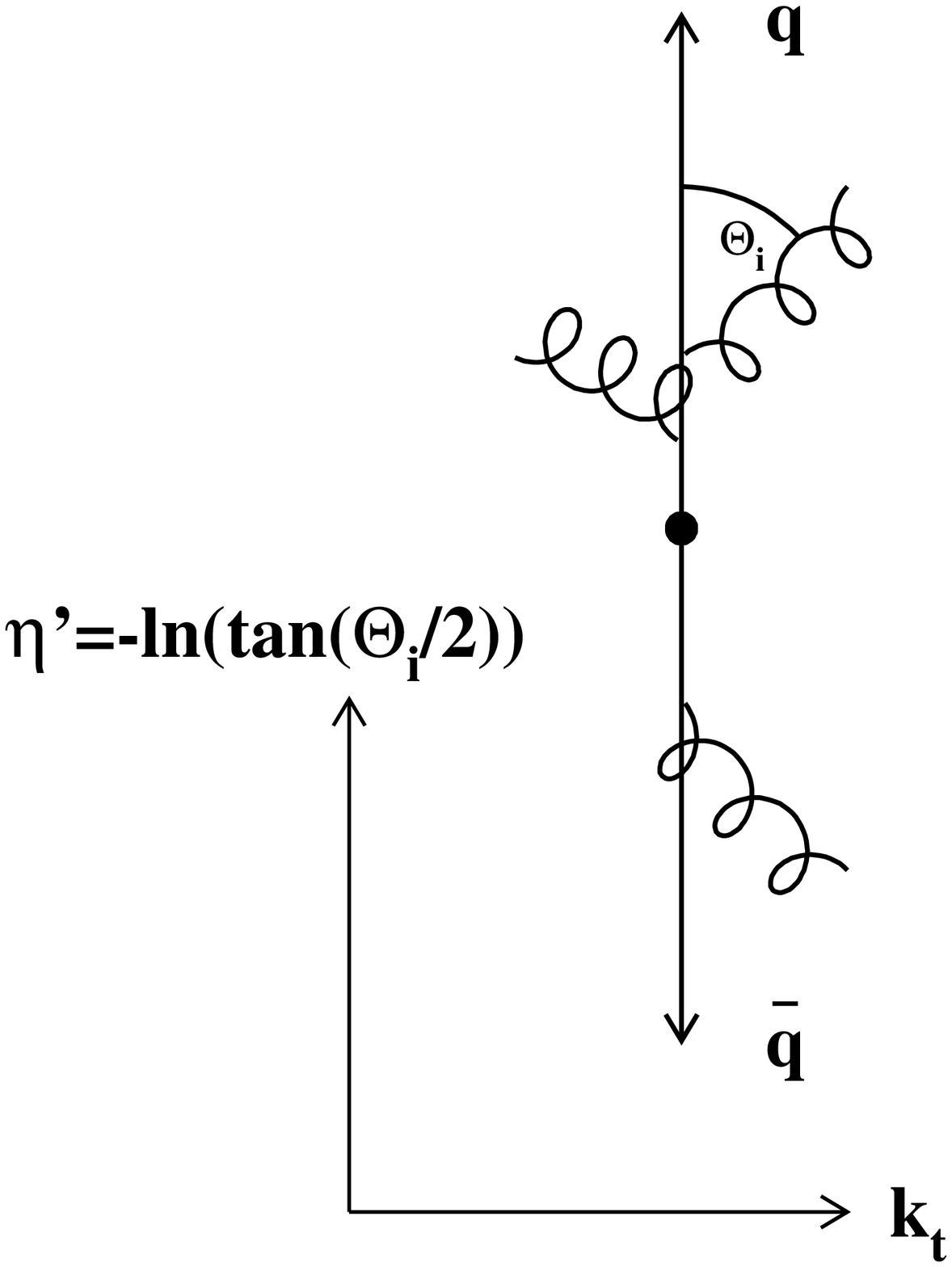}
\caption[ bla ]{ Sketch of a \qqbar\ system in $\eta'-\kperp$ space. }
\label{fig_tube}
\end{center}
\end{figure}

In order to clarify which assumptions are needed a phenomenological
model of hadronisation is considered.  The longitudinal phase space
model or {\em tube model} goes back to Feynman~\cite{salam01a}.  One
considers a \qqbar\ system produced e.g.\ in \epem\ annihilation in
the cms system.  The two primary partons move apart with velocity
$v/c\simeq 1$.  In such a situation the production of soft gluons will
be approximately independent of their (pseudo-) rapidity
$\eta'=-\log(\tan(\Theta_i/2))$, see figure~\ref{fig_tube}. 

The change to e.g.\ the observable Thrust \thr\ due to the production
of a soft gluon at angle $\Theta_i$ with transverse momentum
$k_{\mathrm{t},i}$ is $\Delta(1-T)_i\simeq
k_{\mathrm{t},i}/Qe^{-|\eta'_i|}$.  This observation is generalised to
write the soft or non-perturbative contribution to the mean value of
the \thr\ distribution as follows:
\begin{equation}
  \momone{\thr}_{\mathrm{NP}} = \int\frac{\kperp}{Q}\Phi(\kperp)
                                         \frac{\mathrm{d}\kperp}{\kperp}\cdot 
                                    \int e^{-|\eta'|}d\eta'
\label{equ_tnp}
\end{equation}
The function $\Phi(\kperp)\sim\as(\kperp)$ is the distribution of soft
particles in \kperp.  The first integral in equation~(\ref{equ_tnp}) is
summarised as $\anull/Q$ independent of the observable and the second
as a constant $c_{\thr}$ dependent on the observable.  The quantity
$\anull\sim\int\as(\kperp)\mathrm{d}\kperp$ can only exist when
$\as(\kperp)$ is identified with a non-perturbative strong coupling
$\asnp(\kperp)$ which is assumed to be finite at low \kperp\ around
and below the Landau pole.

A more formal approach to the origin of soft contributions is the
study of infrared renormalons, i.e.\ the divergence of asymptotic pQCD
predictions due to the integration of low momenta in quark loops in
gluon lines~\cite{beneke00a}.  The infrared renormalon divergence of
the \oan\ term in a pQCD prediction is factorial in $n$:
$r_n\alpha_S^{n+1}\sim(2\beta_0/p)^n n!\alpha_S^{n+1}$.  With
Stirlings formula to replace $n!$ one finds
$r_n\alpha_S^{n+1}\sim(2\beta_0\as/p)^n n^ne^{-n}\as$.  The
convergence of the series is optimal for $n=p/(2\beta_0\as)$.  Based
on this relation one finds
\begin{equation}
  r_n\alpha_S^{n+1}\sim\left(\frac{\lmqcd}{Q}\right)^p
\label{equ_renorm}
\end{equation}
where the first order relation between \as\ and \lmqcd\ has been
used.  The result shows that infrared renormalon contributions to pQCD
predictions scale like $Q^{-p}$ similar to the soft contribution
studied in the tube model, see equation~(\ref{equ_tnp}). 

The power correction model of Dokshitzer, Marchesini and Webber (DMW)
extracts the structure of power correction terms from analysis of
infrared renormalon contributions~\cite{dokshitzer95a}.  The model
assumes that a non-perturbative strong coupling exists around and
below the Landau pole and that the quantity
$\anull(\mui)=1/\mui\int_0^{\mui}\asnp(\kperp)\mathrm{d}\kperp$ can be
defined.  The value of \mui\ is chosen to be safely within the
perturbative region, usually $\mui=2$~GeV.  A study of the branching
ratio of hadronic to leptonic $\tau$ lepton decays as a function of the
invariant mass of the hadronic final state supports the assumption
that the physical strong coupling is finite and thus integrable at
very low energy scales~\cite{brodsky03}.

The main result for the effects of power corrections on distributions
$F(y)$ of the event shape observables \thr, \mh\ and \cp\ is that 
the perturbative prediction $\fpt(y)$ is
shifted~\cite{dokshitzer97a,dokshitzer98b,dokshitzer99a}: 
\begin{equation}
  F(y)= \fpt(y-\cy P)
\label{equ_npshift}
\end{equation}
where \cy\ is an observable dependent constant and $P\sim
M\mui/Q(\anull(\mui)-\as)$ is universal, i.e.\ independent of the
observable~\cite{dokshitzer98b}.  The factor $P$ contains the $1/Q$
scaling and the so-called Milan-factor $M$ which takes two-loop
effects into account.  The non-perturbative parameter \anull\ is
explicitly matched with the perturbative strong coupling \as.  For the
event shape observables \bt\ and \bw\ the predictions are more
involved and the shape of the pQCD prediction is modified in addition
to the shift~\cite{dokshitzer99a}.  For mean values of \thr, \mh\ and
\cp\ the prediction is:
\begin{equation}
  \momone{y}= \momone{y}_{\mathrm{PT}}+\cy P
\label{equ_pcmean}
\end{equation}
For \momone{\bt} and \momone{\bw} the predictions are also more
involved due to the modification of the shape of the distributions.

\section{QCD in \epem\ annihilation}
\label{sec_hist}

Studies of QCD in \epem\ annihilation have been carried out for more
than 30 years now.  The earliest experiments were located at small
colliders operating in the \roots\ range of about $1-2$ GeV, ADONE,
ACO and VEPP-2.  A comprehensive review of the early experiments and
results may be found in~\cite{perez-y-jorba77}.

\subsection{Accelerators and experiments}

The development of storage rings with colliding beams of electrons and
positrons was the foundation for the wealth of results we have today
not only about QCD but also about many other aspects of the Standard
Model of particle physics.  Table~\ref{tab_machines} collects the
\epem\ storage rings and their experiments.  The centre-of-mass
energies available to the experiments in the laboratory system were
increased by 2 orders of magnitude.  

\begin{table}[htb!]
\caption[ bla ]{ \epem\ colliders and experiments. }
\label{tab_machines}
\begin{tabular}{lccc} \hline\hline

Facility & Location & $\roots\;[\mathrm{GeV}]$ & Experiments \\ \hline

ACO~\cite{aco71} & LAL Orsay & $\approx 1$ &
M3N~\cite{cosme71,cosme76}\\ 

ADONE~\cite{adone71} & INFN Frascati & $1-3$ & Boson~\cite{bartoli72},
$\mu\pi$~\cite{ceradini73}, $\gamma\gamma$~\cite{bacci73}, \\
& & & $\gamma\gamma2$~\cite{bacci79}, 
MEA~\cite{mea78} \\ 

VEPP-2~\cite{kurdadze72} & Novosibirsk & $1-1.5$ &
VEPP-2~\cite{kurdadze72} \\ 

CEA~\cite{cea71} & Cambridge, MA & 4 & BOLD~\cite{litke73} \\

SPEAR~\cite{spear71} & SLAC Stanford & $2-8$ &
SLAC-LBL~\cite{augustin75a,augustin75b}, \\
& & & MARK~I~\cite{siegrist82}, MARK~II~\cite{schindler81} \\ 

PEP~\cite{pep80} & SLAC Stanford & 29 & MARK~II~\cite{vonzanthier91},
HRS~\cite{bender84}, \\ 
& & & TPC/$2\gamma$~\cite{aihara88a,aihara88b}, MAC~\cite{allaby89} \\ 

DORIS~\cite{doris77,doris80} & DESY Hamburg & $3-11$ &
PLUTO~\cite{criegee82}, DASP~\cite{brandelik79a,albrecht82}, \\
& & & LENA~\cite{niczyporuk82}, DH(HM)~\cite{bartel76,bartel78} \\ 
CESR~\cite{cesr80} & Cornell, Ithaka & $10-11$ &
CLEO~\cite{andrews83,kubota92}, \\ 
& & & CUSB~\cite{bohringer80,finocchiaro80} \\ 

PETRA~\cite{petra80} & DESY Hamburg & $12-47$ & CELLO~\cite{behrend81},
JADE~\cite{naroska87}, \\ 
& & & MARK~J~\cite{adeva84}, PLUTO~\cite{criegee82}, \\
& & & TASSO~\cite{brandelik79a,brandelik80a} \\ 

TRISTAN~\cite{kamada92} & KEK Tsukuba & $50-64$ & TOPAZ~\cite{inoue00},
VENUS~\cite{abe87}, \\ 
& & & AMY~\cite{sagawa88} \\ 

SLC~\cite{seeman91} & SLAC Stanford & $\approx 91$ 
& MARK~II~\cite{vonzanthier91}, SLD~\cite{sldparticles2} \\ 

LEP~\cite{burkhardt96} & CERN Geneva & $88-209$ &
ALEPH~\cite{aleph011,aleph085}, \\
& & & DELPHI~\cite{delphi017,delphi120}, \\ 
& & & L3~\cite{l3det90}, OPAL~\cite{OPALPR021} \\ 

\hline

\end{tabular}
\end{table}

The integrated luminosities of the corresponding data samples range
from $\order{1-100}$/nb of the early experiments to $\order{10-100}$/pb
at PETRA, PEP and TRISTAN.  The experiments at the only recently
decommissioned colliders SLC and LEP collected data samples
corresponding to several hundred 1/pb on the \znull\ peak and, in the
case of LEP, above the \znull\ peak up to cms energies of 209~GeV.

The sizes of the data samples vary from \order{100} events to
\order{10\,000} events per cms energy in experiments running at the
colliders with cms energies below or above the \znull\ peak.  On the
\znull\ peak the LEP experiments accumulated about $5\cdot 10^6$
hadronic events each while the SLC experiment SLD collected
\order{100\,000} hadronic \znull\ decays.

\subsection{Highlights of QCD before the LEP aera}

The first experimental studies of hadron production in \epem\
annihilation where done in the early 1970s when evidence in support of
the quark-parton model~\cite{drell69,drell70} had come from
deep-inelastic scattering experiments.  

A simple prediction of the quark-parton model was that the cross
section for $\epem\rightarrow\mathrm{hadrons}$ should be large
compared to the cross section for $\epem\rightarrow\mpmm$, because of
the additional colour degree of freedom, see section~\ref{sec_incobs}.
Early measurements of
$R=\sigma(\epem\rightarrow\mathrm{hadrons})/\sigma(\epem\rightarrow\mpmm)$
at $\roots\simeq 1-3$~GeV indeed observed $R\approx 2$ as predicted by
the quark-parton model for u, d and s quarks~\cite{siegrist82}.
However, the measurements were difficult to interpret quantitatively
due to the presence of resonances.

The quark-parton model also predicts that hadron production in \epem\
annihilation at sufficiently high energy should show a pattern of two
jets of hadrons recoiling against each other.  The direction of the
hadron jets is given by the direction of the produced quarks and
should thus follow the expected distribution of the angle $\Theta$
between the quark and the beam direction for the production of two
particles with spin $1/2$~\cite{cabibbo70,bjorken70,feynman72}:
\begin{equation}
  \frac{\ddel\sigma}{\ddel\cos\Theta} \sim 1+\cos^2\Theta\;\;.
\end{equation}
The first evidence for jet structure was reported 1975
in~\cite{hanson75} using data recorded at $\roots=6.2$ and
7.4~GeV by the SLAC-LBL magnetic detector at the
\epem\ storage ring SPEAR.  Hadronic events were analysed with the
event shape observable Sphericity~\cite{bjorken70}.  Distributions of
Sphericity values and angular distributions of jet directions given by
the Sphericity axis were compared with jet model and phase space Monte
Carlo simulations.  This evidence was further supported
in~\cite{berger78,brandelik79b,hanson82}.

QCD as the field theory of quark interactions predicts the existence
of gluons as intermediate gauge bosons analogous to the photons of
QED.  It was crucial for the acceptance of QCD that effects directly
connected with gluon activity could be observed in hadronic \epem\
annihilation events.  

The first indirect evidence of the existence of gluons came from
so-called direct decays of the Y(1S) resonance into hadrons.
In~\cite{berger79a} the PLUTO collaboration using data recorded on and
off the Y(1S) at the DORIS storage ring at DESY studied event shape
observables like the Thrust (see section~\ref{sec_shapes}).  In direct
decays of the Y(1S) the Thrust distribution was found to agree with
Monte Carlo simulations based on Y(1S) decays mediated by three gluons
while simple phase space models were ruled out.  A study of the
angular distribution of the Thrust axis w.r.t. the beam direction
provided evidence that the gluons are vector particles, i.e.\ have
spin 1.

The first direct evidence for the existence of gluons was provided by
the observation of planar hadronic events with a clear 3-jet structure
by the experiments at the PETRA \epem\ collider at
DESY~\cite{brandelik79c,barber79a,berger79b,bartel80}.  The cms
energies of the collisions ranged from 17 to 32~GeV.  In these studies
events were classified as deviating from a 2-jet configuration based
on the event shape observables Sphericity, Thrust or Oblateness.
Examination of the energy flow in the event
plane~\cite{brandelik79c,barber79a,berger79b} showed a pattern
consistent with LO QCD based Monte Carlo models including gluon
radiation.  Distributions of event shape observables Oblateness,
Thrust or Planarity were measured in~\cite{barber79a,bartel80} and
were found to be reproduced by Monte Carlo models including gluon
radiation at leading order.  In~\cite{bartel80,berger80,brandelik80c}
the angular and energy distribution of the lowest energy jet are found
to agree with the expected behaviour for radiation of energetic vector
gluons.

The first measurements of the strong coupling constant \as\ were based
on Monte Carlo simulations based on LO QCD and data from the PETRA
experiments recorded at
$\roots\approx30$~GeV~\cite{berger80,brandelik80b,barber79b,bartel80,behrend82}.
The simulations generated $\qqbar\mathrm{g}$ final states according to
LO QCD matrix elements and employed a simple Field-Feynman
fragmentation model.  The value of \as\ together with parameters to
adjust the fragmentation model was varied until the description of
event shape observable distributions or 3-jet production rates was
optimised.  The results were $\as(30 \mathrm{GeV})=0.19\pm0.04$
corresponding to $\asmz= 0.15_{-0.03}^{+0.05}$.

Improved determinations of \as\ using NLO QCD calculations turned out
to be difficult to achieve~\cite{naroska87}.  For event shape or jet
observables the early NLO calculations gave differing results and the
dependence on the fragmentation models resulted in large errors.  For
inclusive quantities like the R-ratio \rhad\ the sensitivity was limited
by uncertainties of the available data.  Global averages shortly
before the start of LEP were $\asmz=0.11\pm0.01$~\cite{altarelli89}
based on most available measurements and $\asmz=0.112\pm0.015$ from
event shape and jet observables alone~\cite{ali88}.  These
determinations had a relative uncertainty of about 10\% where the
dominating uncertainties came from the use of fragmentation models and
missing higher orders in the perturbative QCD calculations.  

The non-abelian nature of QCD becomes manifest in its property of
asymptotic freedom and in the existence of the triple gluon vertex
(TGV).  The first evidence for 4-jet structure was found by studying
event shape observables constructed to be sensitive to non-planar
multi-jet events~\cite{bartel82b}.  The first experimental evidence
for the running of the strong coupling \as, i.e.\ for asymptotic
freedom, was based on comparing data from PETRA runs at several cms
energies between 22 and 47~GeV in terms of 3-jet fractions determined
with the JADE recombination jet
algorithm~\cite{bethke88,braunschweig88}.  The 3-jet fractions were
shown to correspond closely to the underlying parton structure 
assumed in QCD with unimportant dependence on fragmentation models and
thus to reflect the running strong coupling.  In~\cite{park89b} the
evidence was confirmed and in addition angular distributions measured
with 4-jet events were found to prefer a Monte Carlo model based on
\oaa\ QCD including the TGV over an abelian model.

Dedicated studies of properties of hadronic \epem\ events showed that
fragmentation models using independent fragmentation of the hard
partons could not describe the data completely, in particular the
production of soft particles between jets, the so-called
string-effect, see e.g.~\cite{bartel81,bartel85,naroska87}.  This
reduced the set of models to those which used the Lund string model or
the cluster hadronisation model~\cite{saxon88}.  It was shown that the
string effect can be explained in QCD by coherent emission of soft
gluons~\cite{azimov85b}.  In addition it was found that production
rates of multi-jet events were better reproduced by models using a
parton shower algorithm~\cite{bartel86} based on QCD in the LLA, with
further improvements when the parton shower is matched to the
\oa\ QCD matrix element~\cite{bengtsson87,braunschweig88}.  

To summarise, the quark-parton model had been established by the
earliest \epem\ annihilation experiments.  Fundamental predictions of
QCD, the field theory of strong quark interactions, such as the
existence of gluons and their properties, had been successfully
tested.  However, the limited precision of calculations and the lack
of fundamental understanding of the fragmentation process did not
allow experimental tests of the theory with uncertainties better than
about 10\% as reflected in the errors of the combined values of \asmz\
shown above.  In particular the final experimental proof of asymptotic
freedom had not yet been demonstrated although there was already good
positive evidence.  Many detailed studies of the properties of the
hadronic final states in \epem\ annihilation had shown that
fragmentation models implementing a parton shower with coherence
effects together with the Lund string or the cluster hadronisation
model could describe the data most successfully.

\section{Jets and event shapes}
\label{sec_jetsshapes}

One of the most prominent and important features of hadronic final
states produced in \epem\ annihilation is the jet structure, i.e.\ the
presence of a small number of collimated groups of particles recoiling
against each other, see also section~\ref{sec_hist}.  Quantifying the
structure of hadronic final states allows for direct comparison of
experimental data with predictions by QCD and thus for detailed and
stringent tests of the theory.  Many attempts have been made to
quantify the jet structure; these schemes fall in two categories: jet
clustering algorithms and event shape observables.  The jet algorithms
and prescriptions to calculate event shape observables must be
infrared and collinear safe in order for perturbative QCD predictions
to be possible, see section~\ref{sec_pqcd}~\cite{sterman77}.

The following sections discuss commonly used jet clustering algorithms
and event shape observable definitions.  Experimental results will be 
presented and compared to model predictions.  Direct tests of
asymptotic freedom in QCD and measurements of \as\ using jet or event
shape observables will be discussed in separate sections.

\subsection{Jet observables}
\label{sec_jets}

In jet clustering algorithms one tries to group the particles of the
hadronic final state such that the jet structure, which is often
clearly visible, is replicated.  A popular algorithm was introduced by
the JADE collaboration~\cite{bartel86,bethke88}, often referred to as
the JADE algorithm.

The first ingredient of the clustering algorithm is the definition of
the distance $y_{ij}$ in phase space between any two particles $i,j$
in the hadronic final state.  The distance measure is defined as
\begin{equation}
  y_{ij}=\left(\frac{M_{ij}}{\evis}\right)^2
\label{equ_yij}
\end{equation}
where $M_{ij}$ is the invariant mass between
two particles $i,j$ and $\evis=\sum_i E_i$ is the total visible energy
of all particles in the hadronic final state.  The second ingredient
is the prescription for the combination of two particles $i,j$ into a
cluster or jet $k$; in the case of the JADE algorithm this is done by
adding the 4-momenta $p$: $p_k=p_i+p_j$.  The algorithm proceeds
recursively by combining the pair of particles with the smallest
$y_{ij}$, removing the particles $i,j$ from and adding the jet $k$ to
the event until $y_{i,j}>\ycut$ for all remaining $y_{ij}$.  The 
remaining jets are the result of the clustering algorithm.

Many variations of this scheme exist differing in the calculation of
$y_{ij}$ and the combination prescription, see e.g.~\cite{bethke92}.
The original JADE algorithm (JADE E0) defines $M_{ij}^2= 2 E_i E_j
(1-\cos\theta_{ij})$ taking all particles as massless.  The Durham
algorithm~\cite{durham} defines $y_{ij} = 2\min(E_i^2,E_j^2)
(1-\cos\theta_{ij})/\evis^2$, i.e.\ the distance is given by the
relative transverse momentum between one of the particles w.r.t.\
the other.

The JADE E0 and Durham algorithms were shown to be preferable, because
i) they have little dependence on hadronisation models compared to
other algorithms and ii) they were found to have comparatively small
theoretical uncertainties~\cite{OPALPR025,bethke92}.  The Durham and
JADE E0 algorithms have hadronisation corrections of about 5~\% at
$\roots=\mz$~\cite{bethke92}.  The Cambridge
algorithm~\cite{dokshitzer97b} is a variant of the Durham algorithm
where some soft particles or jets are excluded from further jet
clustering in order to reduce hadronisation effects.  However,
in~\cite{bentvelsen98} it was shown that hadronisation effects for 2-,
3- and 4-jet fractions are actually larger for the Cambridge than for
the Durham algorithm in large regions of \ycut, but that these effects
cancel for the more inclusive observable given by the average
multiplicity of jets per event.

The events are usually analysed by studying the fraction of e.g.\ 3-
or 4-jet events as a function of the jet resolution
parameter \ycut.  Alternatively, the values of \ycut\ at which the
number of jets in the events change e.g.\ from 2 to 3 are used to
compute differential distributions~\cite{komamiya90}.  In this case
the observable is analogous to the event shape observables discussed
below.

A large number of results on jet production in \epem\ annihilation has
been published; reviews of earlier work can be found
in~\cite{maettig89,bethke/pilcher,schmelling94}.  Recent and comprehensive
studies of jet production may be found 
in~\cite{OPALPR299,l3290,aleph265,delphi210}. 

\begin{figure}[htb!]
\begin{tabular}{cc}
\includegraphics[width=0.5\textwidth]{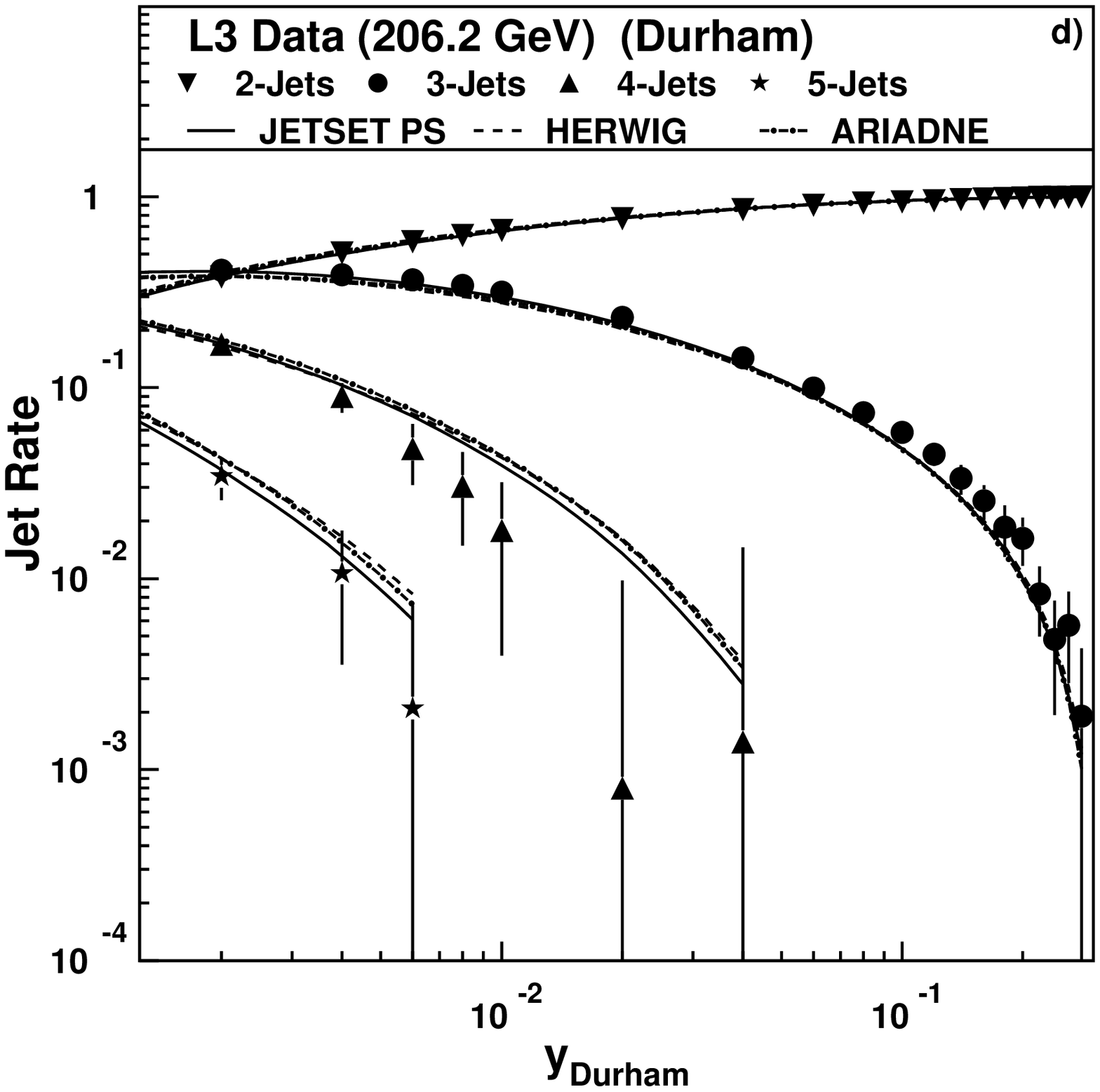} &
\includegraphics[width=0.5\textwidth]{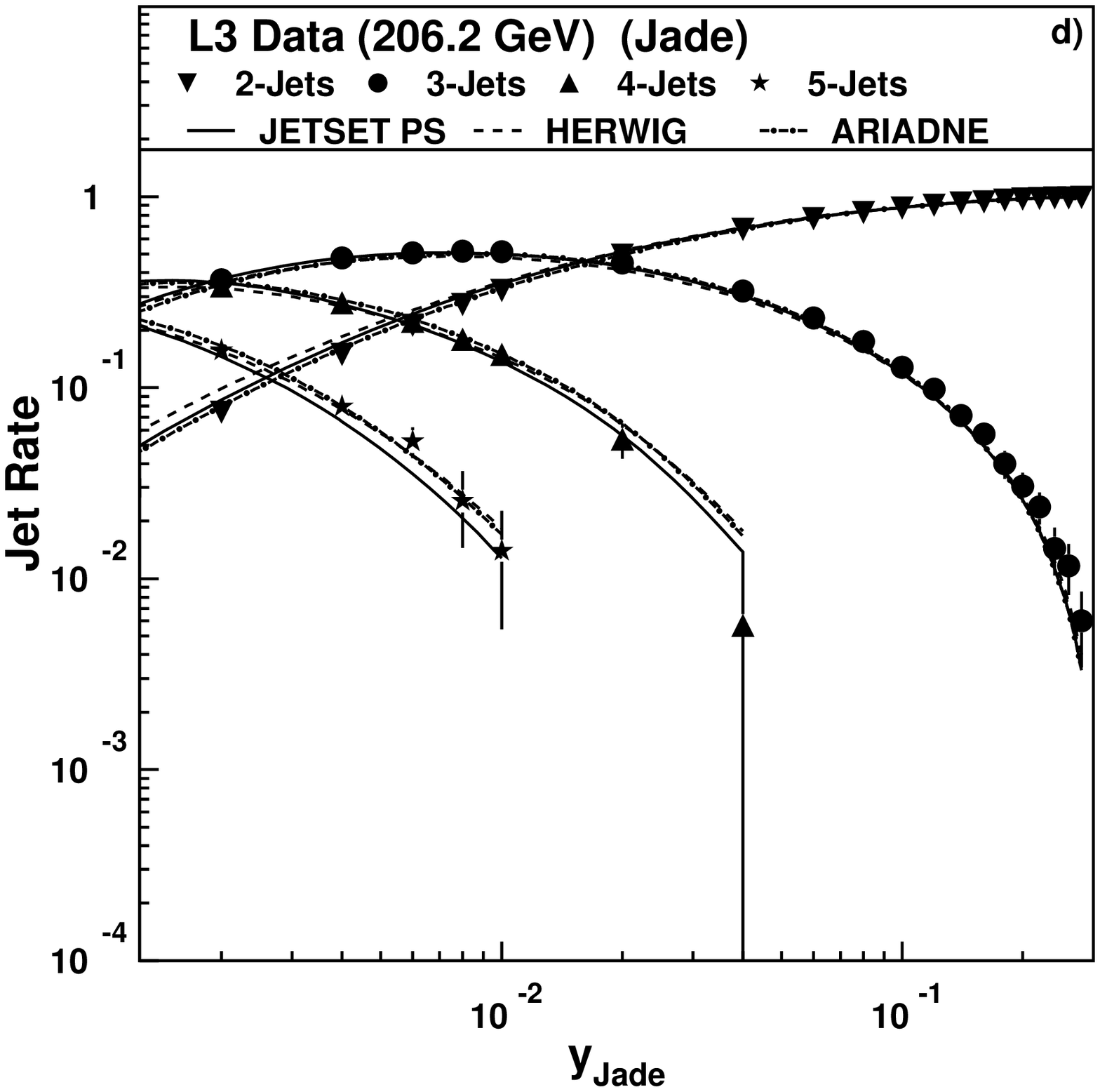} \\
\end{tabular}
\caption[ bla ]{ Jet production rates with the Durham (left) and JADE 
(right) algorithms measured at various values of \ycut\ by
L3~\cite{l3290}.  The 2-, 3-, 4- and 5-jet production rates corrected
for experimental effects are shown as points with error bars; the
lines indicate the corresponding Monte Carlo model predictions as
indicated on the figures.  }
\label{fig_l3_kt_206}
\end{figure}

Figure~\ref{fig_l3_kt_206} shows the fractions of 2-, 3-, 4- and 5-jet
events using the Durham (left) and JADE (right) algorithms measured by
L3 using the highest energy LEP~2 data at an average
$\roots=206$~GeV~\cite{l3290}.  The data are compared with predictions 
by Monte Carlo models and reasonable agreement is found.

\begin{figure}[htb!]
\begin{tabular}{cc}
\includegraphics[width=0.475\textwidth]{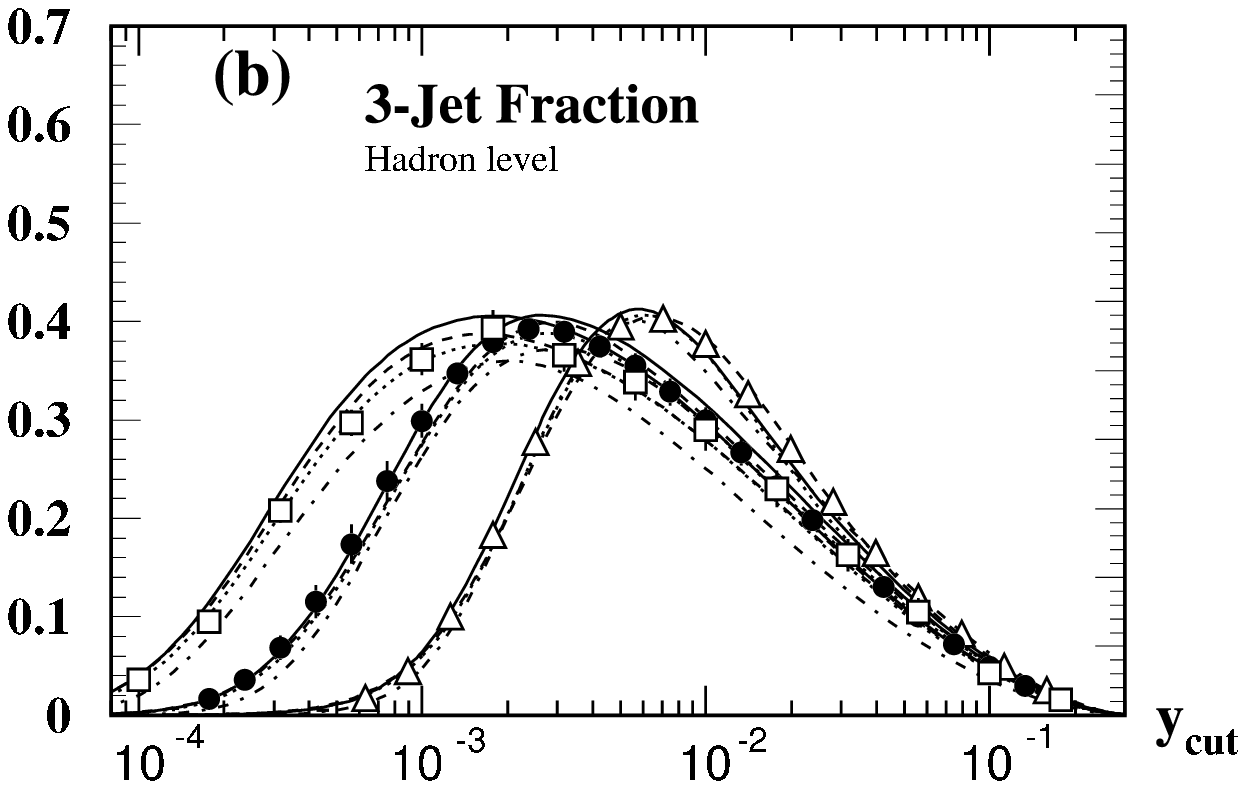} &
\includegraphics[width=0.475\textwidth]{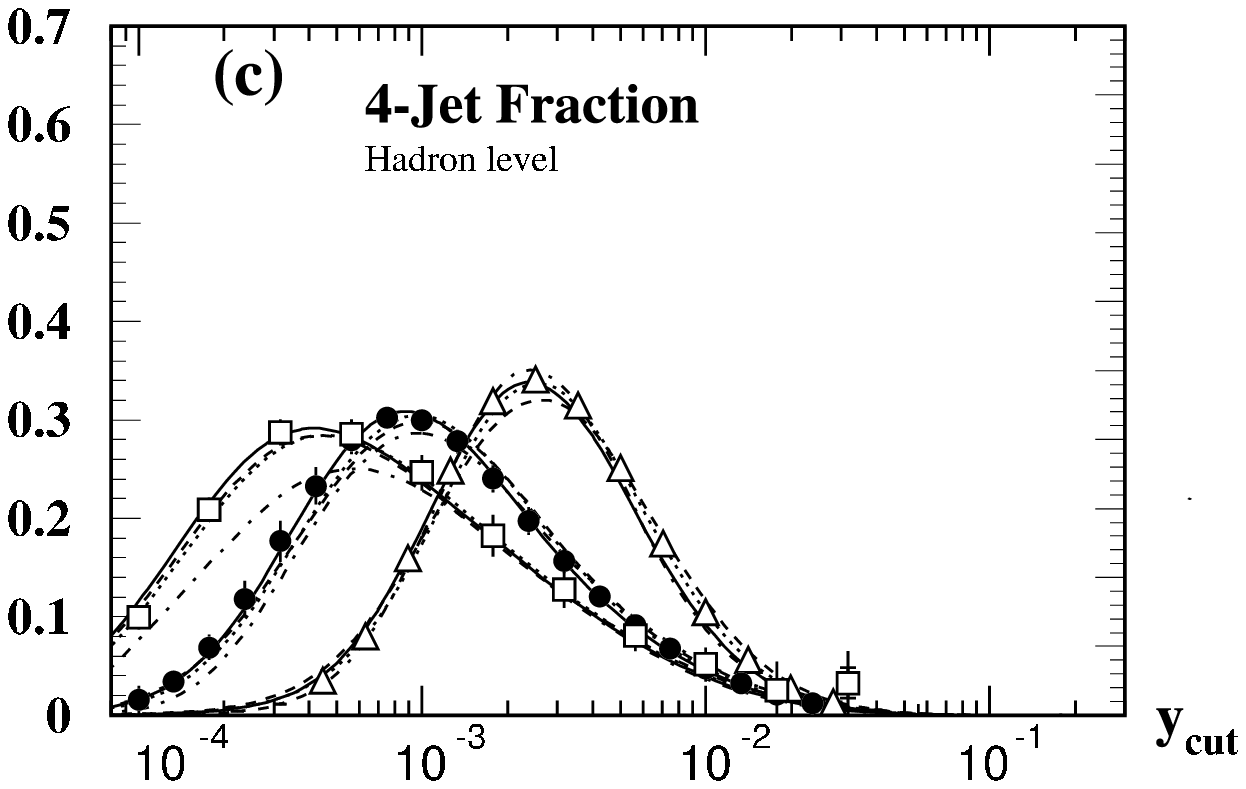} \\
\end{tabular}
\caption[ bla ]{ 3- and 4-jet production rates with the Durham algorithm
measured by JADE at $\roots=35$~GeV (triangles) and OPAL at
$\roots=91$~GeV (points) and 189~GeV (squares)~\cite{OPALPR299}.  The
data are corrected for experimental effects.  The predictions by Monte
Carlo models are indicated by the lines for PYTHIA (solid), HERWIG
(dashed), ARIADNE (dotted) and COJETS (dash-dotted). }
\label{fig_opal_durham}
\end{figure}

Figure~\ref{fig_opal_durham} displays in (b) the 3- and in (c) the
4-jet fraction determined with the Durham algorithm at $\roots=35$, 91
and 189~GeV~\cite{OPALPR299}.  The 35~GeV data stem from a reanalysis
of data from the JADE experiment while the 91 and 189~GeV data are
from OPAL.  These figures show that the same jet fractions are found
for smaller values of \ycut\ at increasing \roots.  In the regions of
large \ycut\ w.r.t.\ the peaks this effect is a consequence of the
running of \as.  The small decrease in the maximum 3-jet and 4-jet
fractions with increasing \roots\ is also due to the running of \as.
The data are compared to various Monte Carlo models tuned to OPAL data
at $\roots\simeq\mz$.  Good agreement is found for all models except
COJETS which fails to describe the data at high energy.

While the jet production rates are generally well described by the
usual Monte Carlo event generators it has been observed that the
detailed kinematics of 4-jet events are not always successfully
modelled~\cite{aleph143,aleph249,OPALPR404}.  The kinematics of 4-jet
events are studied using angular correlations between the jets, e.g.\
the Bengtsson-Zerwas angle~\cite{bengtsson88}
$\chibz=\angle([\vec{p}_1\times\vec{p}_2],
[\vec{p}_3\times\vec{p}_4])$, where the $\vec{p}_i$, $i=1\ldots 4$ are
the momentum vectors of the four jets after energy ordering.
Comparing LEP~1 data with simulations showed deviations of up to about
20\% while the data can be well described by LO (\oaa) or NLO (\oaaa)
calculations corrected for hadronisation
effects~\cite{delphi059,aleph143,OPALPR330,aleph249}.

The jet structure of hadronic events in \epem\ annihilation has also
been studied using a cone based algorithm as commonly used at hadron
colliders~\cite{OPALPR097,OPALPR299}.  An advantage of this study is
that the properties of the jets may be compared more directly to
results from hadron collider experiments.

\subsection{Event shape observables}
\label{sec_shapes}

The event shape observables avoid direct association of particles to
jets and calculate instead a single number which classifies the event
according to its jet topology.  Generally the observables are
constructed such that a value of zero corresponds to an ideal 2-jet
event consisting of only two back-to-back particles, a small value
corresponds to a realistic 2-jet event while increasingly larger
values indicate the presence of one or more additional jets.  A large
number of event shape observables have been proposed and studied.  We
will concentrate here on observables which are infrared and collinear
safe and which have the most complete QCD predictions including
contributions from resummed NLLA calculations.

\begin{description}

\item[Thrust \bm{T}, Thrust major and minor \bm{\tmaj} and \bm{\tmin}:]
  These observables are defined by the expression~\cite{thrust1,thrust2}
  \begin{equation}
    T= \max_{\vec{n}}\left(\frac{\sum_i|\vec{p}_i\cdot\vec{n}|}
                                {\sum_i|\vec{p}_i|}\right)\;\;\;.
  \label{equ_thrust}
  \end{equation}
  where $\vec{p}_i$ is the momentum of particle $i$ in an event.
  The thrust axis $\vec{n}_T$ is the vector $\vec{n}$ which
  maximizes the expression in parentheses.  A plane through the origin
  and perpendicular to $\vec{n}_T$ divides the event into two
  hemispheres $H_1$ and $H_2$.  A value of $T=1$ corresponds to an
  ideal 2-jet event; therefore the thrust observable is often used
  in the form \thr.  Based on the thrust axis $\vec{n}_T$ the thrust
  major \tmaj\ and the thrust major axis $\vec{n}_{\tmaj}$ are defined by 
  equation~(\ref{equ_thrust}) with the constraint 
  $\vec{n}_T\cdot\vec{n}=0$.  The thrust minor \tmin\ and the thrust
  minor axis $\vec{n}_{\tmin}$ are defined by the expression in parenthesis
  of equation~(\ref{equ_thrust}) with the constraint 
  $[\vec{n}_T\times\vec{n}_{\tmaj}]=\vec{n}_{\tmin}$.

\item[Heavy and Light Jet Mass \bm{\mh} and \bm{\ml}:] The hemisphere 
  invariant masses are calculated using  the particles
  in the two hemispheres $H_1$ and $H_2$.  We define
  \mh~\cite{jetmass,clavelli81} as the heavier mass and \ml\ as the
  lighter mass, both normalised to $\roots$.

\item[C- and D-parameter \bm{\cp} and \bm{\dpar}:]
  The linearised momentum tensor $\Theta^{\alpha\beta}$ is defined by
  \begin{displaymath}
    \Theta^{\alpha\beta}= \frac{\sum_i(p_i^{\alpha}p_i^{\beta})/|\vec{p}_i|}
                               {\sum_i|\vec{p}_i|}\;\;\;,
                           \;\;\;\alpha,\beta= 1,2,3\;\;\;.
  \end{displaymath}
  The three eigenvalues $\lambda_j$ of this tensor define
  \cp~\cite{ert} through
  \begin{displaymath}
    \cp= 3(\lambda_1\lambda_2+\lambda_2\lambda_3+\lambda_3\lambda_1)\;\;.
  \end{displaymath}
  An equivalent and numerically more convenient form is~\cite{dasgupta03} 
  \begin{displaymath}
    \cp= \frac{3}{2}\frac{\sum_{ij}|\vec{p}_i||\vec{p}_j|\sin^2\theta_{ij}}
                         {(\sum_i |\vec{p}_i|)^2}
  \end{displaymath}
  where $\theta_{ij}$ is the angle between particles $i$ and $j$.
  The D-parameter is defined by
  \begin{displaymath}
    \dpar= 27\lambda_1\lambda_2\lambda_3\;\;\;.
  \end{displaymath}

\item[Jet Broadening observables \bm{\bt}, \bm{\bw} and \bm{\bn}:] 
  These are defined by computing the quantity
  \begin{displaymath}
    B_k= \left(\frac{\sum_{i\in H_k}|\vec{p}_i\times\vec{n}_T|}
                    {2\sum_i|\vec{p}_i|}\right)
  \end{displaymath}
  for each of the two event hemispheres, $H_k$,  defined above.
  The three observables~\cite{nllabtbw} are defined by
  \begin{displaymath}
    \bt= B_1+B_2,\;\;\;\bw= \max(B_1,B_2)\;\;\;
         \mathrm{and}\;\;\;\bn= \min(B_1,B_2)
  \end{displaymath}
  where \bt\ is the total, \bw\ is the wide and \bn\ is the narrow
  jet broadening.

\item[Transition value between 2 and 3 jets {\boldmath \ytwothreed}:]
 The Durham jet algorithm (see section~\ref{sec_jets}) is applied to
 cluster the particles of an hadronic event into two or three jets.
 The value of \ycut\ at which for an event the transition between a
 2-jet and a 3-jet assignment occurs is called
 \ytwothreed~\cite{komamiya90}.

\end{description}

We will refer to event shape observables in general by the symbol $y$.
The observables \tmin, \ml, \dpar\ and \bn\ belong to the class of
4-jet observables, because for these we have in the cms $y>0$ only for
events with at least four partons in the final state\footnote{Due to
momentum conservation in the cms these observables have $y=0$ for
back-to-back two parton as well as for planar three parton
configurations.}.  We also note that some of the observables take the
whole event into account while others depend only on one selected
hemisphere of the event; the observables \mh, \ml, \bw, \bn\ and
\ytwothreed\ fall into the latter category.

\begin{figure}[htb!]
\begin{tabular}{cc}
\includegraphics[width=0.5\textwidth]{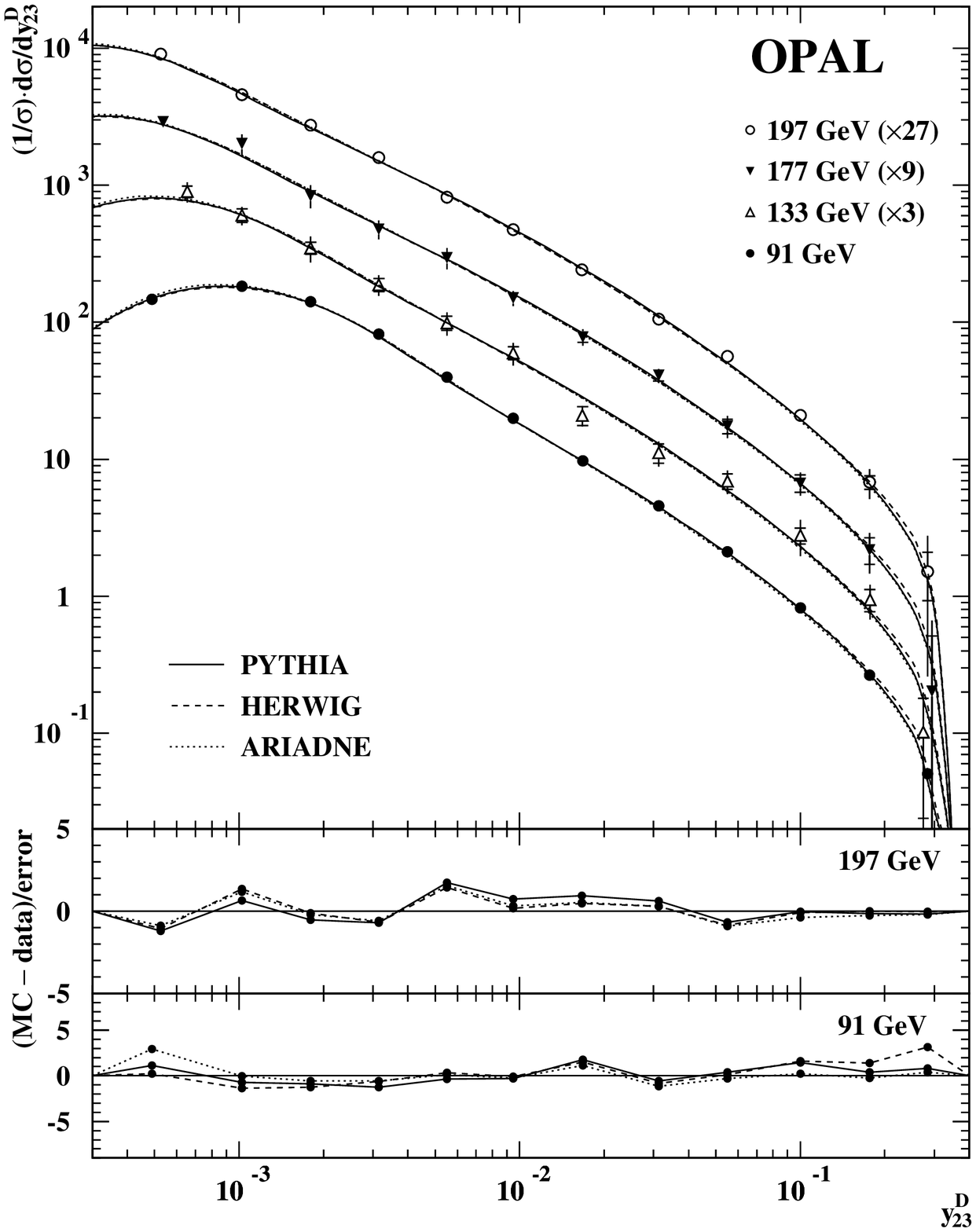} &
\includegraphics[width=0.5\textwidth]{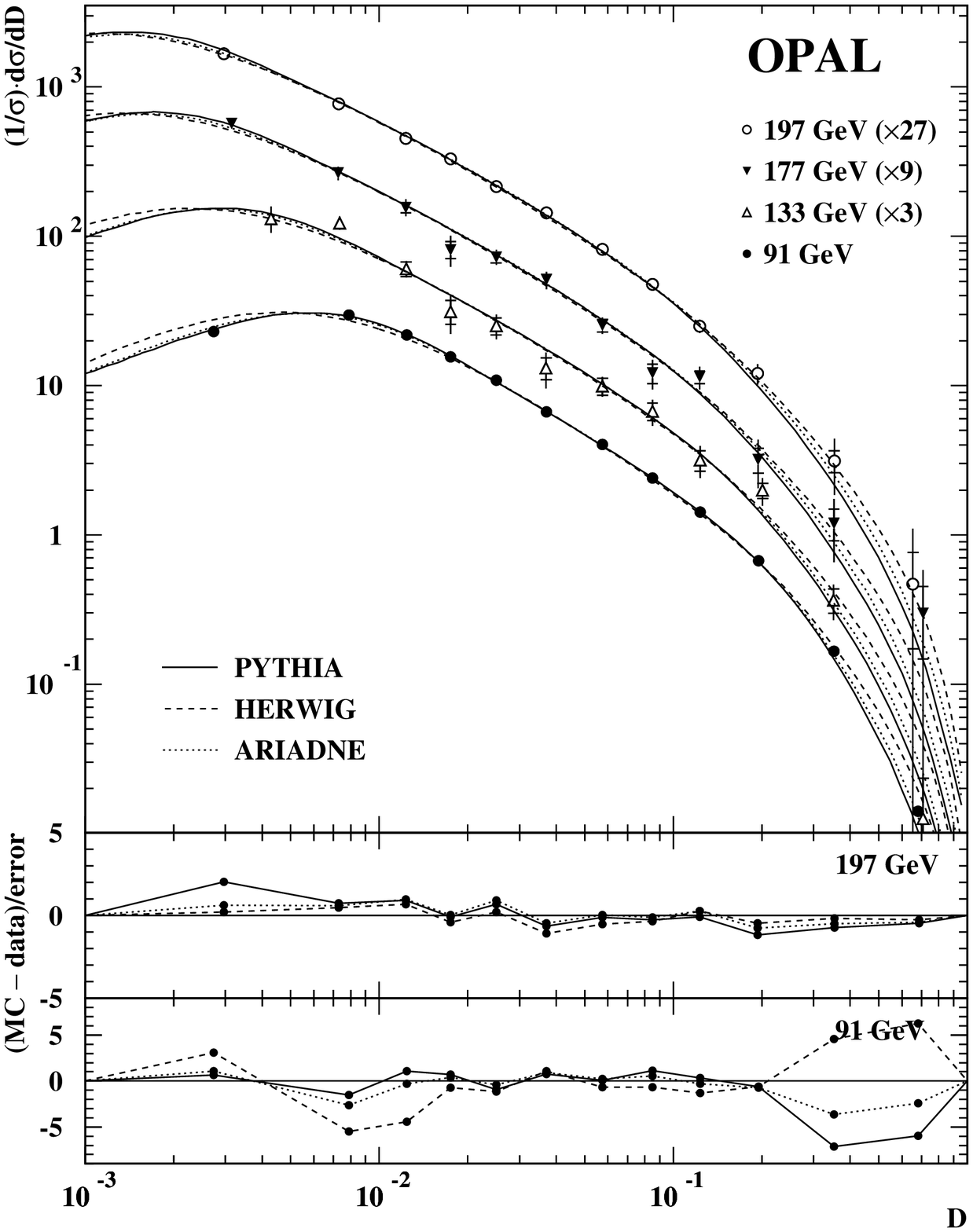} \\
\end{tabular}
\caption[ bla ]{ Distributions of event shape observables \ytwothreed\
(left) and \dpar\ (right) measured by OPAL~\cite{OPALPR404}.  The data
are corrected for experimental effects.  The high energy LEP~2 data
collected at several energy points are combined into two samples at
average $\roots=177$ and 197~GeV.  The data sets are separated
vertically by multiplicative factors as indicated for clarity.  The
lines show the predictions of Monte Carlo models as indicated on the
figures.  The lower parts of the figures present the deviations
between data and Monte Carlo model predictions at $\roots=91$ and
197~GeV. }
\label{fig_opal_evsh}
\end{figure}

Figure~\ref{fig_opal_evsh} presents measurements by OPAL at average
$\roots=91, 133, 177$ and 197~GeV of the event shape observables
\ytwothreed\ (left) and \dpar\ (right) representing 3- and 4-jet
observables~\cite{OPALPR404}.  The data are corrected for experimental
effects and compared to predictions by Monte Carlo models.  One finds
that the data for \ytwothreed\ are well described by the models at all
energies.  The data for \dpar\ are also well described by the models at
high energies $\roots>\mz$ where the statistical and experimental errors are 
comparatively large.  However, at $\roots=91$~GeV and large values of \dpar\
significant deviations between models and data occur; for HERWIG there are
also discrepancies at $\dpar\approx 10^{-2}$.  This observation could be
related to the unsuccessful modelling of the kinematics of 4-jet events
discussed above.

\begin{figure}[htb!]
\begin{tabular}{cc}
\includegraphics[width=0.475\textwidth]{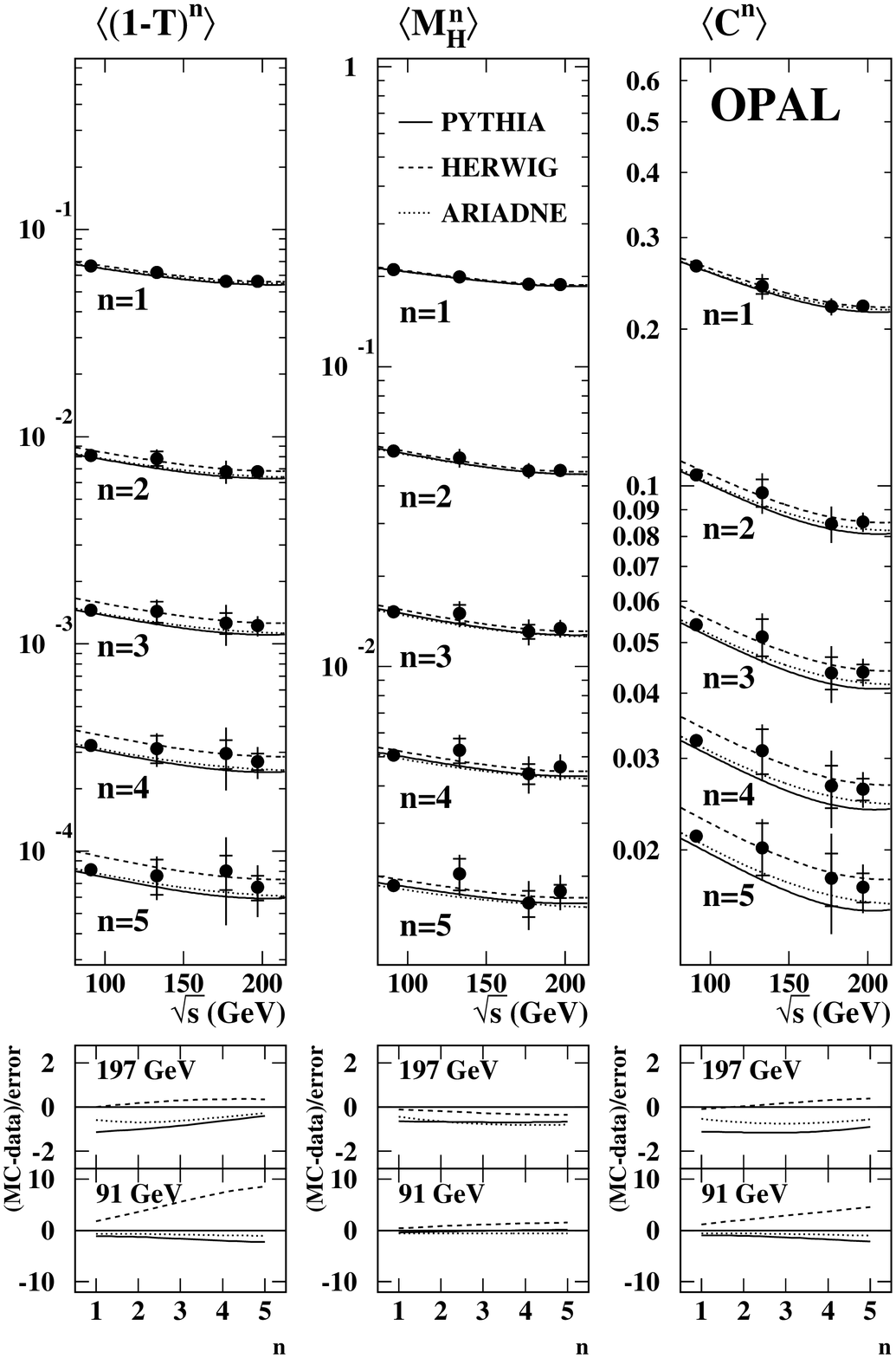} &
\includegraphics[width=0.475\textwidth]{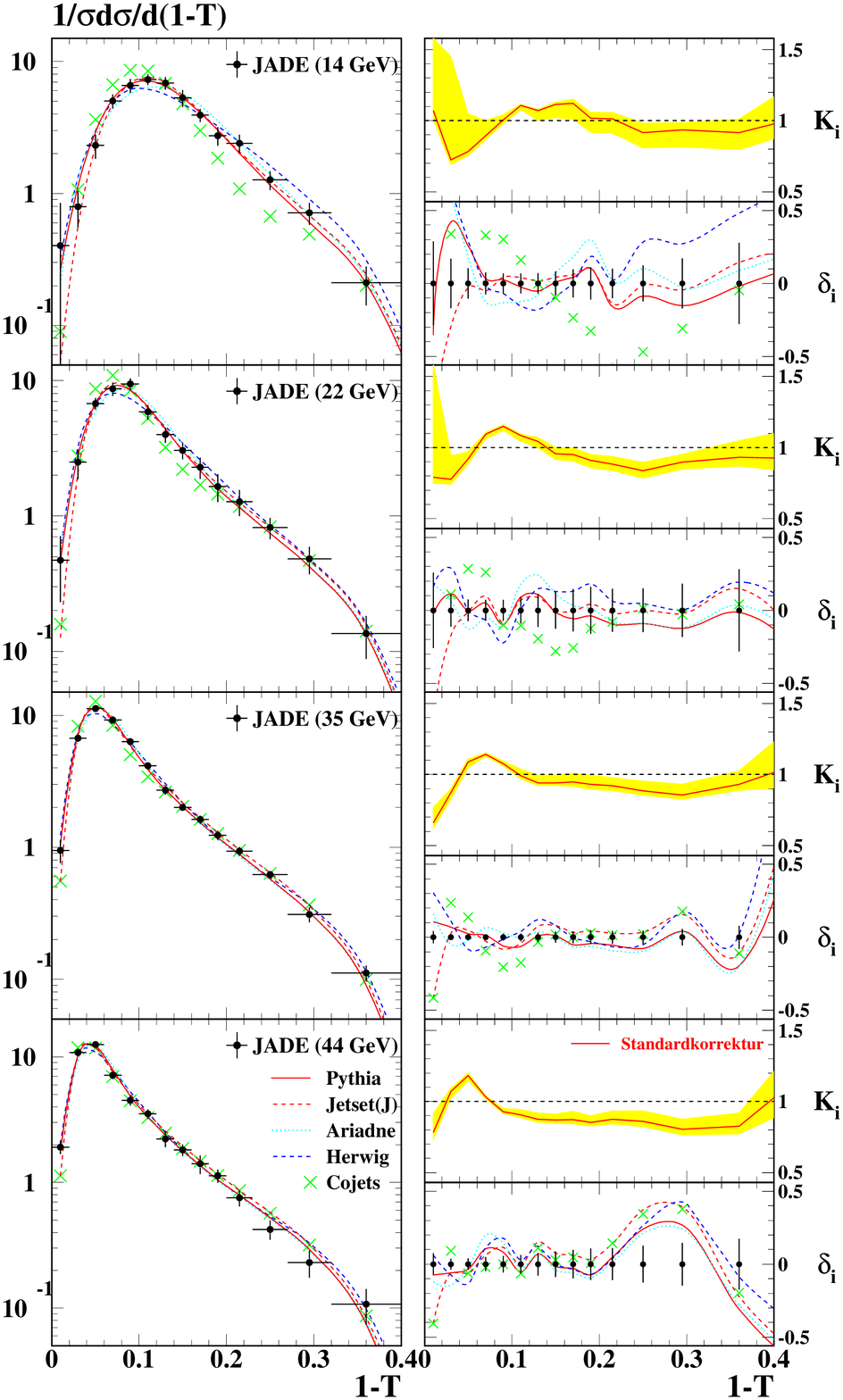} \\
\end{tabular}
\caption[ bla ]{ (left) Moments $n=1,\ldots,5$ of event shape observables
\thr, \mh\ and \cp\ measured by OPAL~\cite{OPALPR404}.  The data are
corrected for experimental effects.  The lines show the predictions by
Monte Carlo models as indicated.  The lower parts of the figure
present the deviations between data and Monte Carlo models at
$\roots=91$ and 197~GeV.  (right) Distributions of \thr\ measured
using JADE data and corrected for experimental effects at $\roots$=14
to 44~GeV~\cite{movilla02a,kluth03}. The lines or crosses show
predictions by Monte Carlo models as indicated.  The set of figures on
the RHS present the correction factors $K_i$ (lines) and their
uncertainties (shaded bands), and the differences $\delta_i$ between
data and Monte Carlo model predictions together with the errors
of the data. }
\label{fig_opal_moms_jade_thr}
\end{figure}

Another way of studying the structure of hadronic events in \epem\
annihilation is to calculate moments of the event shape observable
distributions:
\begin{displaymath}
  \momn{y}{n} = \int_0^{\ymax} y^n 
                \frac{1}{\sigma} \frac{\dd\sigma}{\dd y} \dd y \;,
\end{displaymath}
where \ymax\ is the maximal kinematically allowed value of the
observable.  The moments always sample all of the available phase
space with a different weighting depending on the value of $n$;
moments with small values of $n$ are dominated by 2- and 3-jet events
while for larger values of $n$ multi-jet events become more important.
Figure~\ref{fig_opal_moms_jade_thr} (left) shows the first five
moments of the observables \thr, \mh\ and \cp\ measured by
OPAL~\cite{OPALPR404} at $\roots=91$ to 197~GeV.  The data are
compared to predictions by Monte Carlo models and good agreement is
found, except for HERWIG which shows deviations from the moments of
\thr\ and \cp\ increasing with $n$.

The reanalysis of data from the JADE experiment made it possible to 
compare data for event shape observables at $\roots<\mz$ with the
predictions of current Monte Carlo models~\cite{movilla02a,kluth03}.  
Figure~\ref{fig_opal_moms_jade_thr} (right) presents as an example 
measurements of distributions of \thr\ at $\roots=14$, 22, 35 and 44~GeV.
The data from the JADE experiment are corrected for experimental effects
and the presence of $\epem\rightarrow\bbbar$ events corresponding to
about 9\% of the event samples.  The predictions of Monte Carlo
models tuned by OPAL to LEP~1 data are superimposed and found to
agree well with the data; the exception is COJETS which deviates from
the data in particular at low \roots.

\subsection{Experimental tests of asymptotic freedom} 

Jet rates and event shape observables are well suited to investigate
experimentally one of the most important predictions of QCD, namely
asymptotic freedom of the strong coupling, see sections~\ref{sec_basics} 
and~\ref{sec_hist}.  The large range in cms energies \roots\ provided 
by the LEP data and also the earlier data from previous \epem\
colliders gives direct access to effects caused by the running strong
coupling constant.

\begin{figure}[htb!]
\begin{tabular}{cc}
\includegraphics[width=0.5\textwidth]{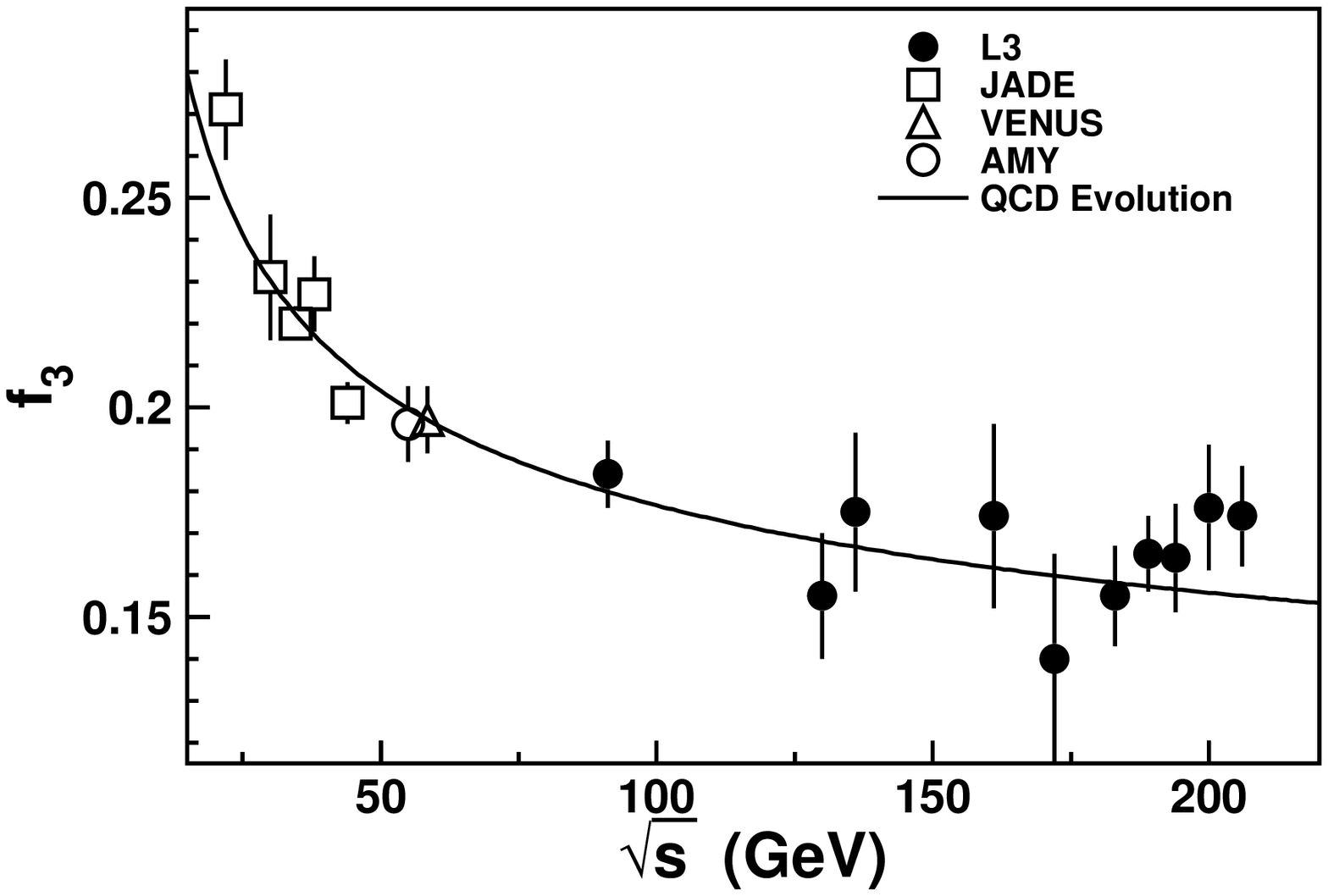} &
\includegraphics[width=0.5\textwidth]{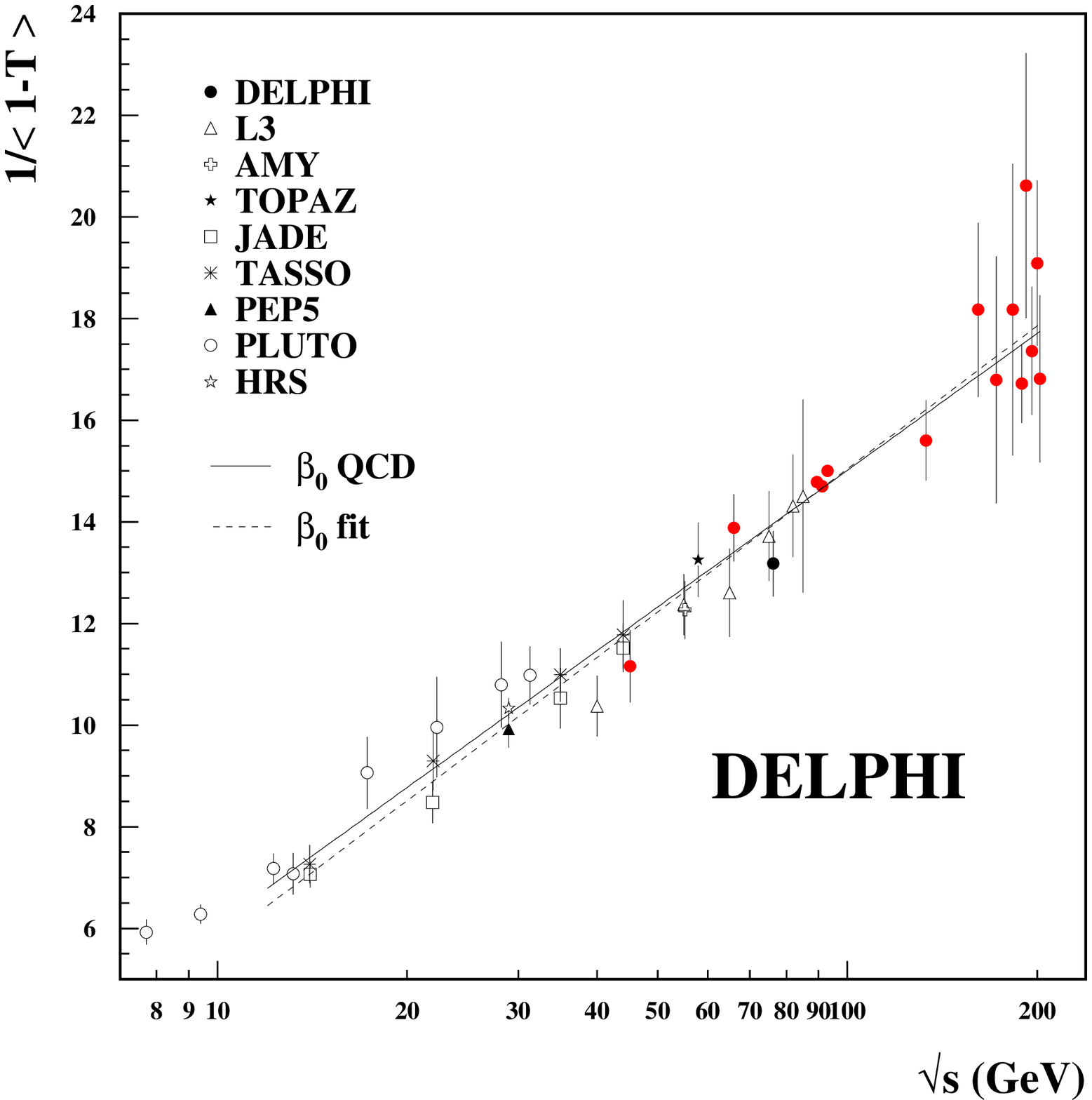} \\
\end{tabular}
\caption[ bla ]{ 3-jet fraction $f_3$ with the JADE E0 algorithm at 
$\ycut=0.08$ (left) and $1/\momone{\thr}$ (right) measured as a
function of the cms energy~\cite{l3290,delphi296}.  The data are
compared with predictions by QCD shown by the solid lines. }
\label{fig_runas}
\end{figure}

Following earlier studies~\cite{bethke88,braunschweig88} the L3
collaboration measured the fraction of 3-jet events using the JADE~E0
algorithm at $\ycut=0.08$~\cite{l3290}.  This measurement is
especially suited to study asymptotic freedom, since the hadronisation
corrections as predicted by Monte Carlo models are known to be small
(less than 10\%) and fairly model
independent~\cite{OPALPR025,bethke92}.  The L3 data and results from
other experiments spanning the range $\roots=22$ to 206~GeV are shown
in figure~\ref{fig_runas} (left) and compared with the \oaa\
perturbative QCD expectation with $\asmz=0.12$.  The prediction
clearly agrees with the data.

Figure~\ref{fig_runas} (right) presents the results of a similar
analysis by DELPHI~\cite{delphi296} using the mean values
\momone{\thr} of the \thr\ distributions measured at energies ranging
from below 10~GeV to 202~GeV.  The figure shows the quantity
$1/\momone{\thr}$ vs.\ \roots\ on a logarithmic scale, because from
equations~(\ref{equ_asrunone}) and~(\ref{equ_3jetnlo}) a logarithmic
dependence in LO is predicted.  This prediction is well confirmed by
the data.  The solid line shows the QCD prediction in NLO in a special
formulation independent of a particular renormalisation
scheme~\cite{dhar84,korner00} (see section~\ref{sec_rgi}) which is
also seen to suppress non-perturbative effects.

\subsection{Measurements of \as} 

\subsubsection{3-jet observables}
\label{sec_lepjadeas}

Early measurements of \as\ before the start of LEP using event shape
or jet observables were briefly summarised in section~\ref{sec_hist}.
The first analyses by the LEP collaborations used \oaa\ QCD
calculations and already a set of Monte Carlo event generators
including QCD coherence effects and tuned to the precise LEP data, see
e.g.~\cite{bethke/pilcher} for a review.  An average value of
$\asmz=0.120\pm0.006$ was determined, where the error is dominated by
theoretical uncertainties estimated by variation of the
renormalisation scale in the QCD calculations.  Compared to the
earlier determinations of \asmz\ the error of the combined LEP
measurements was reduced by a factor of more than two.  This
improvement was mostly due to the use of consistent hadronisation
models as well as reliable \oaa\ QCD calculations~\cite{yellow1qcd}.
The error was also seen to be consistent with the scatter of results
from individual observables; this is an important cross-check of the
methods for estimating the errors.  A reanalysis of low energy data at
$\roots=35$~GeV from the JADE experiment using the methods developed
for LEP obtained $\asmz=0.122^{+0.008}_{-0.005}$ consistent with the
LEP result~\cite{jadenewas}.

A significant improvement in the determinations of \as\ based on event
shape or jet observables was the introduction of resummed calculations
matched to the fixed order predictions (\oaa+NLLA) discussed in
section~\ref{sec_basics}.  These calculations allow the data to be
described over larger regions of phase space in the 2-jet region and
have generally a weaker dependence on the renormalisation
scale~\cite{OPALPR075}.  A combination of values of \asmz\ obtained
from analyses at $\roots\approx\mz$ using \oaa+NLLA calculations was
given in~\cite{bethke00a} as $\asmz=0.121\pm0.006$.  Thus the total
error was not improved compared to \oaa\ based analyses but the result
is considered to be more reliable theoretically due to the reduced
dependence on the renormalisation scale.

Recently the LEP experiments have published their final results on
determinations of \as\ based on data at $\roots=\mz$ up to the highest
LEP~2 energies~\cite{aleph265,delphi327,l3290,OPALPR404}.  These
analyses use a consistent set of observables: \thr, \mh, \bt, \bw\ and
\cp\ and in addition \ytwothreed\ by ALEPH and OPAL.  All analyses
employ the most recent calculations for the resummed NLLA
calculations~\cite{nllathmh,nllabtbw2,nllacp,banfi01} and the modified
\lnr-matching scheme and thus the results may be compared directly.

\begin{figure}[htb!]
\begin{tabular}{cc}
\includegraphics[width=0.475\textwidth]{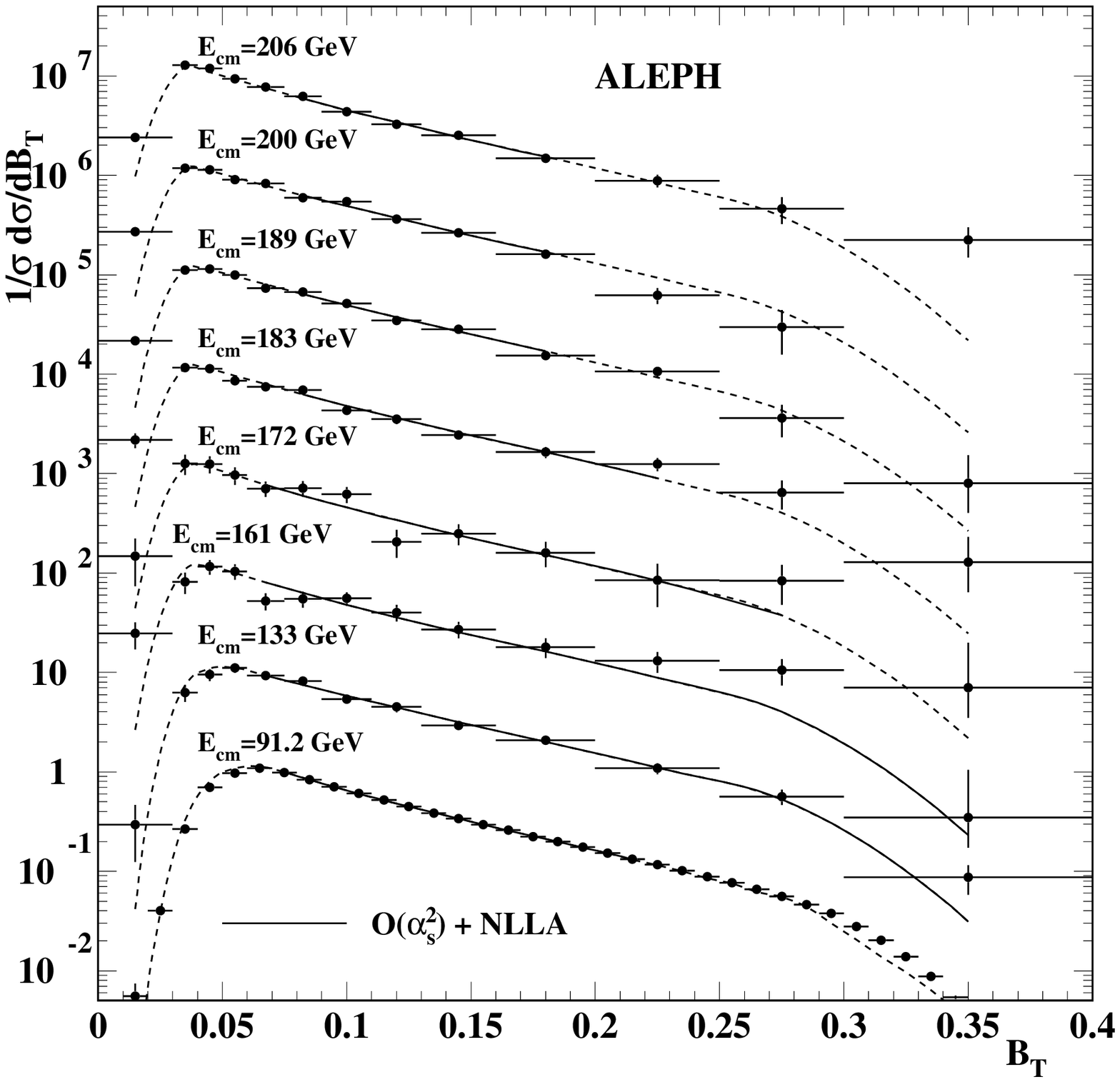} &
\includegraphics[width=0.475\textwidth]{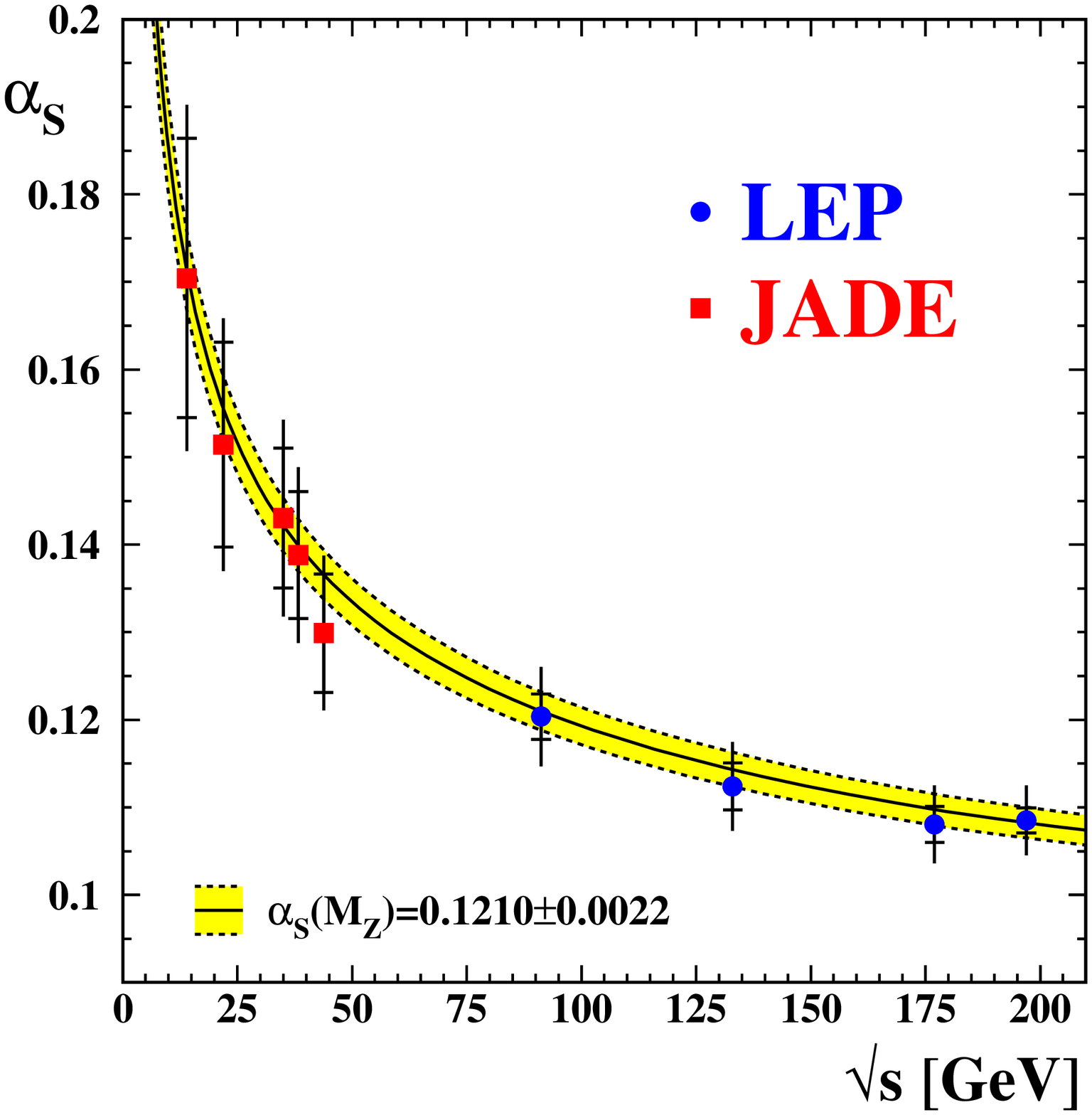} \\
\end{tabular}
\caption[ bla ]{ (left) Distributions of \bt\ measured by ALEPH at
several energy points as indicated~\cite{aleph265}.  Superimposed are
fits of the QCD prediction corrected for hadronisation effects.  The
fits have been performed in the ranges shown by the solid lines while
the dashed lines show an extrapolation of the fitted predictions.
(right) Combined values of \as\ based on the analyses by the LEP
experiments or using JADE data (see text) are shown.  The total
(inner) errors bars show total (experimental) uncertainties.  The
lines and shaded band show the evolution of \as\ based on the value of
\asmz\ given on the figure.  }
\label{fig_lepas}
\end{figure}

Figure~\ref{fig_lepas} (left) shows as a representative example the
results of the analysis by ALEPH using the observable \bt\ at
$\roots=91.2$ to 206~GeV~\cite{aleph265}.  The data are corrected for
experimental effects and the curves present the \oaa+NLLA QCD
prediction using the modified \lnr-matching corrected for
hadronisation.  The fits have been performed over ranges indicated by
the solid lines while the dashed lines are an extrapolation of the
theory using the fitted results.  The agreement between the theory
including hadronisation corrections and the data is good in the fitted
regions and becomes worse in the extreme 2-jet and multijet regions
where the theory is not expected to be able to completely describe the
data.

In order to study asymptotic freedom the individual results on \as\
from the most recent LEP
analyses~\cite{aleph265,delphi327,l3290,OPALPR404} are combined using
a common method, see
e.g.~\cite{aleph265,OPALPR404,cowan98a,mtfordphd}.  The values of \as\
to be combined are evolved to the common scale of the combination if
necessary.  Then the average value \asb\ is computed by minimising
\begin{equation}
  \chi^2 = (\vec{\alpha}-\asb) \cdot V^{-1} \cdot 
           (\vec{\alpha}-\asb)^T
\end{equation}
w.r.t.\ to \asb.  The individual measurements ${\as}_{,i}$ are
collected in the vector $\vec{\alpha}$ and $V$ is the covariance
matrix of the measurements.  The final expression for the average
value is
\begin{equation}
  \asb = \sum_i w_i {\as}_{,i} \,,\;
   w_i = \frac{\sum_j(V^{-1})_{ij}}{\sum_{j,k}(V^{-1})_{jk}}\;\;.
\end{equation}
The correlations between experimental, hadronisation and theoretical
uncertainties are generally not well known, in particular between
different experiments, and thus models for the correlations have to be
constructed.  In these models three levels of correlation $\rho_{ij}$
between the errors $\sigma_i$ and $\sigma_j$ of two measurements $i$
and $j$ may be assumed: i) no correlations $\rho_{ij}=0$, ii) partial
correlation $\rho_{ij}= \min(\sigma_i,\sigma_j)^2/ (\sigma_i\sigma_j)$
or iii) full correlation $\rho_{ij}=1$.  The treatment of the
different systematic uncertainties is as follows:
\begin{description}

\item[Experimental uncertainties]
These are added to the diagonal and off-diagonal 
elements of $V$ using one of the three possible models for
their correlation.

\item[Hadronisation uncertainties]
These are added only to the diagonal elements of $V$.  The
hadronisation uncertainty is evaluated by repeating the combination
with individual ${\as}_{,i}$ values obtained with different hadronisation
models.

\item[Theoretical uncertainties]
These are also added only to the diagonal elements of $V$.  The
theoretical uncertainty is evaluated by varying all individual 
values ${\as}_{,i}$ simultaneously within their theoretical uncertainties 
and repeating the combination.

\end{description}
This procedure suppresses the presence of negative weights $w_i$ which
otherwise appear when large off-diagonal terms are included in
$V$~\cite{mtfordphd}.  Such negative weights are mathematically
correct but difficult to interpret as a physical result.  We therefore
choose a treatment of the correlations of the generally large
hadronisation and theoretical uncertainties which avoids negative
weights as in~\cite{aleph265,OPALPR404}.

In a first step the individual results for each observable and
experiment are combined in the energy ranges $130\le\roots\le
136$~GeV, $161\le\roots\le 183$~GeV and $189\le\roots\le 207$~GeV
corresponding to average centre-of-mass energies $\roots=133, 177$ and
197~GeV.  In these combinations the statistical errors are taken as
uncorrelated and the experimental errors as partially correlated.
OPAL publishes results directly at $\roots=177$ and 197~GeV and also
at points from smaller energy ranges.  Our averages reproduce the OPAL
results within 1\% and with consistent uncertainties. The single
DELPHI result from $\roots=183$~GeV is simply evolved to 177~GeV.

The sets of results for each observable and experiment are further
combined within each experiment.  The statistical correlations between
observables as obtained by ALEPH and OPAL at $\roots=206$~GeV are
taken from~\cite{mtfordphd}.  The experimental uncertainties are again
taken as partially correlated.  Our values for \as\ are shown in
table~\ref{tab_ascomb}.  The corresponding combined results from OPAL
have been calculated using the same method except for a more detailed
treatment of hadronisation uncertainties.  Our combined results again
reproduce the OPAL results within 1\% with consistent errors.

\begin{table}[htb!]
\caption[ bla ]{ Results for average values of \as\ from LEP
experiments~\cite{aleph265,delphi327,l3290,OPALPR404}.  The columns
show the results for $\as(\roots)$, the statistical (stat.),
experimental (exp.), hadronisation (had.) and theoretical (theo.)
uncertainties.  The last two rows give average values of \asmz\ of all
LEP~2 data from $\roots=133$ to 197~GeV and of all LEP data. }
\label{tab_ascomb}
\begin{indented}\item[]
\begin{tabular}{l c@{$\pm$}c@{$\pm$}c@{$\pm$}c@{$\pm$}c } \hline\hline
       & \asmz  & stat.  & exp.   & had.   & theo. \\
\hline
ALEPH  & 0.1201 & 0.0002 & 0.0007 & 0.0020 & 0.0050 \\
DELPHI & 0.1216 & 0.0002 & 0.0021 & 0.0024 & 0.0056 \\
L3     & 0.1211 & 0.0008 & 0.0014 & 0.0035 & 0.0050 \\
OPAL   & 0.1193 & 0.0002 & 0.0007 & 0.0024 & 0.0047 \\
LEP    & 0.1204 & 0.0002 & 0.0006 & 0.0025 & 0.0051 \\
\hline
       & \as(133 GeV) & stat.  & exp.   & had.   & theo. \\ 
\hline
ALEPH  & 0.1156 & 0.0028 & 0.0008 & 0.0014 & 0.0048 \\
L3     & 0.1133 & 0.0023 & 0.0018 & 0.0024 & 0.0048 \\
OPAL   & 0.1091 & 0.0033 & 0.0033 & 0.0016 & 0.0036 \\
LEP    & 0.1124 & 0.0017 & 0.0011 & 0.0018 & 0.0043 \\
\hline
       & \as(177 GeV) & stat.  & exp.   & had.   & theo. \\
\hline
ALEPH  & 0.1106 & 0.0024 & 0.0008 & 0.0010 & 0.0040 \\
DELPHI & 0.1093 & 0.0035 & 0.0020 & 0.0017 & 0.0043 \\
L3     & 0.1077 & 0.0020 & 0.0014 & 0.0018 & 0.0045 \\
OPAL   & 0.1074 & 0.0010 & 0.0011 & 0.0009 & 0.0032 \\
LEP    & 0.1081 & 0.0013 & 0.0009 & 0.0013 & 0.0040 \\
\hline
       & \as(197 GeV) & stat.  & exp.   & had.   & theo. \\
\hline
ALEPH  & 0.1075 & 0.0012 & 0.0009 & 0.0008 & 0.0036 \\
DELPHI & 0.1083 & 0.0013 & 0.0021 & 0.0017 & 0.0042 \\
L3     & 0.1124 & 0.0007 & 0.0013 & 0.0014 & 0.0046 \\
OPAL   & 0.1074 & 0.0010 & 0.0011 & 0.0009 & 0.0032 \\
LEP    & 0.1085 & 0.0006 & 0.0007 & 0.0011 & 0.0038 \\
\hline
       & \asmz & stat.  & exp.   & had.   & theo. \\
\hline
LEP~2  & 0.1200 & 0.0007 & 0.0010 & 0.0016 & 0.0048 \\
{\bf LEP all} & {\bf 0.1201} & {\bf 0.0005} & {\bf 0.0008} 
                             & {\bf 0.0019} & {\bf 0.0049} \\
\hline\hline
\end{tabular}
\end{indented}
\end{table}

Applying the same procedure to the reanalysed data from the JADE
experiment collected at $\roots=14$ to 44~GeV~\cite{pedrophd} yields
the results shown in table~\ref{tab_jadeascomb}\footnote{The
statistical errors at $\roots=14$ and 22~GeV were estimated using the
statistical error at 38~GeV.}.  The analysis of the JADE data was
performed using the same event shape observables, QCD calculations and
essentially the same Monte Carlo programs to calculate the
hadronisation corrections as for the LEP analyses and therefore the
results may be compared directly.

\begin{table}[htb!]
\caption[ bla ]{ Results for average values of \as\ from JADE.  The
last row gives the average value of \asmz\ from all JADE data. }
\label{tab_jadeascomb}
\begin{indented}\item[]
\begin{tabular}{r c@{$\pm$}c@{$\pm$}c@{$\pm$}c@{$\pm$}c } \hline\hline
 \roots & $\as(\roots)$ & stat.  & exp.   & had.   & theo. \\
\hline
 14  & 0.1704 & 0.0044 & 0.0041 & 0.0148 & 0.0117 \\
 22  & 0.1514 & 0.0040 & 0.0030 & 0.0106 & 0.0084 \\
 35  & 0.1430 & 0.0010 & 0.0022 & 0.0076 & 0.0079 \\
 38  & 0.1388 & 0.0035 & 0.0030 & 0.0056 & 0.0070 \\
 44  & 0.1299 & 0.0020 & 0.0031 & 0.0057 & 0.0057 \\
\hline
     & \asmz & stat.  & exp.   & had.   & theo. \\
\hline
{\bf JADE} & {\bf 0.1203} & {\bf 0.0007} & {\bf 0.0017} 
                          & {\bf 0.0053} & {\bf 0.0050} \\
\hline\hline
\end{tabular}
\end{indented}
\end{table}

The average values of \as\ determined at each \roots\ as shown in
tables~\ref{tab_ascomb} and~\ref{tab_jadeascomb} are presented in
figure~\ref{fig_lepas} (right).  The data are compared with the QCD
prediction for the running of \as\ calculated in NNLO based on the
average value $\asmz=0.1211\pm0.0021$ (see section~\ref{sec_as})
which has been determined using only NNLO QCD calculations.  The
agreement with the data is excellent and provides strong evidence for
asymptotic freedom.

Measurements of the first five moments of the six event shape
observables discussed above and as shown partially in
figure~\ref{fig_opal_moms_jade_thr} (left) have been used by OPAL to
determine the strong coupling constant using only NLO (\oaa) QCD
predictions~\cite{OPALPR404}.  The NLLA predictions break down at very
small values of the observables $y\ll 1$ and also become unreliable
close to the kinematic limits \ymax\ and thus cannot be used to derive
moments.  The divergences of the NLO predictions for $y\rightarrow 0$
are not a problem, since these are integrable and stable predictions
for the moments are obtained.  After demanding that the ratio of the
NLO to the LO terms in the perturbative QCD predictions be less
than 0.5 results from 17 observables are considered: \momone{\thr}, 
\momone{\cp}, \momone{\bt}, \momn{\bw}{n} and \momn{\ytwothree}{n}
with $n=1,\ldots,5$, and \momn{\mh}{n} with $n=2,\ldots,5$.  The final
combined result is $\asmz=0.1223\pm0.0005\stat \pm0.0014\expt
\pm0.0016\had ^{+0.0054}_{-0.0036}\theo$ in good agreement with the
results from distributions discussed above.  This analysis constitutes
a valuable cross check of the determination of \as\ from 3-jet
observables, since with moments the complete phase space is used.

\subsubsection{4-jet observables}
\label{sec_r4as}

The study of 4-jet observables for measurements of \asmz\ is
promising, because the sensitivity of these observables to the strong
coupling is doubled compared to 3-jet observables.  The relative
change in \as\ induced by a change of a 3-jet observable $O_3\sim\as$
is given by $\Delta\as/\as=\Delta O_3/O_3$ while for a 4-jet
observable $O_4\sim\as^2$ we have $\Delta\as/\as=\Delta O_4/(2 O_4)$.

The ALEPH collaboration has performed a measurement of \asmz\ using
the fraction of 4-jet events \rfour\ determined using the Durham
algorithm in hadronic \znull\ decays~\cite{aleph249}.  The QCD
prediction is known in \oaaa, i.e.\  LO is \oaa\ and NLO radiative
corrections are \oaaa~\cite{dixon97,nagy98b}.  These calculations are
matched with existing NLLA calculations.

\begin{figure}[!htb]
\includegraphics*[width=0.5\textwidth]{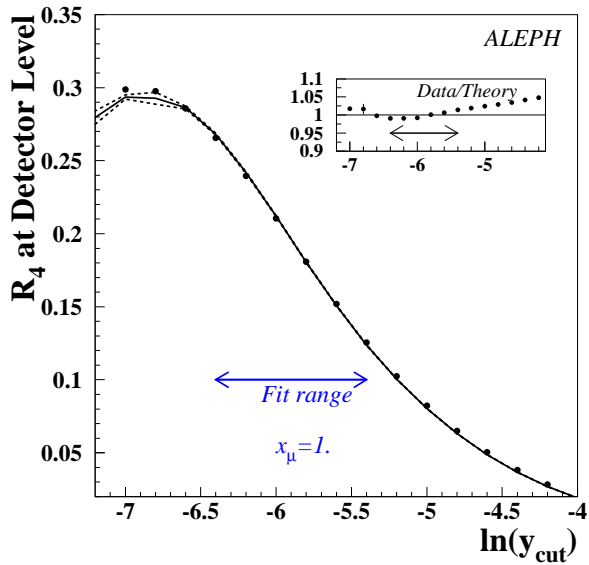}
\caption{ The 4-jet fraction \rfour\ determined using the Durham
algorithm is shown as a function of \ycut\ without corrections.
Superimposed is a fit of \oaaa+NLLA QCD (solid line) corrected for
hadronisation and detector effects~\cite{aleph249}. }
\label{fig_aleph4jet}
\end{figure}

The measurement of \asmz\ based on these predictions is presented
in figure~\ref{fig_aleph4jet}~\cite{aleph249}.  The uncorrected data
for \rfour\ measured with $2.5\cdot 10^6$ hadronic \znull\ decays
using the Durham algorithm are shown as a function of the jet
resolution parameter \ycut.  The line shows the fit of the
\oaaa+NLLA (R-matching) theory corrected for hadronisation and detector
effects.  The fit is performed using only the bins within the fit
range as indicated where the total correction factors deviate from
unity by less than 10\%.  The result of the fit for fixed
renormalisation scale parameter $\xmu=1$ is $\asmz= 0.1170
\pm0.0001\stat \pm0.0022\syst$ using the conservatively estimated
systematic uncertainty from~\cite{aleph249}.  The systematic error has
contributions from experimental ($\pm0.0009$), hadronisation
($\pm0.0010$) and theoretical ($\pm0.0017$) uncertainties and the fit
is seen to describe the precise data fairly well.

A similar analysis by DELPHI~\cite{delphi335} uses only the \oaaa\
(NLO) calculations with experimentally optimised renormalisation scale
and finds $\asmz=0.1175 \pm0.0005\stat \pm0.0010\expt \pm0.0027\had
\pm0.0007\theo$ with $\xmu=0.042$ for the closely related Cambridge
algorithm~\cite{dokshitzer97b} and $\asmz=0.1178\pm0.0036$ with
$\xmu=0.015$ for the Durham algorithm.  A study from OPAL
using \rfour\ measured with LEP~1 and LEP~2 data found $\asmz= 0.1182
\pm0.0003\stat \pm0.0015\expt \pm0.0011\had 
\pm0.0018\theo$~\cite{OPALPR414}.  Measurements of \rfour\ with JADE 
data at \roots\ between 14 and 44~GeV found as a preliminary result
$\asmz= 0.1169 \pm0.0004\stat \pm0.0012\expt \pm0.0021\had
\pm0.0007\theo$~\cite{kluth04}.  The various measurements are in
good agreement with each other.  All measurements using \rfour\
discussed in this section have comparatively small uncertainties and
belong to the group of the most precise determinations of \as.

\subsection{Alternative approaches to soft and hard QCD}

\subsubsection{Tests of power corrections} 
\label{sec_powcor}

Distributions and mean values of event shape observables measured in
\epem\ annihilation at many cms energy points between around 14 and
more than 200 GeV have been used recently by several groups to study
power corrections~\cite{powcor,delphi296,delphi327,aleph265,l3290}.
In all of these studies pQCD predictions in \oaa+NLLA are used for
event shape distributions.  For the mean values only NLO (\oaa) 
predictions are employed as explained in section~\ref{sec_lepjadeas}.
The pQCD predictions are added together in both cases with the power
correction predictions.  The resulting expressions are functions of
the strong coupling \asmz\ and of the non-perturbative parameter
\anull.  The complete predictions are compared with data for event
shape distributions or mean values corrected for experimental effects.
For the mean value of the 4-jet observable D-parameter power
correction predictions have been tested as well~\cite{l3290}.

\begin{figure}[!htb]
\begin{center}
\begin{tabular}{cc}
\includegraphics[width=0.55\textwidth]{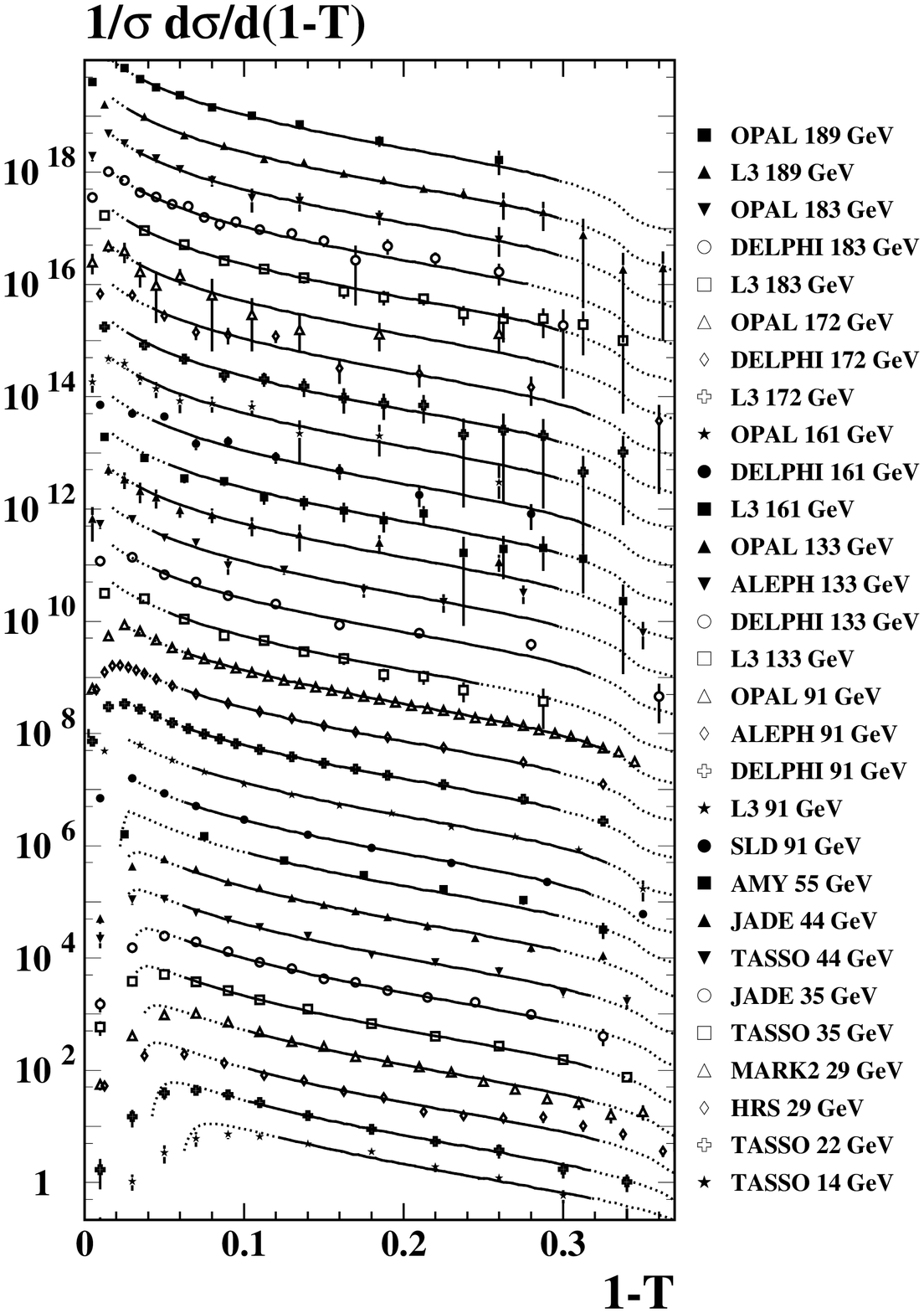} &
\includegraphics[width=0.45\textwidth,clip]{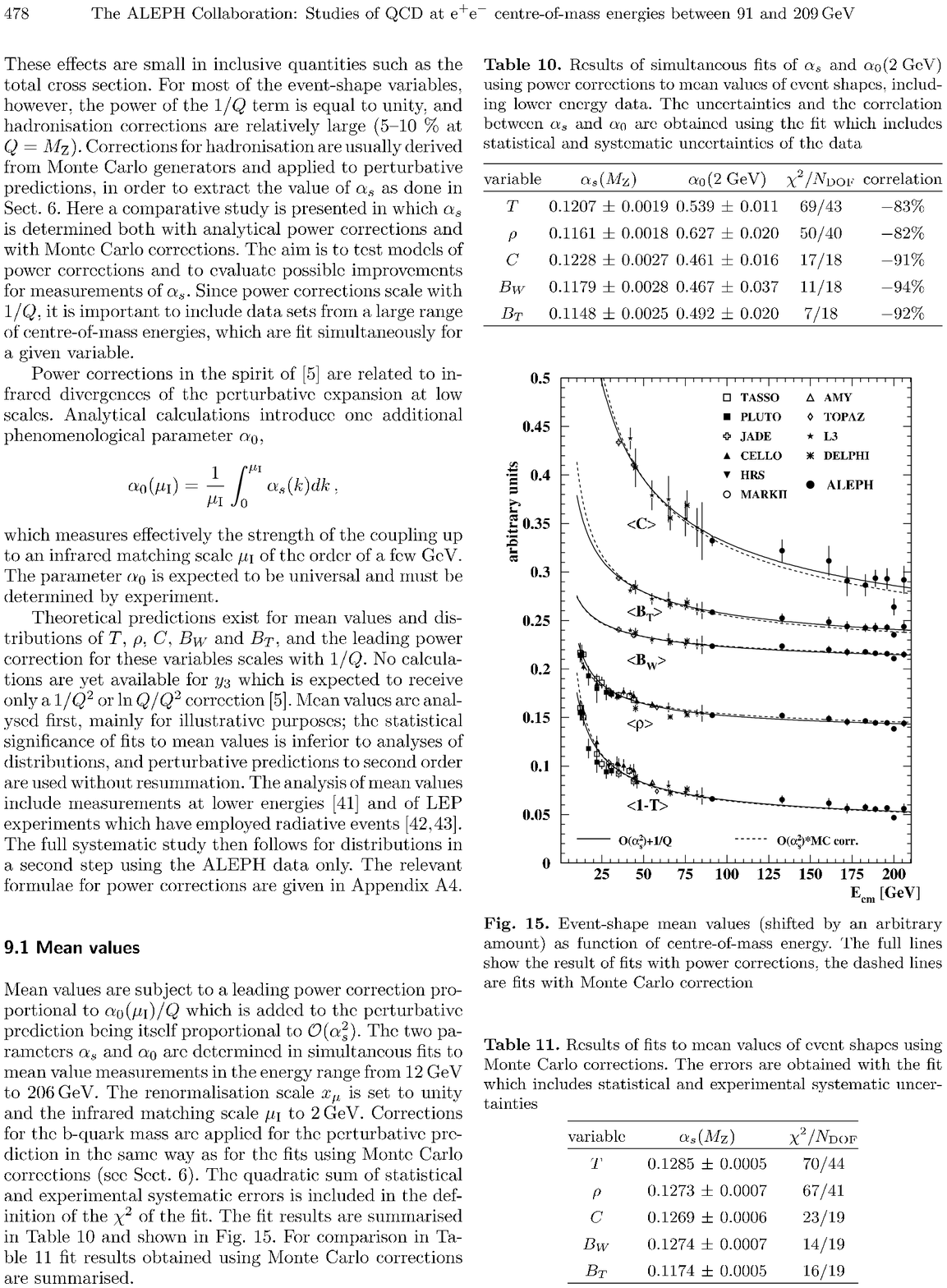} \\
\end{tabular}
\caption[ bla ]{ (left) Distributions of \thr.  The vertical axis
has arbitrary units and the data points have been scaled for clarity.
The solid lines show the result of the fit of the combined pQCD and
power correction prediction.  The dotted lines represent an
extrapolation of the result~\cite{powcor}.  (right) Cms energy
dependence of the mean values of several event shape observables.  The
solid lines show the result of the fit of combined \oaa\ QCD
calculations with power corrections while the dashed lines indicate
results obtained with Monte Carlo based corrections~\cite{aleph265}. }
\label{fig_powcor}
\end{center}
\end{figure}

Figure~\ref{fig_powcor} (left) shows the results of fitting the data for
\thr\ distributions measured at cms energies from 14 to 189~GeV.  The
data from experiments at $\roots<\mz$ are corrected for the effects of
$\epem\rightarrow\bbbar$ events~\cite{powcor}.  The data in the fitted
regions are well reproduced by the theory (solid lines).  The dotted
lines present extrapolations outside of the fitted regions using the
fit results for \asmz\ and \anulltwo\ which describe the data
reasonably well.  The agreement between theory and experiment is
similar for other observables.

Figure~\ref{fig_powcor} (right) presents as solid lines the results of
fits to measurements of mean values of various event shape observables
at $\roots=12$ to 206~GeV~\cite{aleph265}.  The data are well
described by the fitted theory.  The dotted lines show the results of
fits where the power corrections have been replaced by corrections
derived from Monte Carlo simulations.  The Monte Carlo corrections lead
to slighty different theoretical predictions.  

Table~\ref{tab_asa0} collects averaged results for \asmz\ and
\anulltwo\ from the various
analyses~\cite{powcor,aleph265,delphi296,l3290}.  The averages are as
given by the authors and are based upon differing sets of observables,
definitions of systematic uncertainties and averaging methods.  It is 
therefore not useful to compute global averages from these results.  
There is good agreement between the various analyses of mean values
or distributions of event shape observables.

\begin{table}[htb!]
\caption[ bla ]{ Results of power correction analysis using the DMW
model. }
\label{tab_asa0}
\begin{indented}\item[]
\begin{tabular}{lrcccc} \hline\hline
        &           & ALEPH~\cite{aleph265} & DELPHI~\cite{delphi296} 
                    & L3~\cite{l3290}   & \cite{powcor} \\
\hline
        & \asmz     & 0.111      & 0.108      & -- & 0.111 \\
distri- &           & $\pm$0.005 & $\pm$0.002 &    & $\pm$0.004 \\
butions & \anulltwo & 0.50       & 0.55       & -- & 0.58 \\
        &           & $\pm$0.10  & $\pm$0.03  &    & $\pm$0.09 \\
\hline
        & \asmz     & --         & 0.121      & 0.113      & 0.119 \\
mean    &           &            & $\pm$0.006 & $\pm$0.006 & $\pm$0.003 \\
values  & \anulltwo & --         & 0.47       & 0.48       & 0.49 \\
        &           &            & $\pm$0.08  & $\pm$0.06  & $\pm$0.06 \\
\hline\hline
\end{tabular}
\end{indented}
\end{table}

\begin{figure}[!htb]
\begin{center}
\begin{tabular}{cc}
\includegraphics[width=0.5\textwidth,clip]{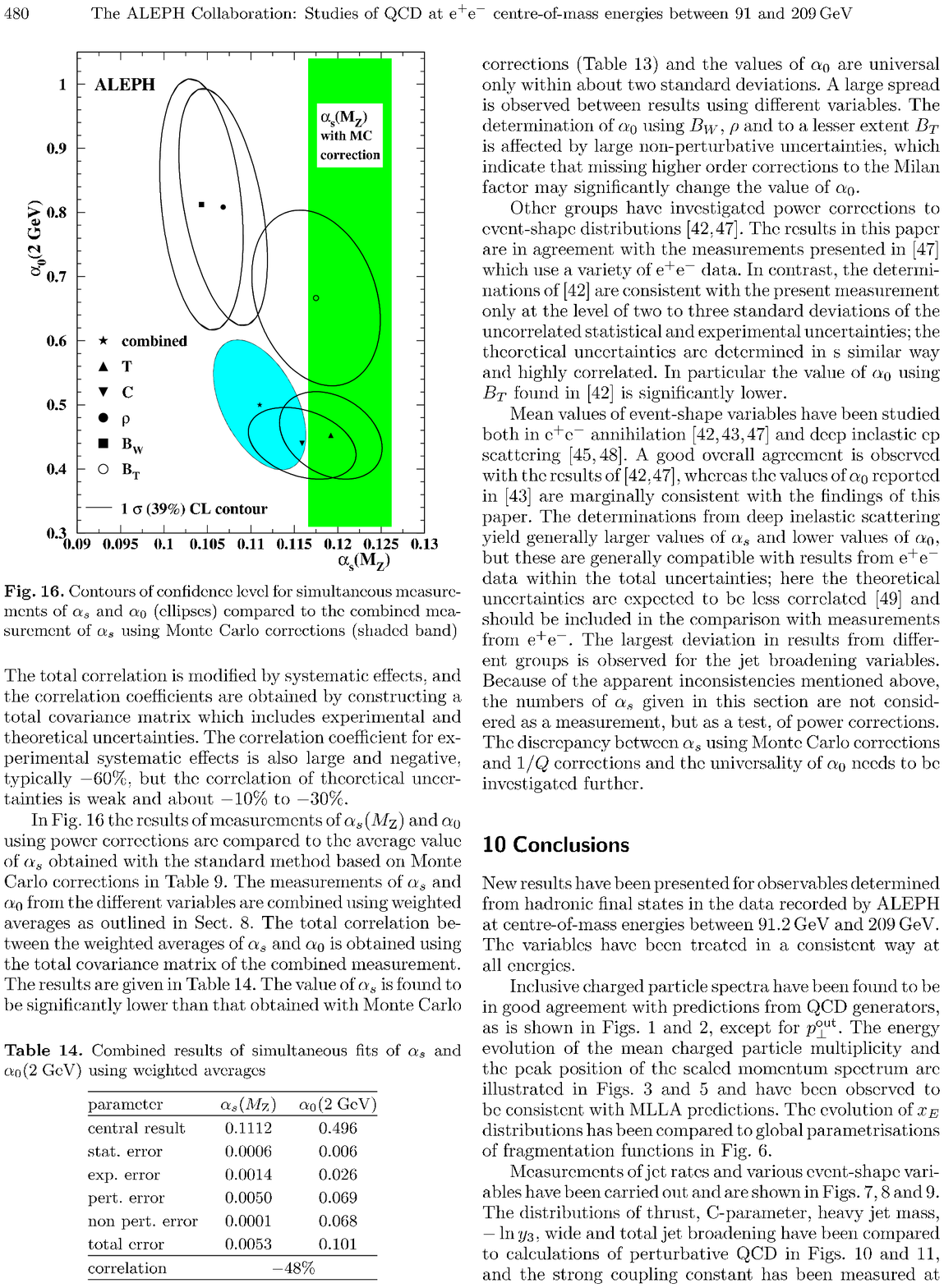} 
\includegraphics[width=0.5\textwidth]{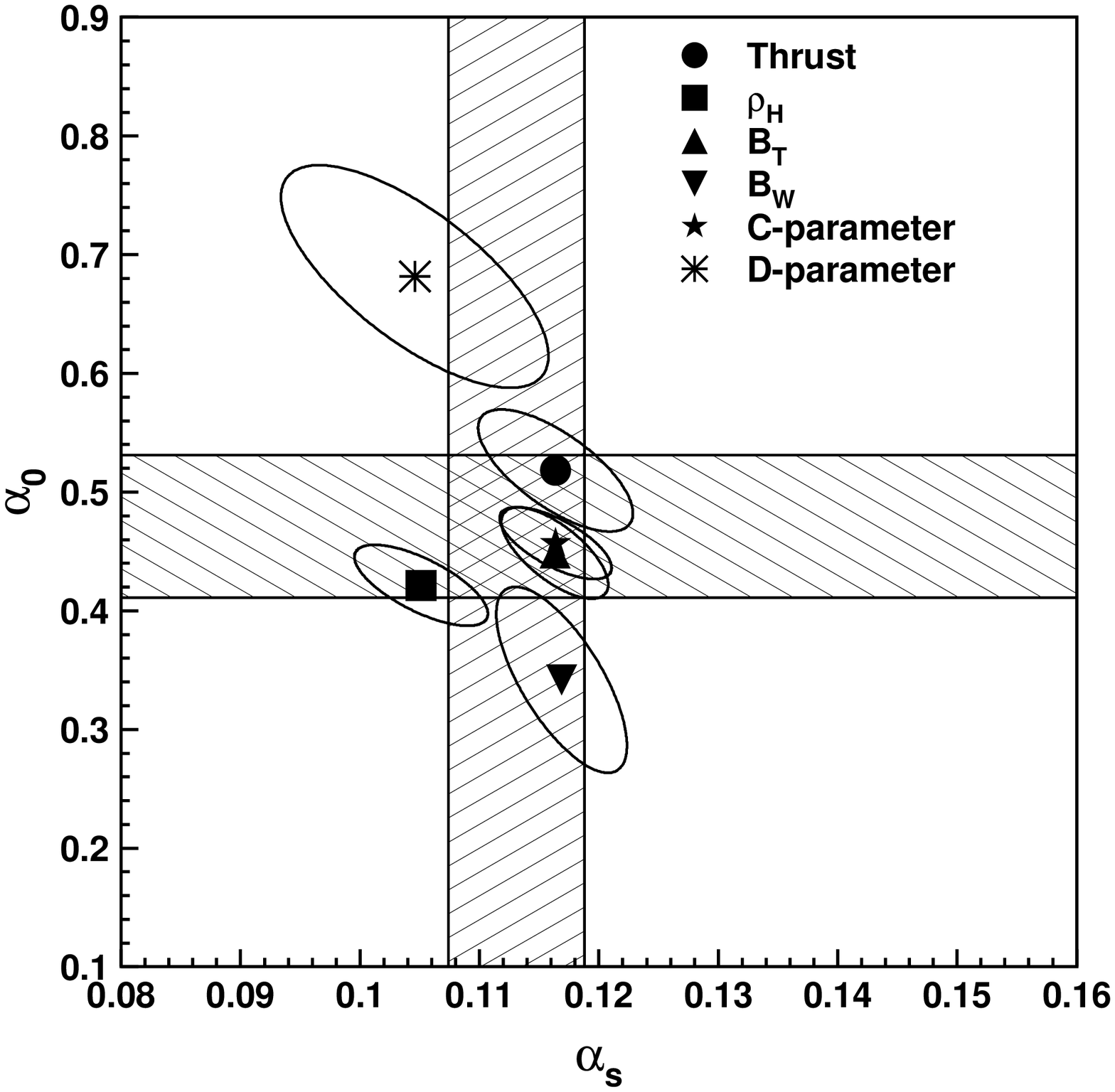} \\
\end{tabular}
\caption[ bla ]{ Summaries of results for \asmz\ and \anulltwo\ for 
distributions (left)~\cite{aleph265} and mean values
(right)~\cite{l3290}.  The error ellipses show one standard deviation
uncertainties (39\% CL).  The shaded band on the left plot gives the
combined result for \asmz\ using Monte Carlo based corrections from
the same analysis.  The bands on the right plot give the unweighted
averages of \asmz\ and \anulltwo\ shown in this figure. }
\label{fig_asa0}
\end{center}
\end{figure}

Figures~\ref{fig_asa0} (left) and (right) give an overview over
results from the analysis of distributions~\cite{aleph265} (left) and
mean values~\cite{l3290} (right).  The results are shown as points
with one-standard-deviation error ellipses.  The results for \asmz\ of
fits to distributions are found to be smaller compared to the results
of fits to the same data using Monte Carlo based hadronisation
corrections; $\asmz\simeq0.111$ compared to $\asmz\simeq0.120$.  There
are systematic differences between the DMW power correction model and
the hadronisation corrections derived from Monte Carlo models leading
to differences in the values of \asmz\ of $\Delta\asmz\simeq0.009$
which are significantly larger than the hadronisation uncertainties
quoted in section~\ref{sec_lepjadeas} as derived from the published
results. In the analysis by ALEPH~\cite{aleph265} a direct comparison
is made using the same data and a consistent average relative
difference $\Delta\asmz/\asmz\simeq10$\% is found.  The results for
\anulltwo\ from $\rho=\mh^2$, \bw\ and \bt\ are fairly large compared
to the averages.  For the observable \mh\ this effect stems from
neglecting particle masses~\cite{salam01a,delphi296}.

\begin{figure}[!htb]
\begin{center}
\includegraphics[width=0.8\columnwidth]{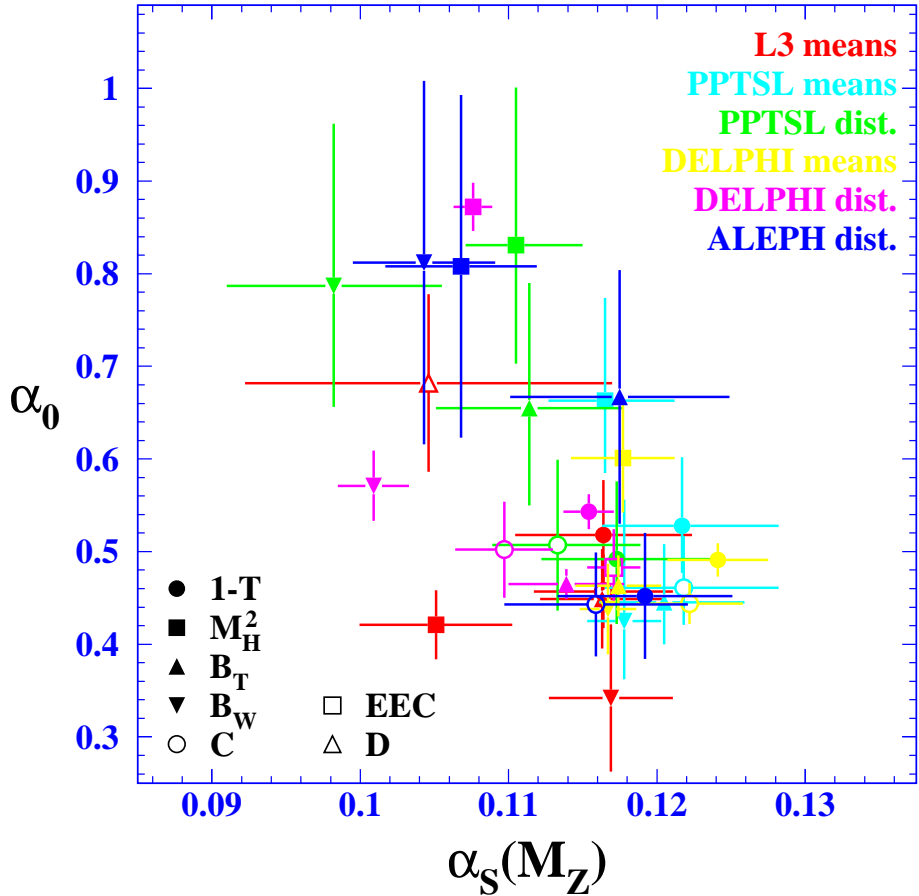}
\caption[ bla ]{ The figure shows a summary of results for \asmz\ and
\anulltwo\ from power correction
studies~\cite{powcor,l3290,delphi296,aleph265}.  The
results labelled PPTSL are from~\cite{powcor}. } 
\label{fig_allasa0}
\end{center}
\end{figure}

Figure~\ref{fig_allasa0} shows a summary of results for \asmz\ and
\anulltwo~\cite{powcor,l3290,delphi296,aleph265}.  One
observes reasonable agreement between the individual measurements.
The non-perturbative parameter \anull\
appears universal within about 20\% as expected~\cite{dokshitzer98b}
while the values for \asmz\ are compatible with world
averages, e.g.~\cite{bethke04} $\asmz=0.1184\pm0.0027$.

\subsubsection{Studies of renormalisation schemes}
\label{sec_rgi}

The ideas about choosing optimal renormalisation schemes (RS)
discussed in section~\ref{sec_rgitheo} have been tested
experimentally.  In~\cite{bethke00a} the predictions for \xmu\ from
the three methods for the hadronic width of the \znull\
$\rz=\Gamma(\znull\rightarrow\mathrm{hadrons})/
\Gamma(\znull\rightarrow\mathrm{leptons})$ were obtained for NLO QCD.
Since for \rz\ the QCD prediction is known in NNLO the resulting
values of \asmz\ from scale optimized NLO QCD may be compared with
the corresponding result from NNLO QCD.  It is found that for this
observable the three methods agree well, but the resulting value of
\asmz\ is significantly lower by $\Delta\asmz\simeq 0.003$ compared
to the values determined with the NNLO calculations.

In~\cite{delphi233} the DELPHI collaboration measured many event shape
observables using precise LEP data from the \znull\ peak and compared
the data to NLO QCD predictions; a similar analysis is discussed
in~\cite{burrows96}.  In fits where the renormalisation scale
parameter \xmu\ and thus effectively the RS was allowed to vary in
addition to \asmz\ consistent results within the errors for \asmz\
were found resulting in an unweighted average of
$\asmz=0.1170\pm0.0025$ where the error is given by the variance of
the set of measurements.  Within the fitted ranges of the
distributions no evidence for a dependence of \xmu\ on the observable
was found, implying that a single value of \xmu\ is indeed sufficient.
As a comparison the corresponding fits with $\xmu=1$ yielded an
unweighted average and variance of $\asmz=0.123\pm0.015$; i.e.\ not
using experimentally optimised scaled results in a significantly
increased variance.  Similar observations were made
in~\cite{OPALPR054}.  However, these results are not supported by the
findings of~\cite{burrows96} where for a similar set of observables as
in~\cite{delphi233,OPALPR054} $\asmz=0.127\pm0.008$ for fits with
$\xmu=1$ and $\asmz=0.117\pm0.007$ for fits with \xmu\ free was
observed.

Applying the scale or RS setting methods to fits of limited ranges of
distributions is in principle problematic, because more than one
energy scale is involved.  However, since it is known experimentally
that a single value of \xmu\ is sufficient to describe the data the
effects of the additional energy scales introduced by the observable
values may be assumed to be small.  In~\cite{delphi233} the ECH and
PMS methods generally lead to similar results for \asmz, consistent
with the theoretical observation that their predictions for the
optimal scale are closely related, see e.g.~\cite{chyla95}.  The
resulting values of \asmz\ are also consistent with each other.  The
weighted averages of the set of observables with the PMS or ECH
methods are $\asmz=0.115\pm0.004$ with $\chisqd\simeq 1$.  For the BLM
method $\chisqd=29/13$ is found indicating less consistency among the
results.  The observation that values of \asmz\ obtained from
theoretically optimised RS tend to be smaller than purely experimental
values may be explained by noticing that in the PMS method (which is
close to the ECH method) a stationary point and usually a minimum of
$\as(\mz,\xmu)$ is chosen as the optimum.

By extending the ECH method it turns out to be possible to relate QCD 
predictions for different physical observables to each other without
explicit dependence on the RS~\cite{dhar84,brodsky95,barclay94}.  The
QCD prediction for an inclusive observable $R(Q)$ depending on only one
energy scale $Q$ is given by the generalised $\beta$-function:
\begin{equation}
Q\frac{\ddel R}{\ddel Q} = -b R^2 (1+\rho_1 R + \rho_2 R^2 + \ldots)
                         = b \rho(R)\;\;.
\label{equ_obsrun}
\end{equation}
with $b=2\pi\beta_0$.
The $\rho_i$ are RS independent but can be derived from the
corresponding RS dependent coefficients.  Equation (\ref{equ_obsrun})
may be solved analogously to the $\beta$-function of \as\ in terms of
a process dependent scale parameter $\Lambda_R$.  Thanks to the
Celmaster-Gonsalves relation~\cite{celmaster79} such parameters for
different observables may be related to each other or e.g.\ to the
scale parameter \lmsbar\ of the running strong coupling \asmu\ in the
\msbar\ renormalisation scheme:
\begin{equation}
\frac{\Lambda_R}{\lmsbar} = e^{r_1/b}\left(\frac{2c_1}{b}\right)^{-c_1/b}\;\;.
\label{equ_celgon}
\end{equation}
The variables are $c_1=2\pi\beta_1/\beta_0$ and $r_1=B_R/(2A_R)$ where
$A_R$ and $B_R$ are the LO and NLO coefficients for the observable
calculated in the \msbar\ scheme.  This procedure may be referred to
as renormalisation group improved perturbative QCD
(RGI QCD)~\cite{grunberg80}.  Non-perturbative effects scaling like
$1/Q$ for most observables may be included with an additional free
parameter $K_0$~\cite{campbell98}.

\begin{figure}[!htb]
\begin{center}
\begin{tabular}{cc}
\includegraphics[width=0.42\textwidth]{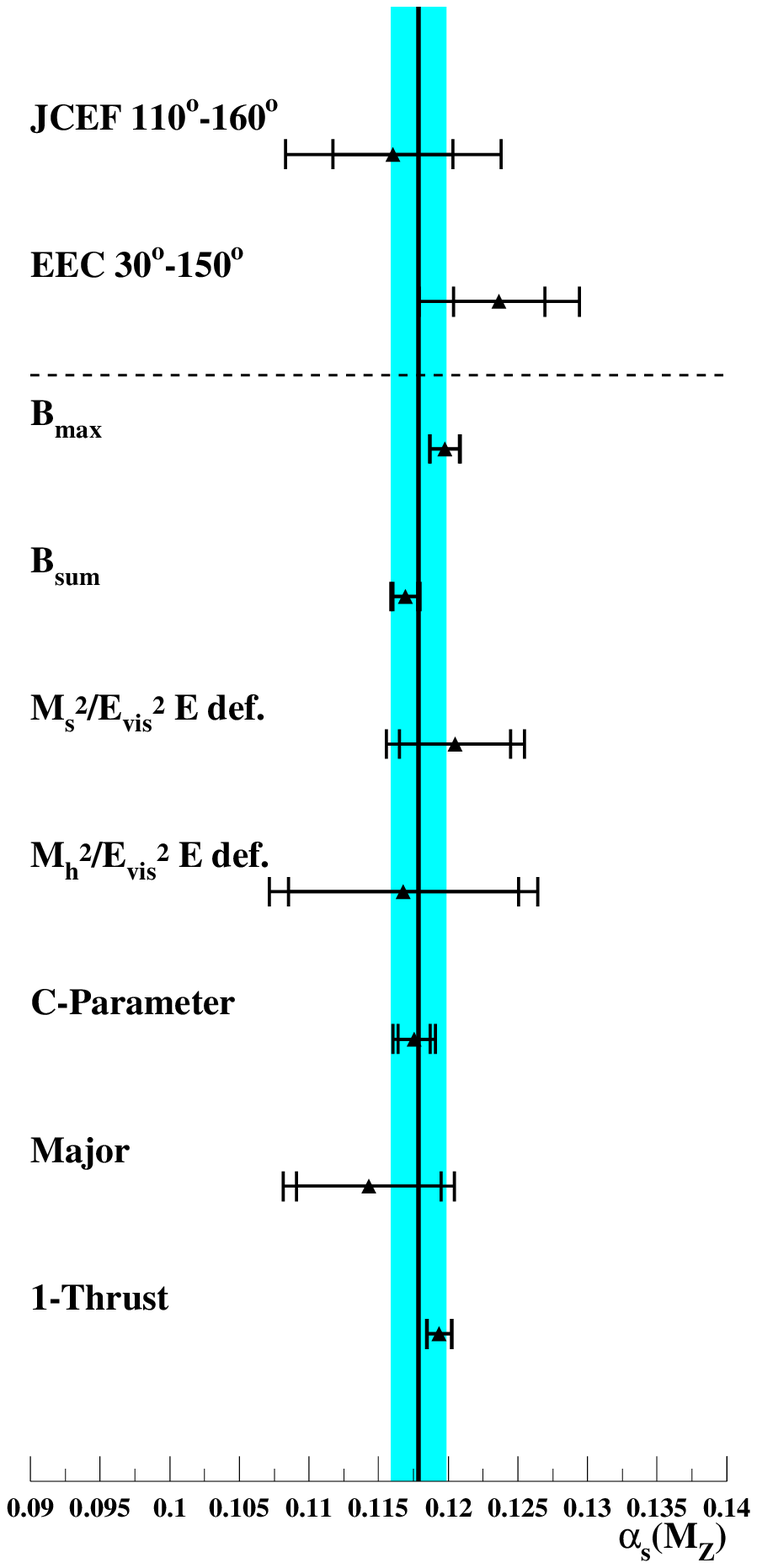} &
\includegraphics[width=0.57\textwidth]{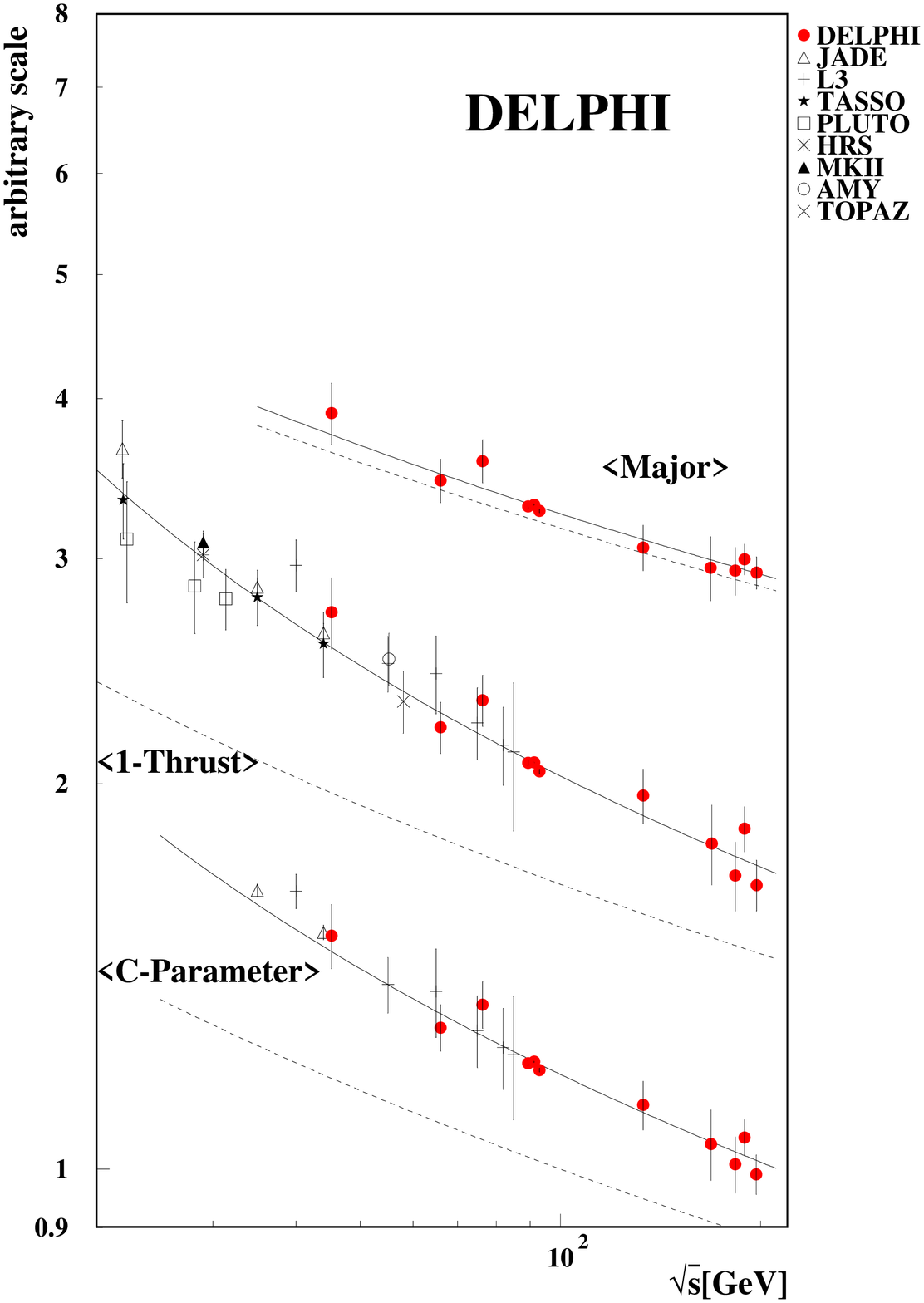} \\
\end{tabular}
\caption[ bla ]{ (left) The figure summarises results for \asmz\ from
RGI QCD fits with power corrections.  The error bars include
statistical and experimental systematic uncertainties.  The solid
lines and grey bands indicate the unweighted mean and variance based
on mean values of event shape observables.  (right) The figure shows
fits of RGI QCD without non-perturbative corrections to mean values of
\tmaj, \thr\ and \cp\ as functions of \roots\ (full lines).  As a
comparison the dashed lines present the corresponding prediction in the
\msbar\ RS~\cite{delphi296}.}
\label{fig_delphirgi}
\end{center}
\end{figure}

The DELPHI collaboration has studied mean values of several event
shape observables and in addition distributions of the event shape
observables EEC and JCEF using the RGI QCD method as explained
above~\cite{delphi296}.  The analysis also uses data from lower energy
experiments to cover values of \roots\ from 22~GeV to 200~GeV.  The
results for \asmz\ from 12 observables are consistent with each other
within their uncertainties, as shown in figure~\ref{fig_delphirgi}
(left).  The variance of the individual values of \asmz\ is about
$\Delta\asmz\simeq 0.002$ while the variance of \asmz\ results of
corresponding fits based on standard NLO QCD combined with power
corrections is observed to be about twice as large.  The values of the
power correction parameter $K_0$ are found to be small and consistent
with zero within their uncertainties.  Figure~\ref{fig_delphirgi}
(right) shows as examples the results of RGI QCD fits without power
corrections, i.e.\ $K_0=0$, for the mean values of \tmaj, \thr\ and
\cp.

The results of the RGI QCD fits are interpreted to imply that a
dominating part of the power corrections or hadronisation corrections
in standard NLO QCD fits in the \msbar\ RS can be absorbed in the RGI
QCD.  In~\cite{delphi296} a systematic shift in \asmz\ of about 2\%
between fits of RGI QCD with and without power corrections was
observed and this shift is conjectured to be an estimate of the true
size of missing higher order contributions or power corrections.

In~\cite{koerner00} the RGI QCD method is applied to extract a value
for \asmz\ from the branching ratio of hadronic to leptonic decays of
$\tau$ leptons.  The result turns out to be in good agreement with the
current world average.  See section~\ref{sec_rhad} for a discussion of
the extraction of \as\ from hadronic $\tau$ decays.

\subsection{Running b quark mass}

The QCD analyses discussed so far have been made assuming massless
partons in the QCD predictions.  The effects of massive quarks on the
perturbative QCD predictions are known in \oaa\ as
well~\cite{rodrigo97,bernreuther97,nason97} (and also in NLLA
\cite{krauss03}).  The mass of the heavy quark \mb\ identified with the
b quark at LEP energies becomes an additional parameter of the theory
and is subject to renormalisation analogously to the strong coupling
constant.  As a result the heavy quark mass \mb\ will depend on the
scale of the hard process in which the heavy quark participates.  In
leading order in the \msbar\ renormalisation scheme the running
b-quark mass is $\mb(Q)=M_{\mathrm{b}}(\as(Q)/\pi)^{12/23}$, where
$M_{\mathrm{b}}$ is the so-called pole mass defined by the pole of the
renormalised heavy quark propagator (see section~\ref{sec_runmq}).
The running \mb\ is known to four-loop accuracy~\cite{vermaseren97}.
Using low energy measurements of \mbmb\ one finds that $\mbmz\simeq
3$~GeV.

The mass of the b quark \mb\ is experimentally accessible in hadronic
\znull\ decays, because i) about 21\% of the decays are
$\znull\rightarrow\bbbar$ and these events can be identified with high
efficiency and purity, ii) event samples of $\order{10^6}$ events are
available and iii) observables like the 3-jet rate $\rthree(\ycut)$ in
the JADE or Durham algorithm can have enhanced mass effects going like
$\mb^2/\mz^2/\ycut$~\cite{bilenky95}.  In the experimental
analyses~\cite{delphi177,OPALPR336,brandenburg99,aleph223,delphiconf644}
the ratio $\bthree=\rthreeb/\rthreel$, \rthreeb (\rthreel) are the 3-jet
rates in b quark (light quark) events, is studied which reduces the
influence of common systematics.

\begin{figure}[!htb]
\begin{tabular}{cc}
\includegraphics*[width=0.475\textwidth]{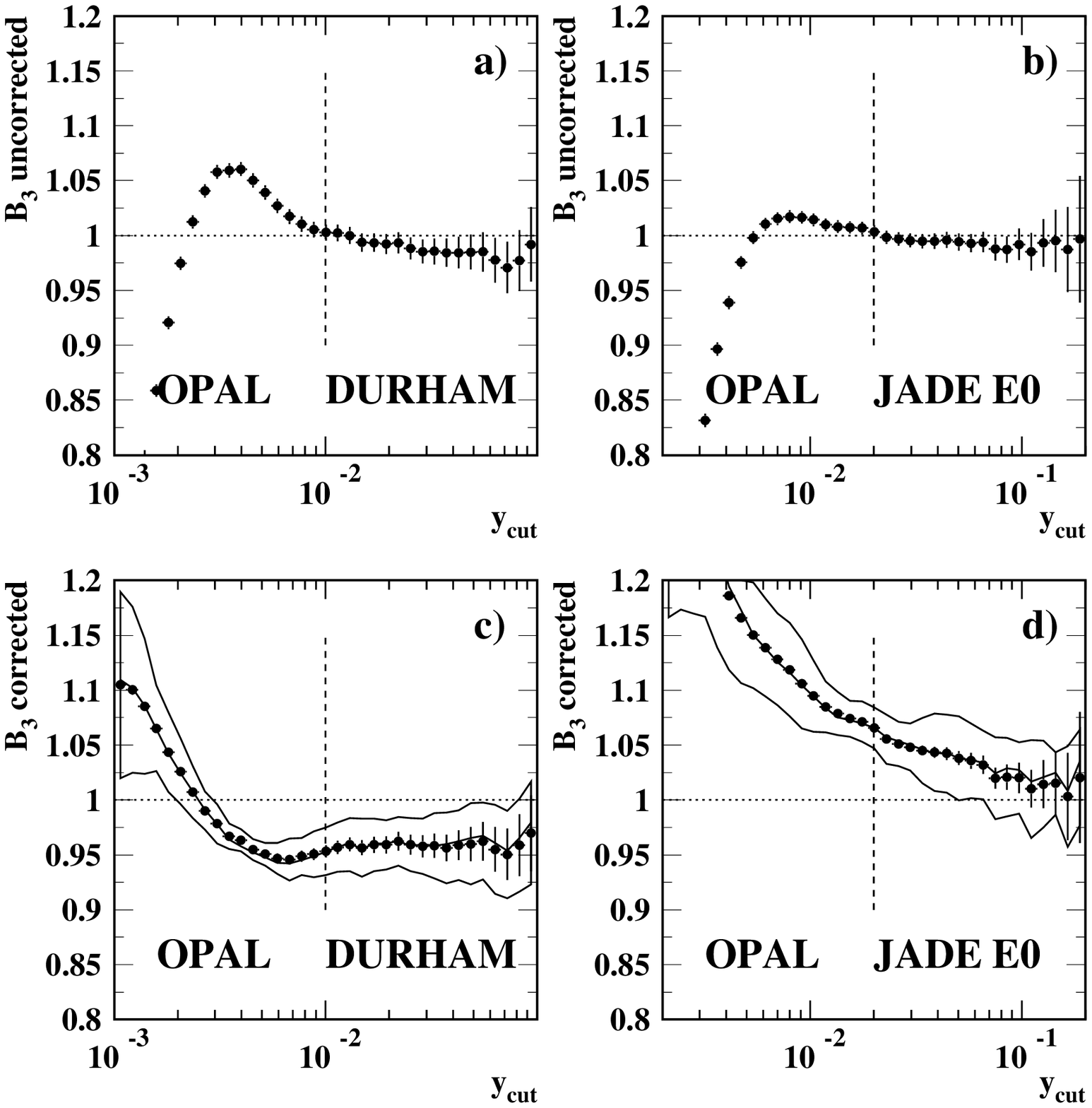} &
\includegraphics*[width=0.475\textwidth]{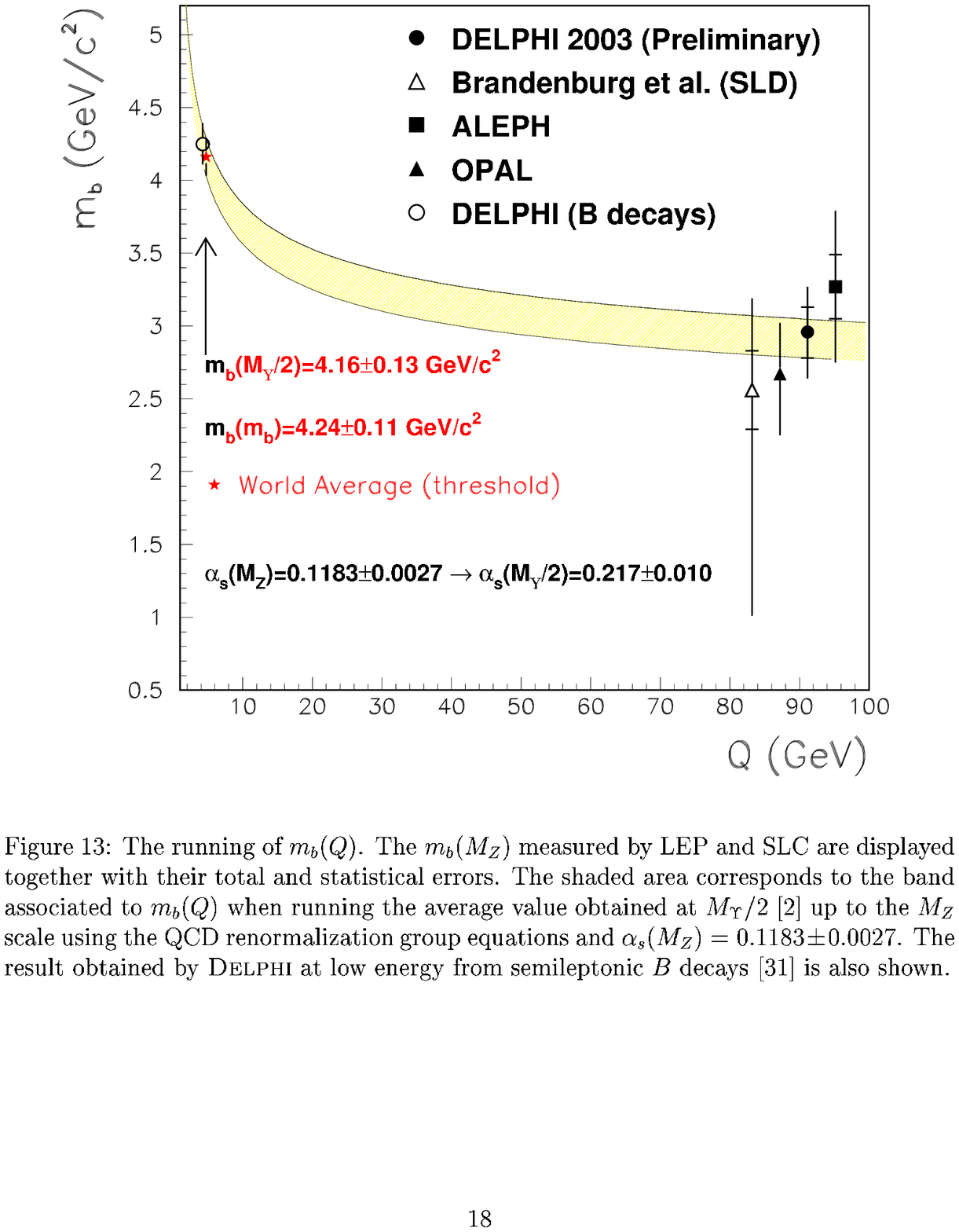} \\
\end{tabular}
\caption{ (left) Distributions of \bthree\ determined using the Durham 
or JADE E0 jet finding algorithms before and after all corrections.
The solid lines in the lower row of figures give the uncertainties on
\bthree.  The vertical dashed lines indicate the values of \ycut\
where the QCD predictions are calculated~\cite{OPALPR336}.  (right)
Measurements of $\mb(Q)$ at low and high energies $Q$ compared with
the QCD expectation for the running of
$\mb(Q)$~\cite{delphiconf644}. }
\label{fig_b3}
\end{figure}

Figure~\ref{fig_b3} (left) shows $\bthree(\ycut)$ measured by
OPAL~\cite{OPALPR336} before and after detector and hadronisation
corrections for the JADE E0 and the Durham algorithm.  One observes
that the effect of the b quark mass is about 5\% on \bthree\ with
reasonable uncertainties as indicated by the bands.  The effect goes
in opposite directions for JADE E0 and the Durham algorithm, because
in the JADE type algorithms based on invariant mass to cluster jets
the large b quark mass causes an enhancement of \bthree\ while in the
Durham algorithm the reduced phase space for hard gluon emission
dominates thus reducing \bthree~\cite{brandenburg99}.  

\begin{table}[htb!]
\caption[ bla ]{ Values of \mbmz\ in units of GeV from various
experiments with symmetrised systematic uncertainties.  The
last row shows the average determined as described in the text. }
\label{tab_mbmz}
\begin{indented}\item[]
\begin{tabular}{lr@{$\pm$}c@{$\pm$}c@{$\pm$}c@{$\pm$}c} 
\hline\hline
Experiment & \mbmz & stat. & syst. & had. & theo. \\
\hline
ALEPH~\cite{aleph223}       & 3.27 & 0.22 & 0.22 & 0.38 & 0.16 \\
DELPHI~\cite{delphiconf644} & 2.96 & 0.18 & 0.14 & 0.19 & 0.12 \\
OPAL~\cite{OPALPR336}       & 2.67 & 0.03 & 0.22 & 0.25 & 0.19 \\
SLD~\cite{brandenburg99}    & 2.56 & 0.27 & 0.33 & 0.84 & 0.44 \\
\hline
{\bf average} & {\bf 2.90} & {\bf 0.12} & {\bf 0.29} & {\bf 0.20} & {\bf 0.14} \\
\hline\hline
\end{tabular}
\end{indented}
\end{table}

Figure~\ref{fig_b3} (right) gives an overview over presently available
measurements of \mbmz~\cite{delphiconf644}.  These are compared with
low energy measurements and the QCD prediction for the running of
$\mb(Q)$.  The predicted range of \mbmz\ is in good agreement with the
measurements.  The measurements of \mbmz\
of~\cite{OPALPR336,brandenburg99,aleph223,delphiconf644} are averaged
using the same method as for averaging values of \as\ (see
section~\ref{sec_lepjadeas}).  The individual results are shown
with symmetrised systematic uncertainties in table~\ref{tab_mbmz}.
The systematic uncertainty of the OPAL result was split into an
experimental and a hadronisation uncertainty according to the average
ratio of these uncertainties obtained from the individual jet
algorithms.  We assume the statistical errors to be uncorrelated and
the experimental and hadronisation uncertainties to be partially
correlated.  The theory uncertainties are evaluated by simultaneous
variation of the input values within their theory uncertainties.  The
average is shown in the last column of table~\ref{tab_mbmz}.
Combining all errors we find $\mbmz=(2.90\pm0.31)$~GeV, which may be
compared with the average of low energy measurements
$\mbmb=(4.24\pm0.11)$~GeV~\cite{el-khadra02}.  The difference becomes
$\mbmz-\mbmb=(1.34\pm0.33)$~GeV which is non-zero by four standard
deviations thus providing strong experimental evidence for the running
heavy quark mass in QCD.

\section{Particle production}
\label{sec_incl}

The direct observation of hadron production in \epem\ annihilation
offers unique ways to test predictions of QCD.  The subject
of multiplicities of hadrons is covered in detail in~\cite{dremin01}.

\subsection{Low energy particles}

\begin{figure}[htb!]
\begin{tabular}{cc}
\includegraphics[width=0.45\textwidth]{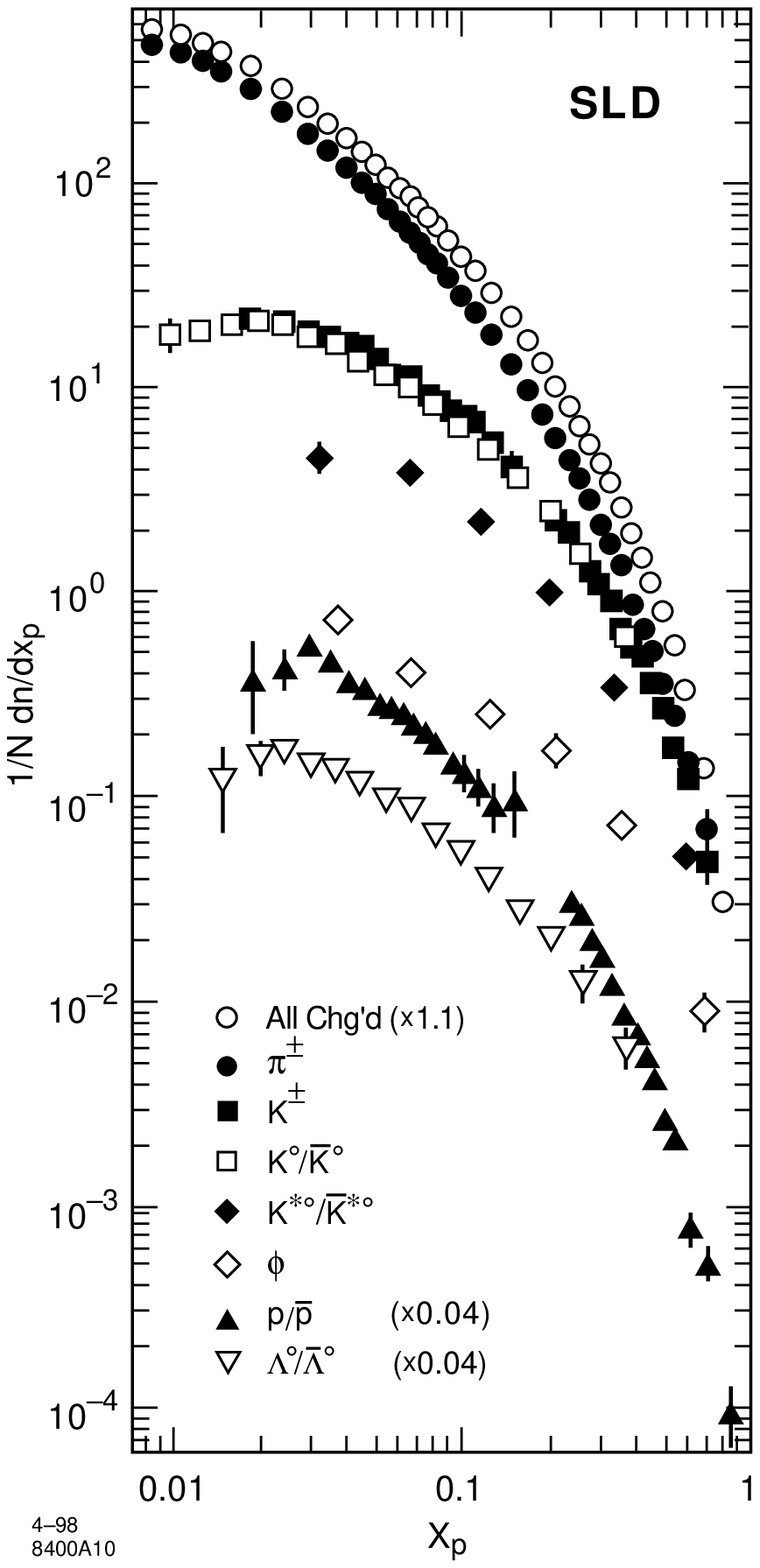} &
\includegraphics[width=0.525\textwidth]{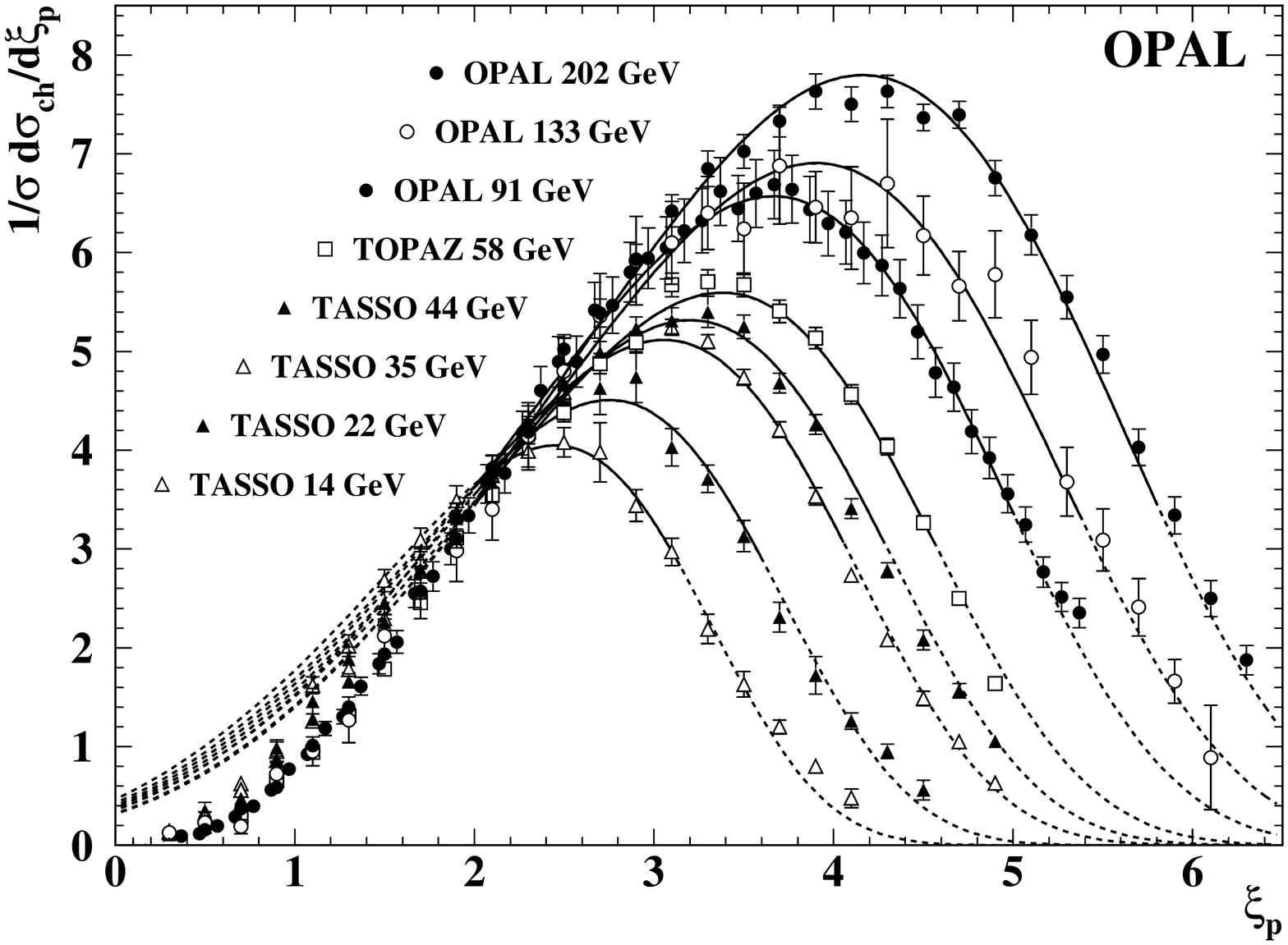} \\
\end{tabular}
\caption[ bla ]{ (left) Spectra of scaled momentum $x$ for identified
particles as indicated and for all charged particles measured by SLD
on the \znull\ peak~\cite{sldparticles1}.  (right) Spectra of
$\ksi=\log(1/x)$ measured by OPAL using LEP~1 and LEP~2 data as well
as lower energy data from TASSO~\cite{OPALPR362}.  The curves show
fits of a NLLA QCD prediction~\cite{fong91} to the data. }
\label{fig_sldparticles}
\end{figure}

Figure~\ref{fig_sldparticles} (left) shows the production rates for
inclusive and for identified charged hadrons ($\pi^{\pm}$, $K^{\pm}$,
p) or hadrons decaying to final states with charged particles ($K^0$,
$\phi$, $\Lambda$) as a function the particles scaled momentum
fraction $x=2p/\roots$ where $p$ is the particle
momentum~\cite{sldparticles1}.  The momentum spectra of the various
particles are fairly similar but their production rates are strongly
suppressed with particle mass.  The production rates for many particle
species are summarised in~\cite{pdg04}.

\begin{figure}[htb!]
\begin{tabular}{cc}
\includegraphics[width=0.475\textwidth]{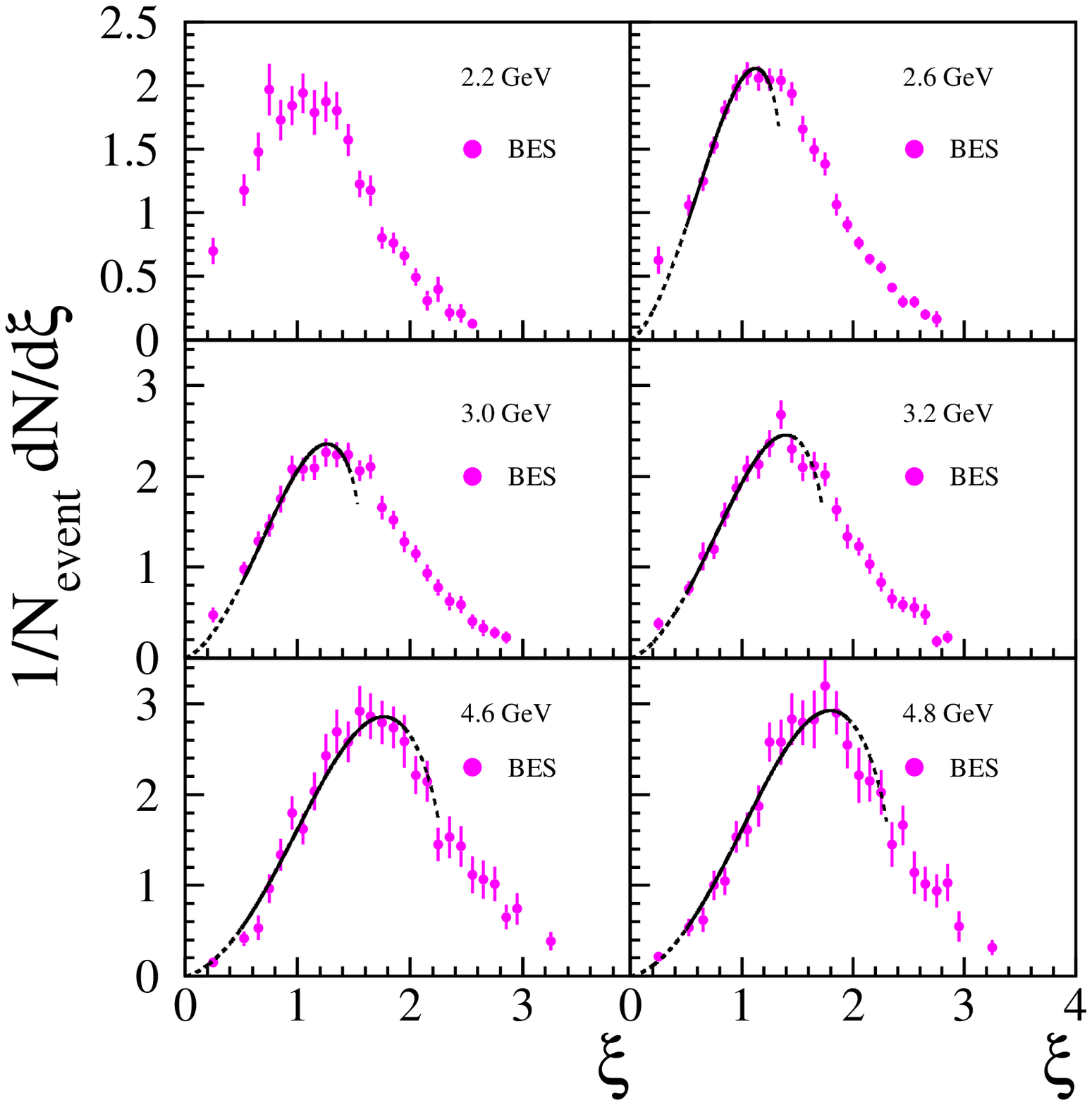} &
\includegraphics[width=0.475\textwidth]{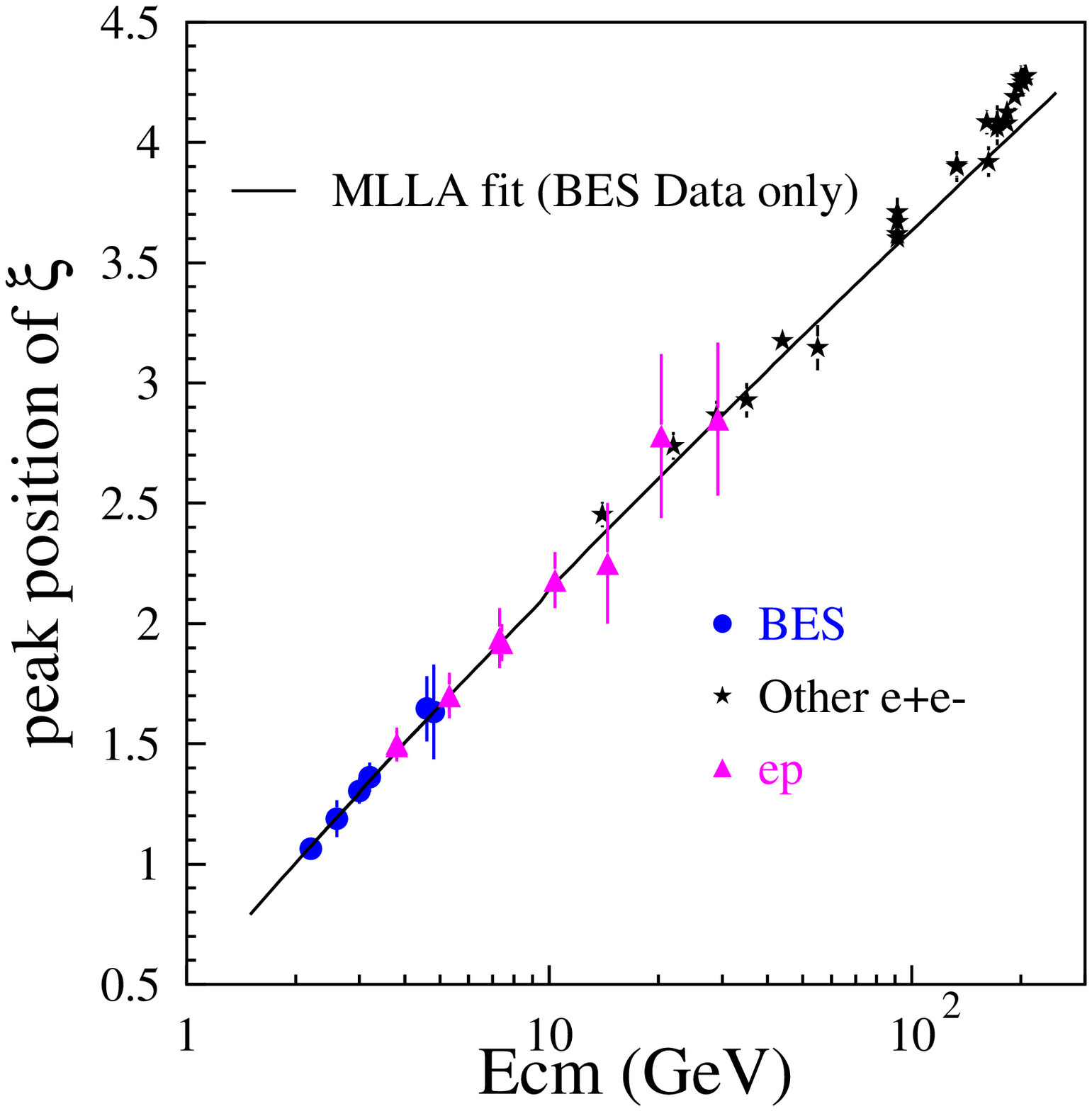} \\
\end{tabular}
\caption[ bla ]{ (left) Measurements of $\ksi=\log(1/x)$ spectra at 
$2.2\leq\roots\leq4.8$~GeV by BES~\cite{bes04}.  (right) Measurements
of the peak position of \ksi\ distributions~\cite{bes04}. }
\label{fig_besxi}
\end{figure}

The study of the observable $\ksi=\ln(1/x)$, where $x$ is the scaled
momentum or energy fraction defined above, allows to focus on the
production mechanisms of low-momentum particles corresponding to large
values of \ksi.  Figure~\ref{fig_sldparticles} (right) shows the
\ksi\ distributions measured by OPAL using all LEP~1 and LEP~2 data
together with data from previous experiments~\cite{OPALPR362}.  A
similar analysis is shown in~\cite{delphi216}.  Figure~\ref{fig_besxi}
(left) presents recent measurements by BES of \ksi\ at $\roots=2.2$ to
4.8~GeV~\cite{bes04}.

On both figures the typical hump-backed plateau~\cite{azimov86} of the
distributions is clearly visible.  The distribution vanishes for small
values of \ksi\ corresponding to large $x$ for kinematic reasons while
the fall for large \ksi\ (small $x$) is due to destructive
interference of coherent gluon radiation.  Essentially, low energy
gluons cannot resolve individual coloured partons anymore and are thus
suppressed since an \epem\ annihilation event is colourless.  This
explanation in the framework of perturbative QCD rests on the
assumption that hadronisation is on average a local process.  This
assumption is referred to as local parton hadron duality
(LPHD)~\cite{azimov85a,azimov86}.  Assuming LPHD predicted spectra
calculated for partons may be directly compared with measured spectra
derived from the observed hadrons where only the relative
normalisation is determined by experiment.  The calculation in the
modified leading-log approximation
(MLLA)~\cite{dokshitzer92a,dokshitzer92b,khoze97} is shown as solid
lines in figure~\ref{fig_besxi} (left) while the related NLLA QCD
prediction~\cite{fong91} is shown as lines in
figure~\ref{fig_sldparticles} (right).

The position of the peaks \ksistar\ of the \ksi\ distributions is seen
to depend on the cms energy.  This effect is predicted by MLLA
QCD~\cite{khoze97} and shown as the line on figure~\ref{fig_besxi} (right).
The fit was performed using only the low energy BES data and the result
agrees well with the data up to LEP~1 energies.

Measurements of \ksi\ and \ksistar\ have also been made for identified
hadrons ($\pi^{\pm}$, K$^{\pm,0}$, p, $\Lambda$) as well as for
identified quark flavours~\cite{sldparticles2} and dependence on
hadron mass and quark flavour has been observed.  Distributions of
\ksi\ at different \roots\ may be related directly via perturbative
evolution equations in the MLLA~\cite{albino04}.

\subsection{High momentum particles and scaling violation}
\label{sec_sv}

The fragmentation of partons produced in the hard interaction
$\epem\rightarrow\qqbar(g)$ into hadrons is investigated using the
scaled momentum of hadrons $x$.  In the quark parton model, i.e.\
without strong interactions, one expects the distributions of $x$
measured at different cms energies to be identical.  This prediction
of a scaling behaviour is similar to the situation in deep inelastic
scattering (DIS) of leptons on hadrons where $x_{Bj}=Q^2/(2m(E-E'))$
with transferred 4-momentum $Q$, parton mass $m$, and lepton energies
$E$ and $E'$ before and after the scattering.  Distributions of $x_{Bj}$
are independent of $Q^2$ in the quark parton model.  

The scaling behaviour is modified or broken by strong interaction
processes~\cite{callan74}.  The probability for radiation of gluons
from quarks depends on the energy scale of the quark and therefore due
to the running of \as\ fewer hard gluons are produced at high energy.
The scaled momentum $x$ spectra are expected to be softer at high
energies.  This is formally described by the theory with the
Dokshitzer-Gribov-Lipatov-Altarelli-Parisi (DGLAP) evolution
equations, see~\cite{nason94} and references therein.  The normalised
distribution of $x$ is described as the differential cross section
$\sigma^h$ for production of a hadron with scaled momentum $x$:
\begin{equation}
  \frac{1}{\sigmah}\frac{\ddel\sigma^h}{\ddel x} = 
  \int_x^1 \sum_{f\in\{udscbg\}} C_f(z,\as(\mur),\xmu) 
           D_f(\frac{x}{z},\mur) \frac{\ddel z}{z} \;\;.
\end{equation}
The coefficient functions $C_f(x,\as(\mur),\xmu)$ correspond to the
probability to obtain a parton $f$ with scaled momentum $x$, where $f$
is either one of the five active quark flavours u, d, s, c, or b or a
gluon g.  The parameter $\xmu=\mur/Q$, with $Q=\roots$, gives the
renormalisation scale $\mur$ where the coefficient and fragmentation
functions are evaluated.

\begin{figure}[htb!]
\begin{tabular}{cc}
\includegraphics[width=0.45\textwidth]{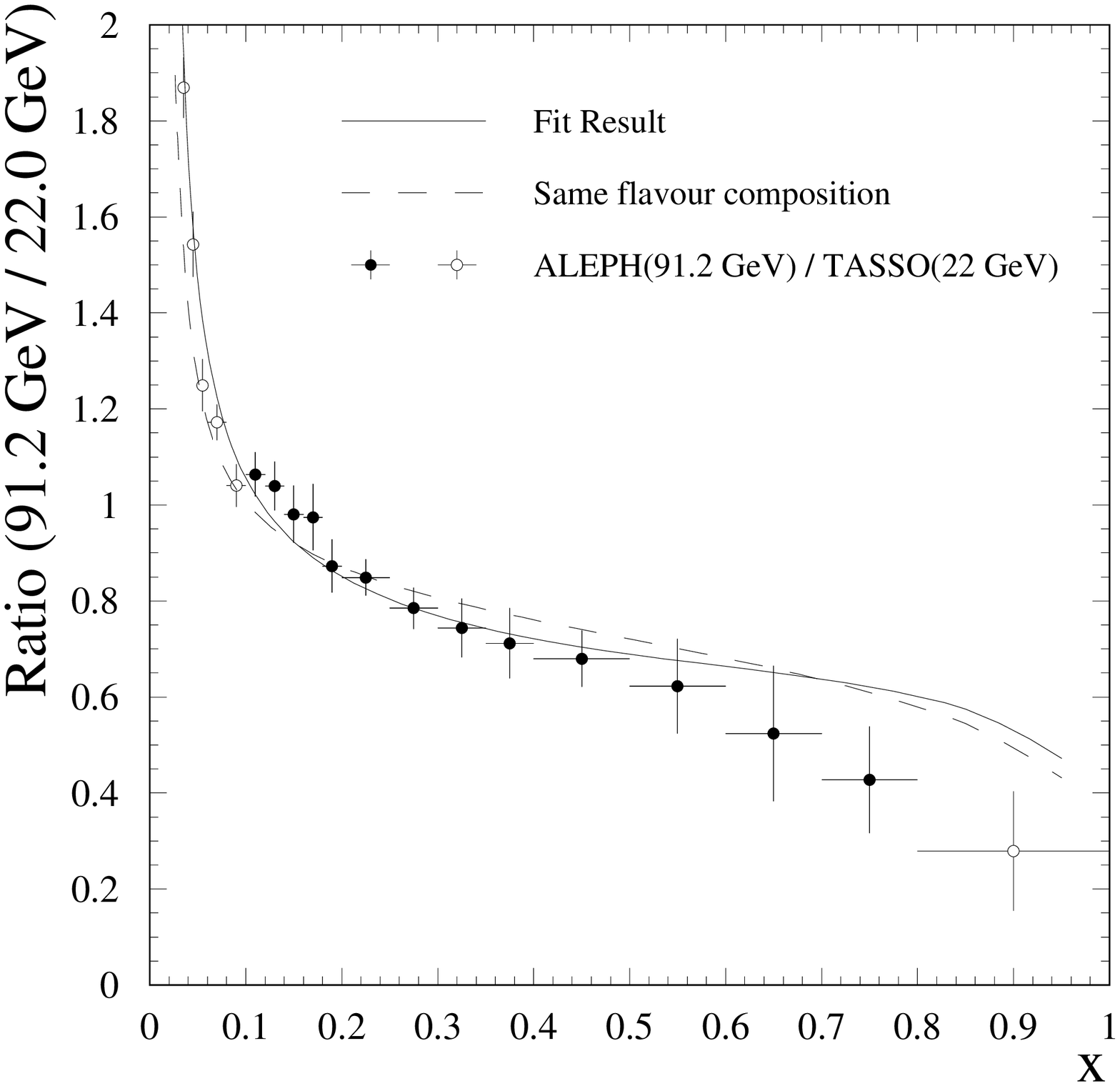} &
\includegraphics[width=0.5\textwidth]{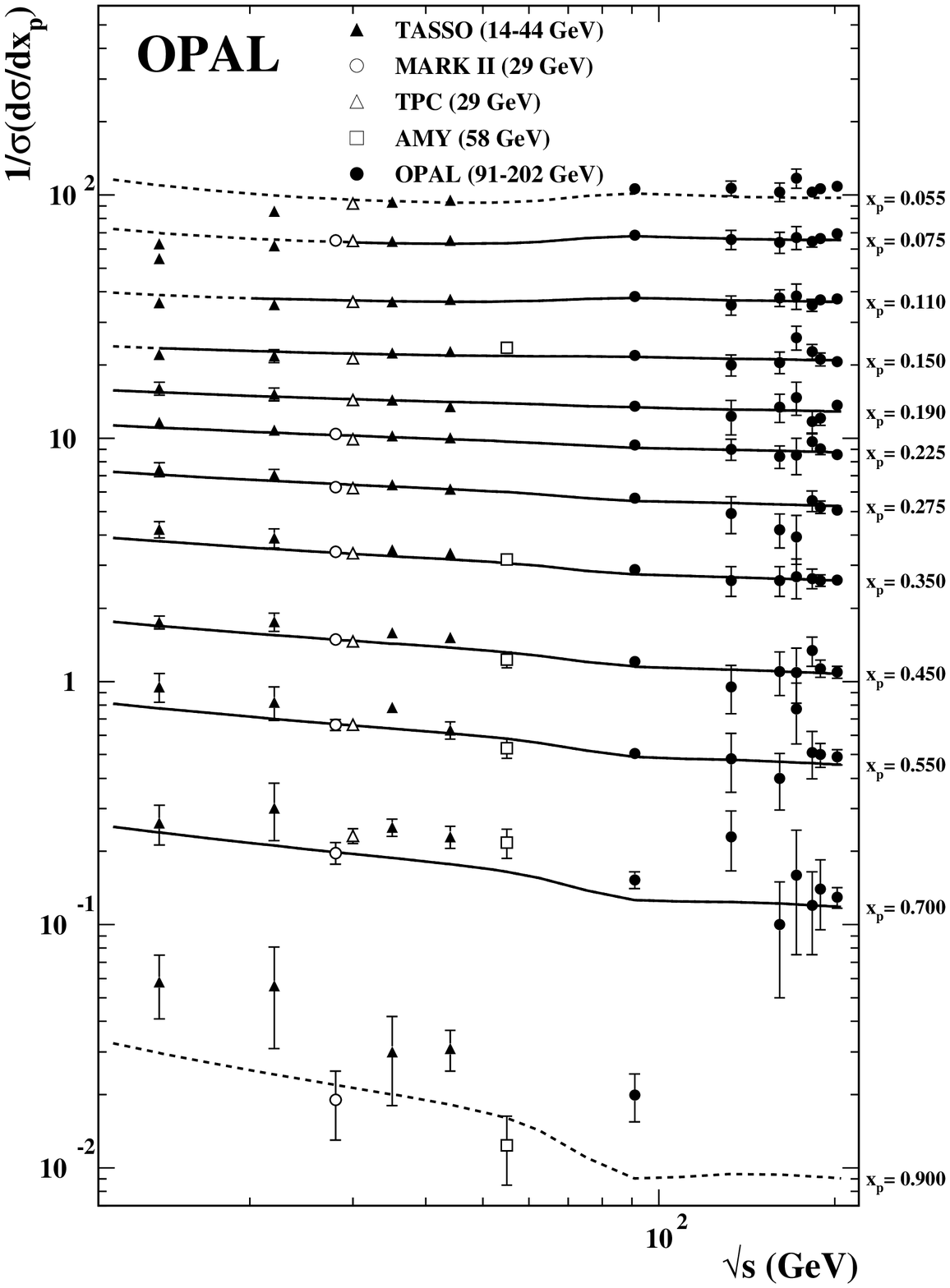} \\
\end{tabular}
\caption[ bla ]{ (left) Ratio of the inclusive charged particle $x$
spectra $1/sigmah\ddel\sigma^h/\ddel x$ measured by ALEPH at
$\roots\simeq 91$~GeV to data from TASSO measured at
$\roots=22$~GeV~\cite{aleph102}.  The lines present the result of a
fit of the NLO QCD prediction (see text); the dashed shows the
prediction with constant quark flavour composition.  (right)
Measurements of the normalised cross section for charged particle
production in bins of $x$ as a function of \roots~\cite{OPALPR362}.
The solid lines show the result of a fit of the NLO QCD prediction
(see text) while the dashed show an extrapolation of the fit result. }
\label{fig_xspectra}
\end{figure}

The fragmentation functions $D_f(x,\mu)$ correspond to the probability
to obtain a hadron with momentum fraction $x$ from a parton $f$ at the
momentum scale of the hard interaction $\mu$.  The coefficient
functions are known in NLO~\cite{rijken96,rijken97}.  The
fragmentation functions are non-perturbative and thus cannot be
calculated; they are the \epem\ analog of parton density functions in
DIS.  However, the rate of change (evolution) of the $D_f(x,\mu)$ with
changing hard scale $\mu$ is predicted by the DGLAP equations:
\begin{equation}
  \frac{\ddel D_f(x,\mu)}{\ddel\ln\mu^2} = \sum_{i\in\{udscbg\}} \int_x^1 
  P_{if}(z,\as(\mur),\xmu) D_i(\frac{x}{z},\mu) \frac{\ddel z}{z}
\end{equation}
where the splitting functions $P_{ij}$ are also known in NLO QCD,
see~\cite{nason94} and references therein.  With these predictions one
measurement at a reference scale $Q_0$ may be used to predict the
$x$ distribution at another scale $Q$ with the strong coupling constant
\as\ as a free parameter.  

Figure~\ref{fig_xspectra} (left) shows the ratio of $x$ distributions
measured by ALEPH at $\roots=\mz$ and by TASSO at
$\roots=22$~GeV~\cite{aleph102}.  The smaller (larger) fraction of
hadrons at large (small) $x$ at LEP~1 energies compared to
$\roots=22$~GeV is clearly visible.  The solid line indicates the
result of a fit of the NLO QCD prediction for this observable to
several measurements of $x$ distributions at different energy scales.
Figure~\ref{fig_xspectra} (right) presents the measurements by OPAL
using LEP~1 and LEP~2 data together with measurements by previous
experiments~\cite{OPALPR362}.  The solid lines show the QCD fit as
explained above with $\as=0.1182$ fixed and variable parametrisations
of the fragmentation functions.  The description of the data by the
fit is adequate within the uncertainties.  The result of a fit with
\asmz\ as an additional free parameter is shown in 
table~\ref{tab_assvcomb}. 

\begin{table}[htb!]
\caption[ bla ]{ Values of \asmz\ determined from analyses of
scaling violation of $x$ spectra in \epem\ annihilation.  Experimental
errors also include statistical uncertainties and uncertainties from
parametrisation of non-perturbative fragmentation functions.  The
calculation of the average value is explained in the text.  }
\label{tab_assvcomb}
\begin{indented}\item[]
\begin{tabular}{l c @{$\pm$} c @{$\pm$} c } 
\hline\hline
Experiment & \asmz & expt. & theo. \\
\hline
ALEPH~\cite{aleph102}   & 0.126 & 0.007 & 0.006 \\
DELPHI~\cite{delphi146} & 0.124 & 0.007 & 0.009 \\
OPAL~\cite{OPALPR362}   & 0.113 & 0.005 & 0.007 \\
{\bf average}           & {\bf 0.1192} & {\bf 0.0056} & {\bf 0.0070} \\
\hline\hline
\end{tabular}
\end{indented}
\end{table}

Table~\ref{tab_assvcomb} collects results for \asmz\ from analysis of
scaling violation in $x$ spectra from \epem\ annihilation to hadrons.
All analyses are based on the same NLO QCD calculations for the
coefficient and splitting functions and thus the results may be
compared directly.  We observe good agreement between the three values
of \asmz\ within the experimental uncertainties.  The average value of
\asmz\ from analyses of scaling violation in \epem\ annihilation is
derived assuming the experimental uncertainties to be partially
correlated while the theoretical uncertainties are accounted for by
varying the input values simultanously within their theoretical
errors.  The result is shown in the last row of
table~\ref{tab_assvcomb} and is in good agreement with other
determinations of \asmz\ discussed in this report.  The weights of the
individual results in the average are 0.35 (ALEPH), 0.15 (DELPHI) and
0.50 (OPAL).  This determination of
\asmz\ does not have explicit hadronisation uncertainties, because the
inclusive measurement of hadron production results in a supression of
non-perturbative effects with $1/Q^2$~\cite{dasgupta96}.  Remaining
hadronisation uncertainties enter as uncertainties of the
parametrisation of the fragmentation functions $D_f$ and are thus part
of the experimental errors.

Measurements of $x$ distributions for identified
hadrons~\cite{sldparticles1} have also been studied in global QCD
analyses using perturbative evolution~\cite{kniehl00,kretzer00} and
good agreement between data and theory was found.  The scaling
violations of identified light and heavy (b) quark and gluon jets
using LEP~1 and LEP~2 data has been measured and compared to NLO QCD
predictions in~\cite{OPALPR397}.  The QCD predictions for $x$ spectra
have been observed to be generally in good agreement with the data.

\subsection{Longitudinal and transverse cross section}
\label{sec_flft}

The measurement of the polar angle dependence of hadron production
provides insight into the spin structure of the process
$\epem\rightarrow\mathrm{hadrons}$~\cite{nason94}.  The normalised
double differential cross section for hadron production in
the process $\epem\rightarrow h+X$ is written as
\begin{equation}
  \frac{1}{\sigtot}\frac{\ddel^2\sigma^h}{\ddel x\ddel\cos\theta} =
   F_T(x)\frac{3}{8}(1+\cos^2\theta) +
   F_L(x)\frac{3}{4}\sin^2\theta +
   F_A(x)\frac{3}{4}\cos\theta 
\label{equ_sigtla}
\end{equation}
where $\sigma^h$ is the cross section for producing a hadron $h$, 
$x$ is the scaled momentum fraction of the hadron and $\theta$ is
the angle between the hadron and electron beam directions.  The
fragmentation functions (FFs) are given by
$F_P(x)=1/\sigtot\ddel\sigma^h_P/\ddel x$, $P=T, L, A$.  The first two
terms stem from the polarisation states (transverse or longitudinal)
of the intermediate vector boson ($\gamma$ or \znull) while the third
term comes from parity violation of the electroweak interaction.

After integrating out the $\theta$ dependence only the $F_T(x)$
and $F_L(x)$ terms remain.  As a result of energy conservation the
integral
\begin{equation}
  \frac{1}{2}\int_0^1 x \frac{1}{\sigtot}\frac{\ddel\sigma^h}{\ddel x} dx =
  \frac{1}{2}\momone{x}\momone{n} = 1 =
  \frac{\sigt}{\sigtot} + \frac{\sigl}{\sigtot}
\end{equation}
where $\momone{n}$ is the average hadron multiplicity and $1/2\int x
F_P(x)dx = \sigma_P$, $P=T, L$.  The transverse and longitudinal cross
sections \sigt\ and \sigl\ fulfil $\sigt+\sigl=\sigtot$.

Measurements of the FFs $F_P(x)$, $P=T, L, A$ for charged hadrons have
been performed by ALEPH, DELPHI and OPAL using large samples of hadronic
\znull\ decays~\cite{OPALPR133,aleph102,delphi124}.  The associated cross
section ratios $\sigt/\sigtot$ and $\sigl/\sigtot$ for all hadrons are
derived from the $F_P$ for charged hadrons assuming that the ratio of
charged to neutral hadrons is identical for \sigt\ and \sigl.  This
assumption is supported by studies of simulated
events~\cite{OPALPR133,delphi124}.  A re-analysis of JADE data gave
results for \sigt\ and \sigl\ at
$\roots=36.6$~GeV~\cite{blumenstengel01a}.

\begin{figure}[htb!]
\begin{tabular}{cc}
\includegraphics[width=0.5255\textwidth]{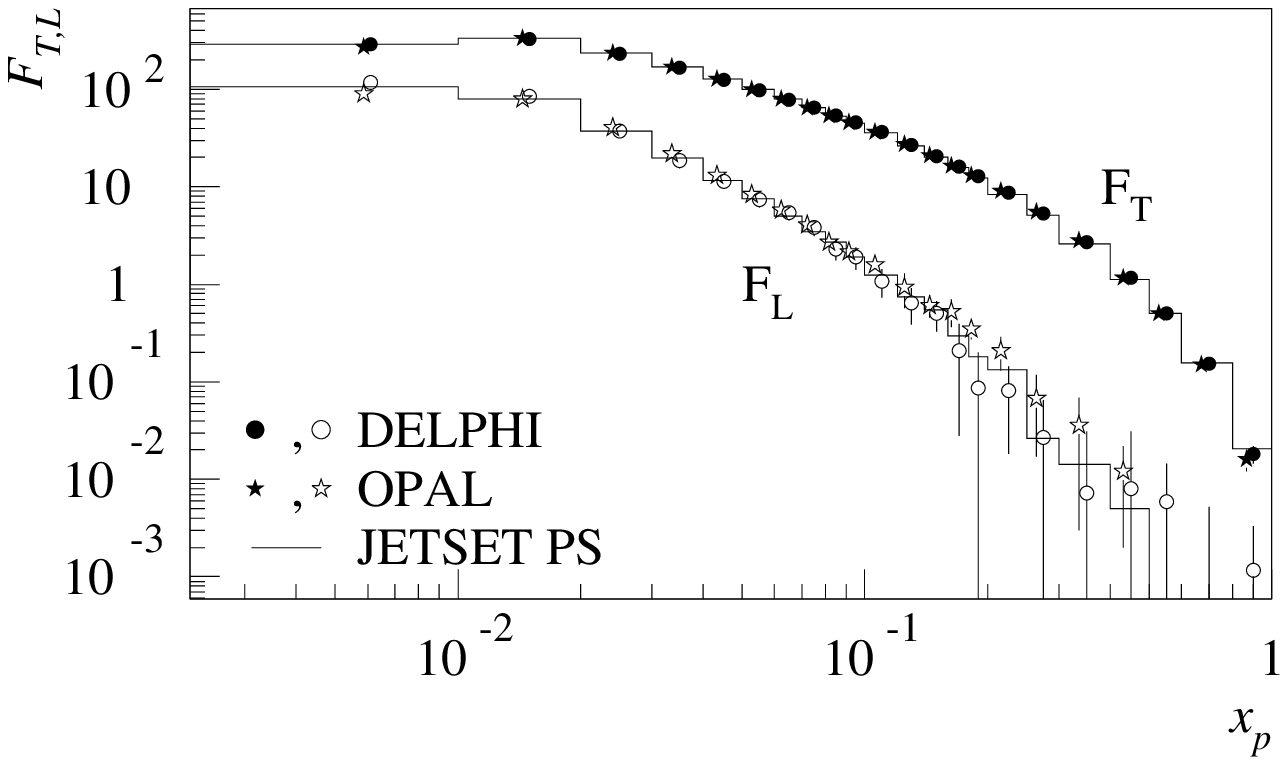} &
\includegraphics[width=0.425\textwidth]{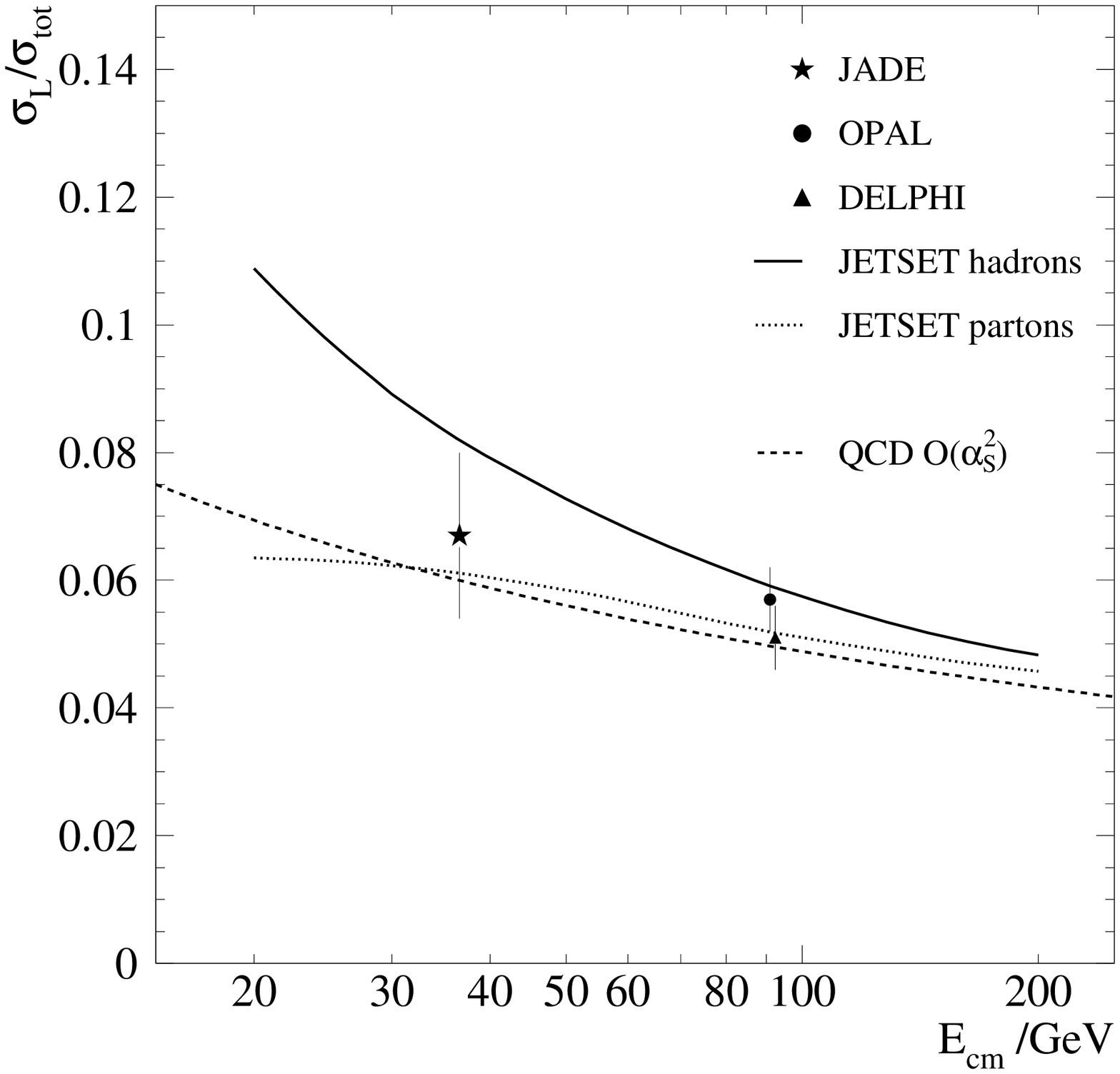} \\
\end{tabular}
\caption[ bla ]{ (left) Fragmentation functions $F_T$ and $F_L$ measured
by DELPHI and OPAL~\cite{delphi124}.  The lines show the prediction of
the JETSET Monte Carlo model.  (right) Measurements of $\sigl/\sigtot$
by DELPHI, OPAL and JADE compared with parton- and hadron-level
predictions by the JETSET Monte Carlo model and a NLO QCD
prediction~\cite{blumenstengel01a} (see text for details).  }
\label{fig_ftfl}
\end{figure}

Figure~\ref{fig_ftfl} (left) shows the FFs determined by DELPHI and
OPAL~\cite{delphi124}.  The analyses agree well with each other and a
Monte Carlo model prediction (JETSET) is also found to agree with the
data.  The longitudinal FF $F_L$ is suppressed w.r.t.\ to the
transverse FF $F_T$ reflecting the preferred polarisation of the
intermediate gauge bosons.  The presence of $F_L$ is understood as
due to the possible presence of gluons in the final state.  The
results for \sigl\ are shown in figure~\ref{fig_ftfl}
(right)~\cite{blumenstengel01a} and compared with the JETSET Monte
Carlo model prediction for partons and hadrons.  The difference
between the two lines for partons and hadrons indicates the presence
of non-perturbative corrections.  These power corrections have been
calculated~\cite{dokshitzer95,dasgupta96} and were found to agree with
the data~\cite{blumenstengel01a}.

The extraction of \asmz\ from the data can be done using the NLO
prediction for the observable
$\sigl/\sigtot$~\cite{rijken96,delphi124}.  In order to obtain the
best possible measurement we average the results for $\sigl/\sigtot$
from ALEPH, DELPHI and OPAL.  The ALEPH measurements of $F_L(x)$ and
$F_T(x)$~\cite{aleph102} are used to extract a value for
$\sigl/\sigtot$ after estimating the missing contributions at
$0<x<0.008$ using the ALEPH measurement of
$\nch=21.91\pm0.22$~\cite{aleph098} and the DELPHI measurement of
$F_L(x)/(F_L(x)+F_T(x))=0.286\pm0.021$ for
$0<x<0.01$~\cite{delphi124}.  The estimate uses the relation $\int_0^1
F_L(x) + F_T(x) dx = \nch$.  The result is
$(\sigl/\sigtot)_{\mathrm{ALEPH}}= 0.0561 \pm0.0003\stat
\pm0.0023\syst$.  The average of the ALEPH, DELPHI and OPAL results is
computed assuming the experimental uncertainties to be uncorrelated
with the result $\sigl/\sigtot= 0.0558 \pm0.0021$.  The value of
\asmz\ is extracted based on the NLO QCD
prediction~\cite{rijken96,delphi124} combined with a hadronisation
correction using JETSET derived from~\cite{OPALPR133}.  We find
\begin{equation}
  \asmz= 0.1169\pm0.0035\expt\pm0.0018\had\pm0.0072\theo\;\;.
\end{equation}
The hadronisation uncertainty is given by the difference of values for
\asmz\ when HERWIG instead of JETSET is used to derive the hadronisation
correction~\cite{OPALPR133}.  The theoretical uncertainty is evaluated
by varying the renormalisation scale parameter in the range
$0.5<\xmu<2.0$ according to equation~(\ref{equ_rcoeff}) and is
consistent with~\cite{delphi124}.

The connection of the asymmetric FF $F_A(x)$ with parity violation in
the electroweak interaction involving quarks has been investigated
in~\cite{nason94a}.  A precise measurement of $F_A(x)$ could be
converted into a measurement of the electroweak mixing angle
$\sin\theta_W$ assuming that hadrons at high $x$ are likely to carry
a quark produced in the \epem\ annihilation.  This prediction has
been found to agree with data in~\cite{OPALPR133,delphi124} but
the precision of the measurements was insufficient to extract a value
for $\sin\theta_W$.

\subsection{Fragmentation of b quarks}

The fragmentation process with heavy quarks is of special interest,
because the large mass of the heavy quark provides a natural cutoff in
perturbative QCD calculations (see e.g.~\cite{mele91,cacciari03}).  As
a consequence reliable predictions of the perturbative part of the
fragmentation process become possible.  Assuming the b quark
fragmentation process to be universal, precise measurements of the
fragmentation of b quarks into b-flavoured hadrons with \epem\
annihilation data can have consequences for the interpretation of
hadron collider data for B meson production~\cite{cacciari02}.

\begin{figure}[htb!]
\begin{tabular}{cc}
\includegraphics[width=0.425\textwidth]{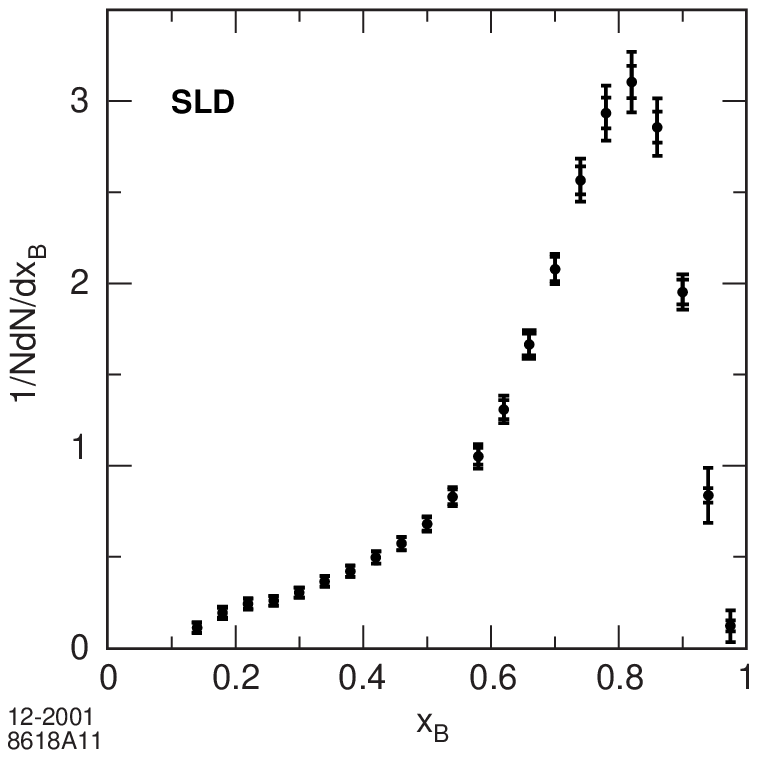} &
\includegraphics[width=0.525\textwidth]{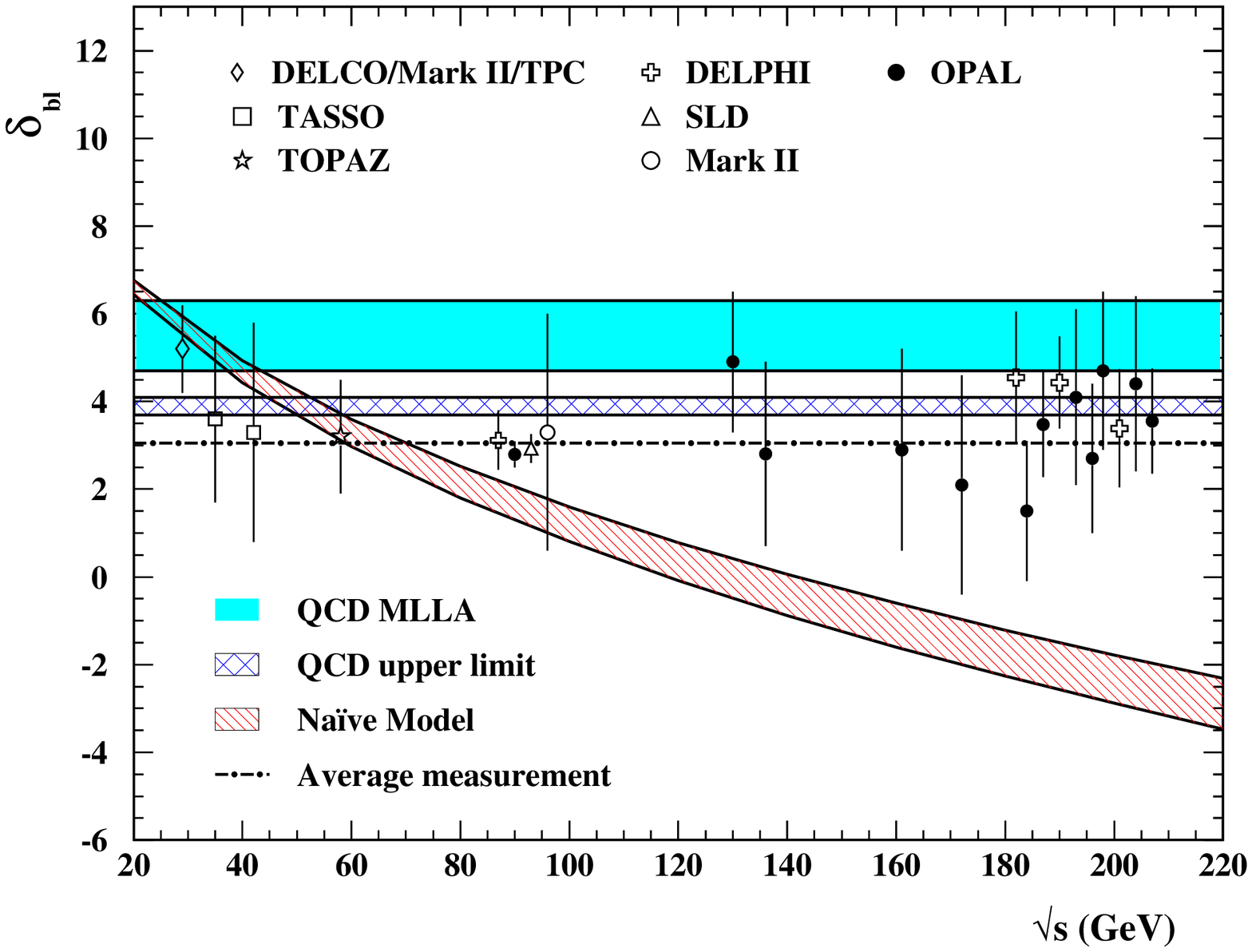} \\
\end{tabular}
\caption[ bla ]{ (left) Normalised $x$ spectrum of B hadrons produced
in \epem\ annihilation on the \znull\ peak~\cite{sldbfrag}.  (right)
Difference of charged particle multiplicity between light and b quark
events \delbl\ measured by OPAL and other experiments at various
cms energies~\cite{OPALPR359}.  The data are compared with several
predictions as explained in the text. }
\label{fig_bfrag}
\end{figure}

Figure~\ref{fig_bfrag} (left) presents as a representative example the
measurement by SLD of the $x$ spectrum of B hadrons produced in \epem\
annihilation at $\roots\simeq\mz$~\cite{sldbfrag}.  The variable
$x=x_B=2E_B/\roots$ corresponds to the scaled momentum variable
discussed above, but due to the large B hadron masses the B hadron
energy $E_B$ is used.  Other recent measurements
are~\cite{aleph232,OPALPR359}.

The measurement of the $x$ spectrum of B hadrons is model dependent,
because correcting the measured B hadron energies for experimental
effects uses Monte Carlo simulations with models for the b quark
fragmentation.  The first step in Monte Carlo simulations is the
parton shower which corresponds to a description of the perturbative
part of the transition of a heavy quark into a observable heavy
hadron.  After the parton shower has stopped hadronisation models such
as the Lund string model describe the formation of hadrons from the
remaining quarks and gluons of the parton shower.  The heavy quark
fragmentation models specify how the energy of the heavy quark is
transferred to a heavy hadron within the hadronisation process.

All recent measurements studied the consistency of various models for
b quark fragmentation to B hadrons~\cite{aleph232,OPALPR359,sldbfrag}.
The best description of the observed $x$ spectra of B hadrons is
obtained with the Bowler and with the Lund symmetric models
implemented within the JETSET Monte Carlo event
generator~\cite{OPALPR359,sldbfrag,jetset3} (see the references for
details on the fragmentation models).  In particular the commonly used
Peterson heavy quark fragmentation model is found to be disfavoured by
the data.  

This result is confirmed in a theoretical study~\cite{benhaim04} which
unfolds the perturbative and the non-perturbative part of the
description of the $x$ spectrum of B hadrons using the Mellin
transform technique.  Essentially, moments of the distribution are
calculated and the unfolding is performed in moment space where it
simplifies to a simple multiplication.  The perturbative component is
taken from JETSET or from a NLL perturbative QCD calculation.  In both
cases the Lund and the Bowler fragmentation functions are found to
agree reasonably well with the extracted non-perturbative component
while other fragmentation functions are disfavoured.

The radiation of soft gluons from heavy quarks with mass $m$ and
energy $E$ is predicted to be suppressed within a cone around the
heavy quark flight direction of opening angle $\alpha=m/E$ if
$E>>M>>\lmqcd$.  Assuming direct correspondence of parton and hadron
distributions (LPHD as discussed above) allows to make predictions in
MLLA QCD e.g.\ for the difference in charged particle multiplicities
of light and heavy quark events,
\delbl~\cite{dokshitzer91,khoze97,dokshitzer06}.  The MLLA QCD
prediction is independent of cms energy and its latest value is
$\delbl=4.4\pm0.4$~\cite{dokshitzer06}.  In contrast, the so-called
naive prediction for \delbl\ considers only the reduction in phase
space due to the heavy quark mass.  Thus the observable \delbl\ becomes
energy dependent and decreases with \roots.

Figure~\ref{fig_bfrag} (right) shows a summary of measurements of
\delbl\ at various cms energies~\cite{OPALPR368}.  The bands
indicate the predictions from perturbative QCD and the naive model.
The predictions become significantly different at high energies.  The
recent measurements using LEP~2 data by DELPHI~\cite{delphi256} and
OPAL allow to discriminate between the predictions and the naive
model is clearly disfavoured by the data.  The dash-dotted line
presents an average of all measurements assuming energy independence:
$\delbl=3.05\pm0.19$ which is significantly lower than the
MLLA QCD prediction.  This apparent discrepancy is of the order of
important corrections beyond the MLLA~\cite{dokshitzer06}.

\subsection{Gluon splitting into heavy quarks}
\label{sec_gqq}

The gluon may split into a quark-antiquark pair; the process
$\mathrm{g}\rightarrow\qqbar$ is one of the fundamental processes of
QCD.  When the virtual mass of the gluon is sufficiently large the
splitting process can produce a pair of heavy (charm or bottom)
quarks.  In \epem\ annihilation this process is rare, because it
occurs only at \oaa.  Neglecting quark masses one expects as a rough
estimate at $\roots=\mz$ a rate per hadronic \znull\ decay
$g_{\qqbar}\approx 1$~\% where $q=b,c$.

The study of gluon splitting to heavy quarks is interesting for
several reasons.  The quark masses provide a natural infrared cutoff
in a QCD calculation thus avoiding a usually present source of
divergence and allowing for reliable predictions.  The rates of gluon
splitting into \bbbar, \gbb, and into \ccbar, \gcc, are an important
source of uncertainty in the measurement of the partial widths for
$\znull\rightarrow\bbbar$ or $\znull\rightarrow\ccbar$ decays.  In
particular the $\znull\rightarrow\bbbar$ vertex is sensitive to
possible effects of new physics coupling to the
b-quark~\cite{lepewwg}.  At high energy hadron colliders a large
fraction of the observed B hadrons are thought to stem from gluon
radiation via $\mathrm{g}\rightarrow\bbbar$.  Such processes are a
major background to many searches for new particles decaying to heavy
quarks like e.g.\ the Standard Model Higgs boson.  In order to have
precise predictions for the production rates of b-flavoured jets from
QCD processes at hadron colliders the value of \gbb\ must be known
well.  The most complete calculation~\cite{miller98} includes
resummation of soft gluon contributions and predicts $\gcc=2.0$\% and
$\gbb=0.18$\% for $\asmz=0.118$, $m_c=1.2$~GeV and $m_b=5.0$~GeV.

\begin{figure}[htb!]
\includegraphics[width=0.75\textwidth]{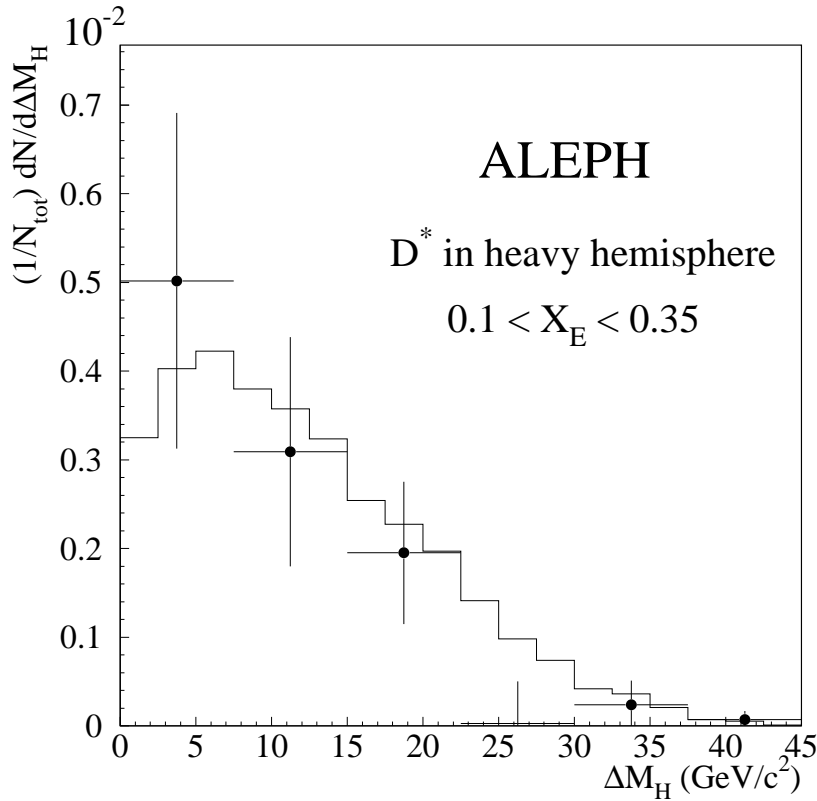}
\caption[ bla ]{ Distribution of \dmh\ of events with a \dstar\ in the
heavy hemisphere after subtraction of the measured contribution from
primary quarks~\cite{aleph206}.  The histogram shows the Monte Carlo
prediction for the process $\mathrm{g}\rightarrow\ccbar$. }
\label{fig_alephgcc}
\end{figure}

For the measurement of \gcc\ the ALEPH collaboration~\cite{aleph206}
classifies hadronic \znull\ decays tagged by the presence of a \dstar\
meson using the observable $\dmh=\mh-\ml$, the difference between the
scaled invariant masses of the heavy (\mh) and light (\ml) hemispheres
of an event.  It is expected that events containing the process
$\mathrm{g}\rightarrow\ccbar$ tend to have a larger \dmh\ due to the
necessary presence of an energetic gluon jet.
Figure~\ref{fig_alephgcc} shows the \dmh\ spectrum of events where the
\dstar\ was found in the heavy hemisphere after subtraction of the
measured contribution from primary charm production.  There is clear
evidence for the process $\mathrm{g}\rightarrow\ccbar$ consistent with
expectations.  The other analyses from L3~\cite{l3194} and
OPAL~\cite{OPALPR284} use the JADE jet clustering algorithm to find
3-jet events.  The gluon jet is identified as the lowest energy jet or
alternatively by OPAL also as the jet which splits into two at the
largest value of \ycut\ using the JADE algorithm.  Events with a charm
tag in the gluon jet are counted as $\mathrm{g}\rightarrow\ccbar$.  The
results are summarised in table~\ref{tab_gccgbb}.

The measurements of \gbb\ of ALEPH~\cite{aleph186},
OPAL~\cite{OPALPR313} and SLD~\cite{sldgbb} select 4-jet events in
hadronic \znull\ decays using the Durham algorithm; between about 10\%
to 15\% of the events are selected.  The jets are searched for the
presence of displaced secondary vertices indicating B-hadron decays.
Events with two such b-tags in two jets lying closely together in
angle are considered as candidates originating from
$\mathrm{g}\rightarrow\bbbar$.  The DELPHI analysis~\cite{delphi226}
measures the rate of \bbbar\bbbar\ events in hadronic \znull\ decays
and the value of \gbb\ is extracted using a theoretical calculation of
$BR(\znull\rightarrow\qqbarg,
g\rightarrow\bbbar)/BR(\znull\rightarrow\bbbar\bbbar)$.  All results
are shown in table~\ref{tab_gccgbb}.

\begin{table}[htb!]
\caption[ bla ]{ Results from  various experiments for \gcc\ and \gbb\
with statistical errors and the uncorrelated and correlated systematic
uncertainties as extracted from the references.  The weights found in
calculating the averages are also shown. }
\label{tab_gccgbb}
\begin{indented}\item[]
\begin{tabular}{lccccc} 
\hline\hline
 & \gcc\ & $\pm$stat. & $\pm$syst. & $\pm$syst. & weight \\
 & $\cdot 10^{-2}$ & & (uncorr.) & (corr.) & \\
\hline
ALEPH~\cite{aleph206} & 3.23 & $\pm$0.48 & $\pm$0.36 & $\pm$0.39 & 0.12 \\
L3~\cite{l3194}       & 2.45 & $\pm$0.29 & $\pm$0.32 & $\pm$0.43 & 0.20 \\
OPAL~\cite{OPALPR284} & 3.20 & $\pm$0.21 & $\pm$0.18 & $\pm$0.35 & 0.68 \\
{\bf average}         & {\bf 3.05} & {\bf$\pm$0.16} & {\bf$\pm$0.14} & {\bf$\pm$0.36} & \\
\hline
 & \gbb\ & $\pm$stat. & $\pm$syst. & $\pm$syst. & weight \\
 & $\cdot 10^{-3}$ & & (uncorr.) & (corr.) & \\
\hline
ALEPH~\cite{aleph186}   & 2.77 & $\pm$0.42 & $\pm$0.28 & $\pm$0.49 & 0.38 \\
DELPHI~\cite{delphi226} & 3.30 & $\pm$1.00 & $\pm$0.41 & $\pm$0.67 & 0.06 \\
OPAL~\cite{OPALPR313}   & 3.07 & $\pm$0.53 & $\pm$0.41 & $\pm$0.88 & 0.09 \\
SLD~\cite{sldgbb}       & 2.44 & $\pm$0.59 & $\pm$0.13 & $\pm$0.32 & 0.47 \\
{\bf average}           & {\bf 2.67} & {\bf$\pm$0.33} & {\bf$\pm$0.13} & {\bf$\pm$0.38} & \\
\hline
\end{tabular}
\end{indented}
\end{table}

The averages are calculated based on uncorrelated and correlated
systematic errors~\cite{hfcomb} extracted from the references given in
table~\ref{tab_gccgbb}.  The off-diagonal elements of the covariance
matrix $V$ for the average are calculated from the correlated errors
$\sigma_i$ and $\sigma_j$ of experiments $i$ and $j$ as
$V_{ij,i\neq j}=\min(\sigma_i,\sigma_j)^2$.  The resulting weights of
the individual measurements in the averages are shown in the last
column of table~\ref{tab_gccgbb}.  The results shown here are
consistent with those given in~\cite{hfinput} but have slightly larger
errors.  The averages $\gcc= 3.05\pm 0.42$\% and $\gbb= 0.267\pm
0.052$\% are larger by two to three standard deviations than the
theoretical predictions given above.

\section{Inclusive observables}
\label{sec_inclobs}

Inclusive observables are independent of the topology of
the hadronic events.  Prominent examples are
$\rhad=\sigma(\epem\rightarrow\mathrm{hadrons})/\sigma(\epem\rightarrow\mpmm)$,
the analogous observable for Z-decays
$\rz=\Gamma(\znull\rightarrow\mathrm{hadrons})/\Gamma(\znull\rightarrow\mathrm{leptons})$
or the hadronic branching ratio in $\tau$-decays
$\rtau=\Gamma(\tau\rightarrow\mathrm{hadrons})/\Gamma(\tau\rightarrow\mathrm{leptons})$.
Inclusive observables depend on only one energy scale and are
effectively QCD corrections to processes predicted by the electroweak
theory involving quarks.  The three observables \rhad, \rz\ and \rtau\
describe the QCD corrections to inclusive hadronic final states
produced in decays of all electroweak gauge bosons $\gamma$, \znull\
and \wpm.  A comparison of the QCD corrections for these three different
electroweak gauge boson decays to hadrons is an important consistency
test of the theory of strong interactions.

Since QCD corrections to electroweak decays to hadrons generally scale
in LO like $(1+\as/\pi)$ at high energies careful measurements with
uncertainties below \order{1\%}\ are needed.  The LEP experiments now
provide precise measurements of electroweak observables such as cross
sections from the line shape of the \znull\ resonance or the decay
widths of the \znull\ boson.  Another important set of results from
LEP are precise measurements of hadronic and leptonic $\tau$ decays
which form the basis for detailed studies of the large QCD corrections
to the electroweak $\tau$ decay.  Data for precision studies of \rhad\
come from the low energy experiments~\cite{davier98,menke01}.

\subsection{\znull\ properties}
\label{sec_zpeak}

The LEP experiments have published their final results on electroweak
observables and a consistent combination of the data is
available~\cite{lepewwg}.  The line shape of the \znull\ resonance
assuming lepton universality is described by the mass of the \znull\
boson \mz, its total decay width \gammaz, the total hadronic cross 
section \sigmah, the ratio of hadronic and leptonic decay width of
the \znull\ boson \rz\ and the leptonic pole forward-backward 
asymmetry $A_{\mathrm{FB}}^{\ell}$.  In addition the partial decay
widths to leptons, hadrons and invisible particles (neutrinos) 
\gammal, \gammah\ and \gammainv\ are measured. 

The results of the combination are preliminary but since the input
values and the combination procedure~\cite{lepewcombmeth} are final
the results are not expected to change.  The important QCD corrections
to the electroweak observables derived from the \znull\ resonance are
known in NNLO QCD and rather precise determinations of the strong
coupling \asmz\ become possible.

We concentrate here on a derived set of four electroweak observables
with high sensitivity to QCD corrections~\cite{lepewwg}.  These are
\gammah, \rz\ and the total cross sections for lepton and hadron
production in \epem\ annihilation \sigmal\ and \sigmah\ on the peak of
the \znull\ resonance.  These observables are related to \mz\ and the
partial and total decay widths:
\begin{equation}
\rz = \frac{\gammah}{\gammal} \;,\;\;
\sigmah = \frac{12\pi}{\mz^2}\cdot\frac{\gammal\gammah}{\gammaz^2} 
\;\;\mathrm{and}\;\;
\sigmal = \frac{12\pi}{\mz^2}\cdot\frac{\gammal^2}{\gammaz^2}\;\;. 
\end{equation}
Results for the partial decay widths are known with
correlations~\cite{lepewwg} while correlations to \mz\ are neglected.
Using this information we derive the correlations between \gammah, \rz,
\sigmah\ and \sigmal\ as shown in table~\ref{tab_zobs}.

\begin{table}[htb!]
\caption[ bla ]{ Electroweak observables with high sensitivity to
QCD corrections.  The correlations were derived from correlations for
the partial decay width of the \znull\ assuming lepton
universality~\cite{lepewwg}. }
\label{tab_zobs}
\begin{indented}\item[]
\begin{tabular}{lc@{$\pm$}l|cccc} 
\hline\hline
 & \multicolumn{2}{c|}{} & \multicolumn{4}{c}{correlations} \\
Observable & \multicolumn{2}{c|}{Result} & \gammah & \rz & \sigmah & \sigmal \\
\hline
\gammah [MeV] & 1744.4 & 1.5    &  1.0  &       & &          \\
\rz           & 20.767 & 0.025  &  0.62 &  1.0  & &          \\
\sigmah [nb]  & 41.540 & 0.037  &  0.24 &  0.18 & 1.0 &      \\
\sigmal [nb]  & 2.0003 & 0.0027 & -0.40 & -0.77 & 0.48 & 1.0 \\
\hline\hline
\end{tabular}
\end{indented}
\end{table}

In order to extract values of \asmz\ from measurements of the
electroweak observables we compare with the complete predictions
including electroweak and QCD radiative corrections calculated with
the electroweak library ZFITTER~6.41~\cite{zfitter}.  The important
QCD corrections are known to \oaaa, e.g.\ for
\gammah~\cite{larin94,chetyrkin94}.  The value of
\asmz\ is varied between 0.091 and 0.14 and the resulting predictions
are interpolated until the experimental value is matched\footnote{The
other input values were $m_t=178$~GeV and $\mz=91.1875$~GeV.}.  The
mass of the Higgs boson is taken to be $\mhiggs=
114^{+69}_{-45}$~GeV~\cite{lepewwg}.  Theoretical uncertainties are
estimated by a consistent variation of the renormalisation
scale~\cite{stenzel05} in the range $0.5<\xmu<2$.  In addition
\mhiggs\ is varied within its uncertainties.  The four results for
\asmz\ are shown in table~\ref{tab_assmobs} with experimental,
theoretical and \mhiggs\ uncertainties and their experimental
correlations.

The theoretical uncertainties for \gammah\ are consistent with an
estimate based on the difference between the standard approach and an
improved theoretical treatment which takes so-called $\pi^2$-terms
into account~\cite{soper96}.  An earlier estimate of the theoretical
uncertainties when using \rz\ is also consistent with our observed
theoretical uncertainty~\cite{hebbeker94}.

A combined value of \asmz\ is derived from the individual values as a
weighted average using the experimental covariance matrix with the
theoretical and \mhiggs\ errors added to the diagonal.  The
experimental error of the combined result is obtained from the
experimental covariance matrix while the theoretical and \mhiggs\
uncertainties are found by varying simultanously the renormalisation
scale or \mhiggs\ and repeating the combination. The combined result
is shown in the last row of table~\ref{tab_assmobs}.  The individual
values of \asmz\ are consistent within their experimental
uncertainties with the average and the combination yields a value of
$\chisqd=1.1$.  The weights of the combination are 0.29 for \gammah,
0.25 for \rz, 0.17 for \sigmah\ and 0.29 for \sigmal.  Changing the
combination procedure to use only the experimental covariance matrix
or including fully correlated theoretical and \mhiggs\ uncertainties
also in the off-diagonal elements of the covariance matrix yields
deviations of the average smaller than 1\%.  The average value is in
agreement with the final result for \asmz\ from jet and event shape
observables discussed in section~\ref{sec_jetsshapes} and also with
current world averages.

\begin{table}[htb!]
\caption[ bla ]{ Results for \asmz\ from electroweak observables with
high sensitivity to QCD corrections.  In each row the first error is
experimental, the second is from variation of \xmu\ and the third is
from variation of \mhiggs.  The measurements are correlated as shown in
table~\ref{tab_zobs}.  The average value was obtained as described in
the text. }
\label{tab_assmobs}
\begin{indented}\item[]
\begin{tabular}{l c@{$\pm$}c@{$\pm$}c@{$\pm$}c}
\hline\hline
 \multicolumn{5}{c}{} \\
Observable & \asmz & exp. & scale & \mhiggs  \\
\hline
\gammah & 0.1221 & 0.0037 & 0.0020 & 0.0015 \\
\rz     & 0.1231 & 0.0037 & 0.0013 & 0.0005 \\
\sigmah & 0.1075 & 0.0069 & 0.0006 & 0.0001 \\
\sigmal & 0.1187 & 0.0030 & 0.0011 & 0.0004 \\
{\bf average} & {\bf 0.1189} & {\bf 0.0027} & {\bf 0.0013} & {\bf 0.0007} \\
\hline\hline
\end{tabular}
\end{indented}
\end{table}

This result may be viewed as one of the most reliable measurements of
\asmz, because the theoretical uncertainties are small due to the
underlying NNLO QCD calculations and the inclusive nature of the
observables.  Uncertainties from measurements of the luminosity used
to obtain cross sections are included in the experimental
uncertainties.  Uncertainties from non-perturbative contributions are
expected to be negligible compared to the experimental errors as e.g.\
the leading power correction to the closely related observable \rhad\
was shown to scale like $1/Q^6$ where $Q=\mz$ is the hard scale of the
process~\cite{dokshitzer95a}.  The uncertainty due to the unknown
\mhiggs\ may be as large as $\Delta\asmz\simeq0.003$ if ranges of
\mhiggs\ up to 1~TeV are considered~\cite{bethke04}.  However, such
large Higgs boson masses would have consequences for the formulation
of the electroweak theory and the whole procedure of extracting a
precise value of \asmz\ would have to be reevaluated.

\subsection{Decay of the $\tau$ lepton}
\label{sec_rtau}

Hadronic decays of the $\tau$ lepton offer unique possibilities to
study the physics of strong interactions at low energy scales
given by the mass of the $\tau$ lepton $\mtau=1.777$~GeV.  The
interplay between perturbative and non-perturbative QCD effects
is expected to play an important r\^ole.  The ratio of decay widths
of the $\tau$ lepton to hadronic and purely leptonic final states 
$\rtau=\Gamma(\tau\rightarrow\mathrm{hadrons}\,\nu_{\tau})/\Gamma(\tau
\rightarrow\ell\nu_{\ell})$ is an observable like \rhad\ and \rz.  
Leading electroweak effects cancel in the ratio and only QCD 
and rather small electroweak corrections remain. 

A particularity of $\tau$ decays is the presence of a W boson in the
weak decay instead of a \znull\ or $\gamma$ in the case of \rz\ or \rhad.
This leads to the possibility of hadronic final states with even
or odd number of pions connected to vector (V) or axial-vector (A) currents
which has interesting consequences for the size of non-perturbative
contributions as discussed below.  The contributions from
perturbative and non-perturbative processes may be separated by
studying the spectrum of invariant masses of the hadronic final 
states~\cite{diberder92b}.

\begin{figure}[htb!]
\begin{tabular}{cc}
\includegraphics[width=0.5\textwidth]{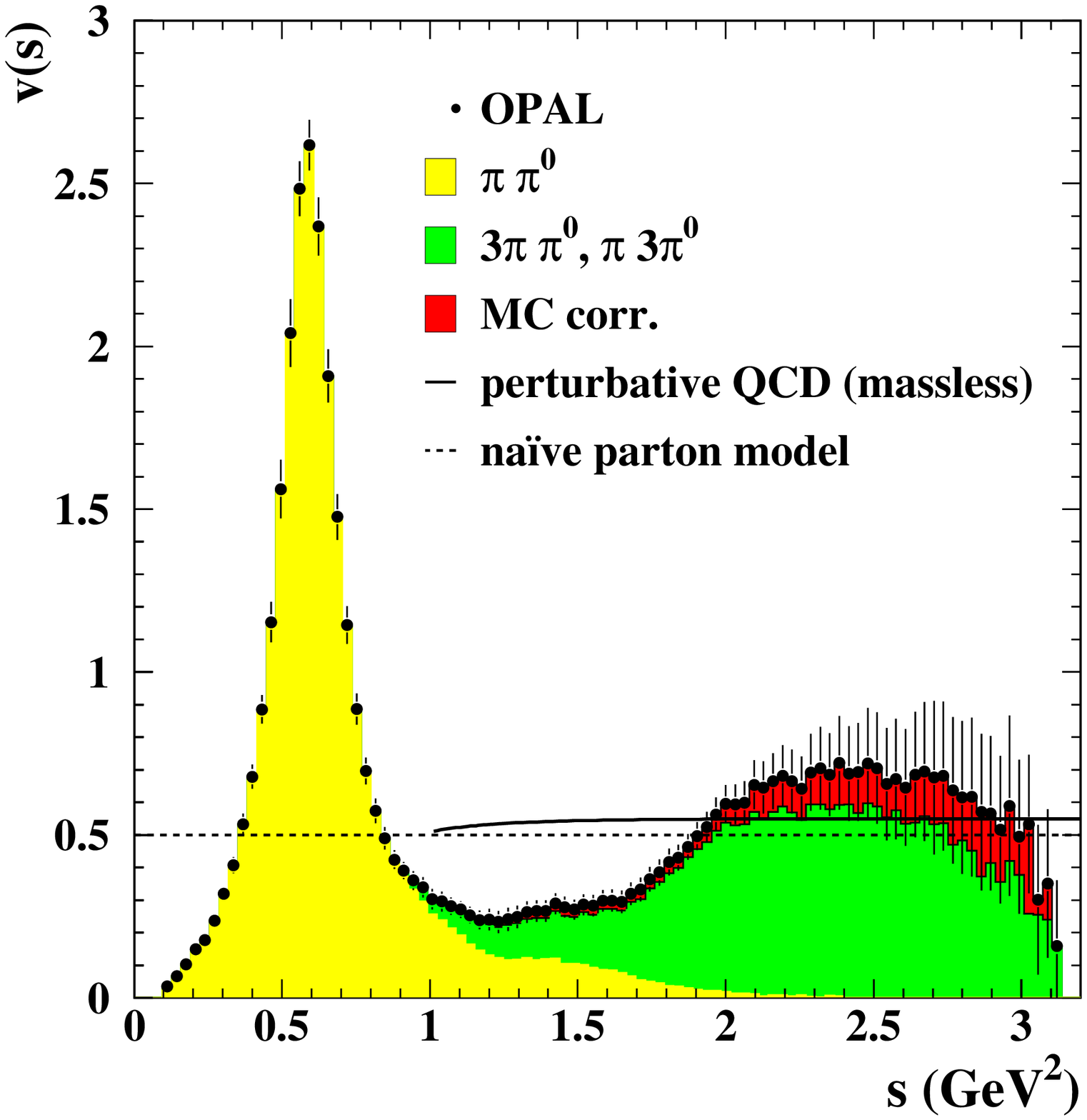} &
\includegraphics[width=0.5\textwidth]{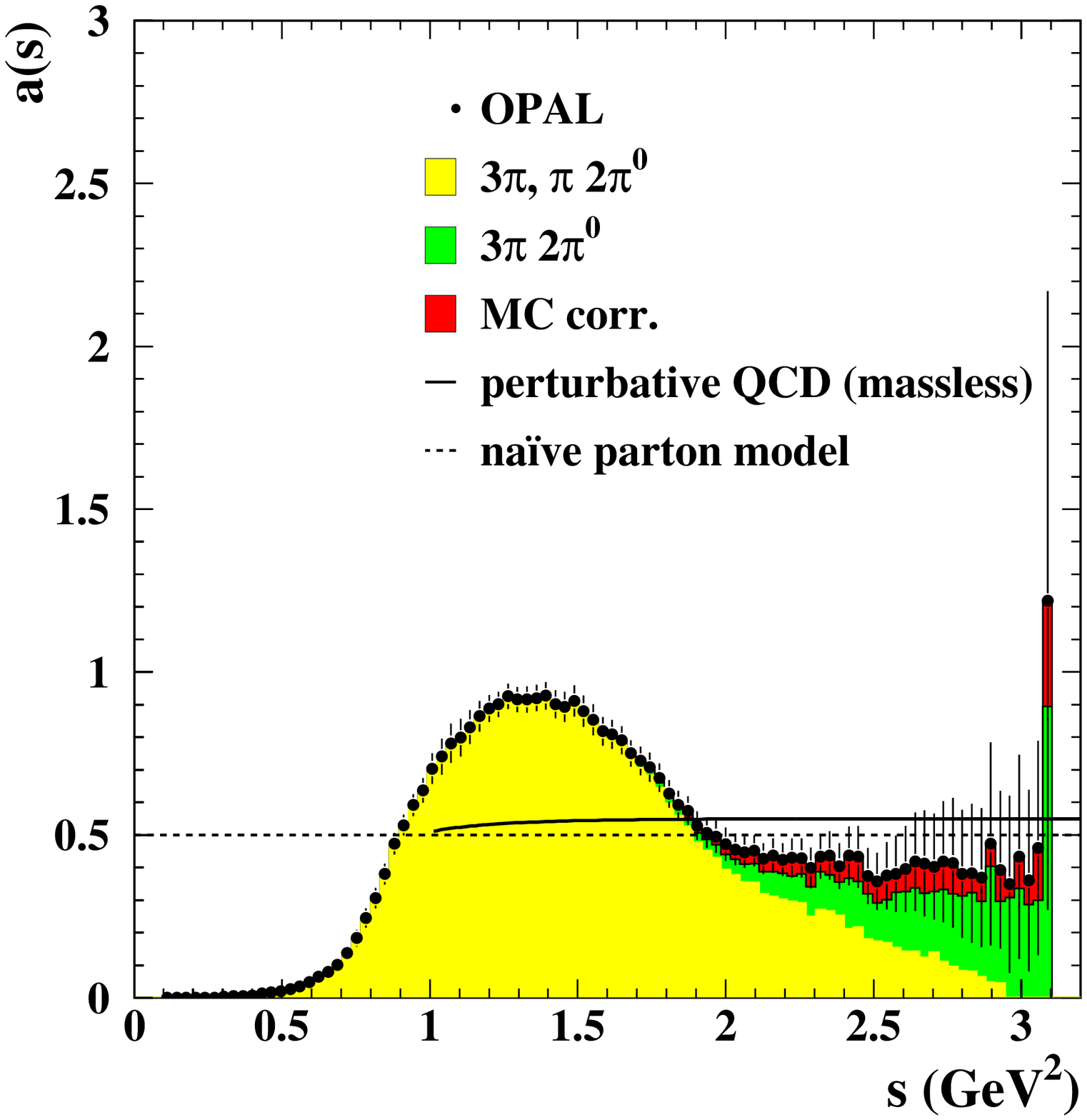} \\
\includegraphics[width=0.475\textwidth]{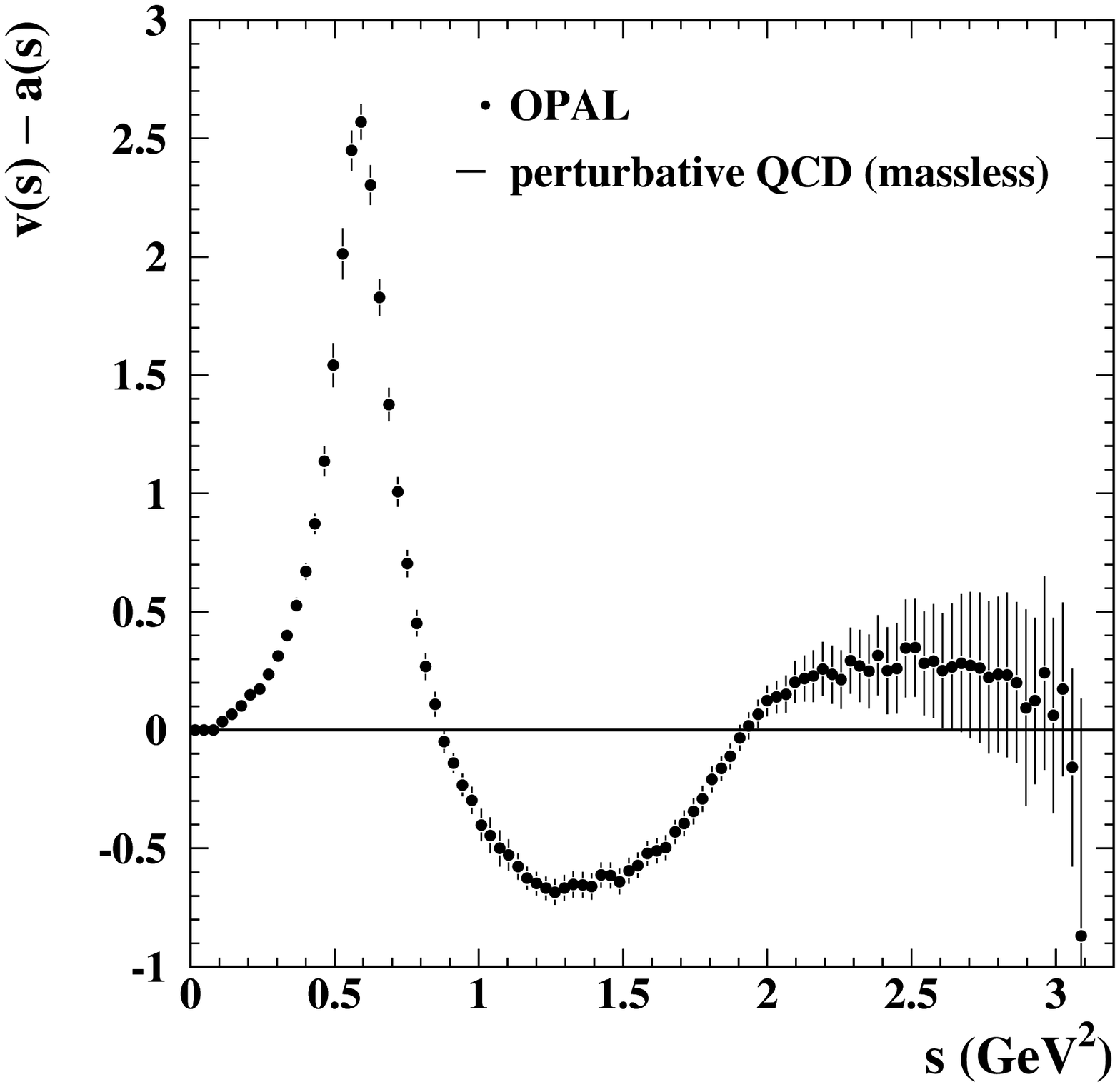} &
\includegraphics[width=0.475\textwidth]{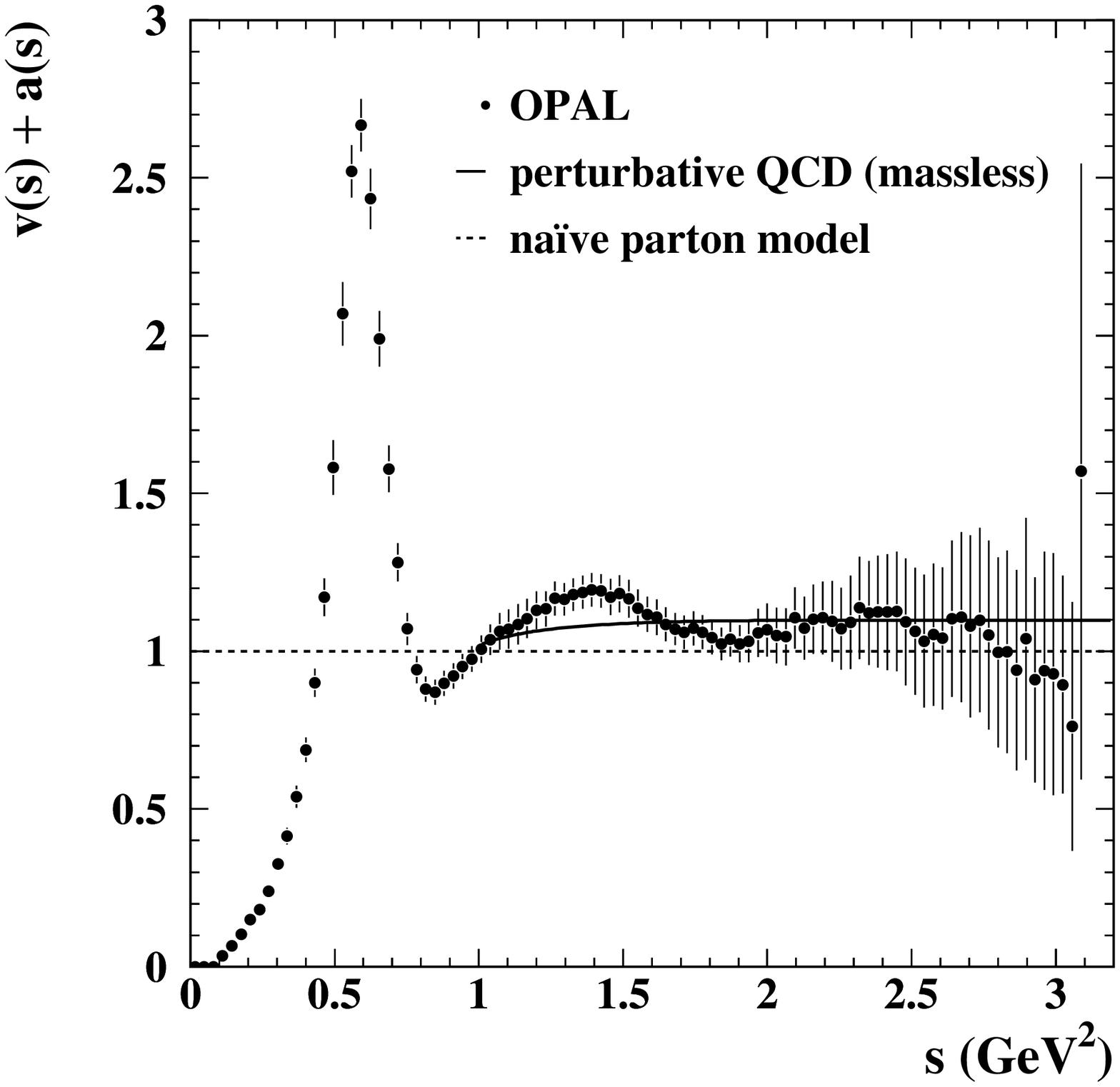} \\
\end{tabular}
\caption[ bla ]{ (upper row) Spectral functions for hadronic $\tau$ 
decays to final states with even (v(s)) or odd (a(s)) number of pions.
The shaded areas show contributions from hadronic final states as
indicated and the Monte Carlo (MC) based correction for unobserved
final states. (lower row) Sum or difference of the spectral functions.
The lines show predictions as indicated on the figures. All figures
from~\cite{OPALPR246}. }
\label{fig_tausf}
\end{figure}

The LEP experiments ALEPH and OPAL have analysed hadronic final states
from $\tau$ decays using large samples of \znull\ decays produced on
resonance~\cite{aleph173,OPALPR246}.  In the ALEPH (OPAL) analysis
124k (150k) $\tau$ pair candidate events are used.
Figure~\ref{fig_tausf} (upper row) shows the spectral functions
$v(s)=\dd R_{\tau,v}/\dd s$ and $a(s)=\dd R_{\tau,a}/\dd s$ with
$R_{\tau,v/a}(s)=\Gamma(\tau\rightarrow\mathrm{pions}\,\nu_{\tau})/
\Gamma(\tau\rightarrow\e\nu_e\nu_{\tau})$ of hadronic $\tau$ decays 
into final states with an even or odd number of pions.  The variable
$s$ denotes the invariant mass of the hadronic final state.  In the
figures the small corrections based on Monte Carlo simulation for
unobserved final states are also indicated.  In the lower row the sum
and the difference of the vector and axial vector spectral functions
$v(s)$ and $a(s)$ are shown.

The dashed and solid lines in figure~\ref{fig_tausf} show the
predictions by the naive parton model and perturbative QCD with
massless partons and $\asmz=0.122$.  The pQCD predictions are seen to
agree well with the inclusive $v(s)+a(s)$ spectral function while for
$v(s)$ and $a(s)$ the prediction lies below or above the data.  This
observation already indicates that there are significant
non-perturbative contributions to $v(s)$ and $a(s)$ which, however,
largely cancel in the sum $v(s)+a(s)$.

For the detailed analysis moments of the spectral functions are
determined~\cite{diberder92b}:
\begin{equation}
  R_{\tau,v/a}^{kl}(s_0) = \int_0^{s_0} \left( 1-\frac{s}{s_0} \right)^k
                           \left( \frac{s}{\mtau^2} \right)^l 
                           \frac{\dd R_{\tau,v/a}}{\dd s} ds
\label{equ_rtau}
\end{equation}
for $kl=00, 10, 11, 12, 13$.  These moments are constructed such that
they are theoretically and experimentally well under control.  The
moment $R_{\tau,v/a}^{00}(s_0=\mtau^2)=\rtau$ corresponds to the
usual hadronic branching ratio in $\tau$ decays.

The theoretical prediction is formulated as a sum of perturbative and
non-perturbative QCD contributions:
\begin{equation}
  R_{\tau,v/a}^{kl}(s_0) = \frac{3}{2} S_{EW} |V_{ud}|^2 
    \left( 1 + \delta_{EW}^{kl} + \delta_{\mathrm{pert}}^{kl} +
           \delta_{\mathrm{non-pert},v/a}^{kl} \right)
\label{equ_dpert}
\end{equation}
where $S_{EW}=1.0194$, $V_{ud}=0.9745\pm0.0004$~\cite{pdg04} and
$\delta_{EW}^{00}=0.001$.  The term $\delta_{pert}^{kl}$ contains the
perturbative QCD part known to
\oaaa~\cite{gorishnii91,samuel91,braaten92}.  The original calculation
of~\cite{braaten92} has been improved in~\cite{diberder92a} by a more
accurate treatment of the running strong coupling in the calculation
using the full $\beta$-function instead of an approximation.  A third
calculation includes a resummation of the effects of renormalon
chains~\cite{neubert96}.  The non-perturbative contribution can be
analysed in the framework of the Operator Product Expansion (OPE) as a
power series in terms of $1/m_{\tau}^2$~\cite{braaten92}
\begin{equation}
  \delta_{\mathrm{non-pert},v/a}^{kl} =
           \sum_{D=2,4,6,\ldots} \delta_{v/a}^{D,kl} \;\;\;.
\label{equ_dpow}
\end{equation}
The first term ($D=2$) in equation~(\ref{equ_dpow}) is a correction
for finite parton masses and can be calculated perturbatively.  The
$D=4,6,8$ terms are studied as non-perturbative corrections in the
analyses.

As already indicated by figure~\ref{fig_tausf} the non-perturbative
corrections turn out to be consistent with zero when moments of 
the inclusive spectral function $v(s)+a(s)$ are analysed.  The
correction terms are found to be of roughly the same size
but opposite sign when moments of $v(s)$ and $a(s)$ are studied
individually~\cite{aleph173,OPALPR246}.  

The analysis of the perturbative term $\delta_{pert}^{kl}$ allows a
precise determination of the strong coupling constant at energy scales
given by the mass of the $\tau$ lepton
$\mtau=1777.0\pm0.3$~GeV~\cite{pdg04} essentially free of
non-perturbative uncertainties.  

Following~\cite{aleph173,OPALPR246} the determination of \as\ from
hadronic $\tau$ decays using $\rtau=R_{\tau}^{00}$ is
updated\footnote{Code and help by S. Menke is greatly appreciated.}
for recent values of the $\tau$ lifetime $\tau_{\tau}=290.6\pm1.1$~ps
and the branching ratio $B_{\mu}=
B(\tau\rightarrow\mu\nu_{\mu}\nu_{\tau})=
0.1736\pm0.0006$~\cite{pdg04}.  The contribution of strange hadronic
$\tau$ decays is subtracted using $R_{\tau,S}=
0.1677\pm0.005$~\cite{OPALPR396}.  We find for the non-strange
branching ratio $\rtau=3.468\pm0.011$.  A similar
study~\cite{davier05} using preliminary ALEPH data~\cite{aleph273}
finds $\rtau=3.471\pm0.011$ in agreement with our result.  The
improved calculation of~\cite{diberder92a} is chosen to extract a
value for \as.  The \order{\as^4} term is set to zero, because for
consistency with other determinations of \as\ discussed in this report
we prefer to extract a value of \as\ without the additional
assumptions needed to estimate the \order{\as^4} term~\cite{baikov03}
\footnote{ In~\cite{baikov03} the $\nf^2$ and $\nf^3$ contributions to
the \order{\as^4} term are given.  Further parts of the \order{\as^4}
term are estimated using the renormalisation scheme optimisation
techniques PMS and ECH (see section~\ref{sec_rgi}). }.  We find
\begin{equation}
  \asmt = 0.347\pm0.005\expt\pm0.003\nonp\pm0.018\theo
\end{equation}
The non-perturbative correction was set to
$\delta_{\mathrm{non-pert},v+a}= -0.0024\pm0.0025$~\cite{OPALPR246}
(consistent with $\delta_{\mathrm{non-pert},v+a}=
-0.0048\pm0.0017$~\cite{aleph273}) and varied within its errors in
order to obtain the uncertainty due to non-perturbative effects.  The
theoretical error was determined by a variation of the renormalisation
scale in the range $0.5<\xmu<2.0$.  We take the larger variation from
changing \xmu\ to define the theoretical uncertainty.  We did not
consider a variation of the renormalisation scheme as done
in~\cite{OPALPR246} in order to be consistent with other
determinations of \as\ discussed in this report.  After evolving to
the \znull\ mass scale using the 3-loop $\beta$-function and 2-loop
matching at the crossing of heavy quark thresholds~\cite{chetyrkin97b}
we find
\begin{eqnarray}
\label{equ_asmztau}
  & \asmz= 0.1221 & \pm0.0006\expt \pm0.0004\nonp \\ \nonumber
  &               & \pm0.0019\theo \pm0.0003\evol \;\;. \nonumber
\end{eqnarray}
The last error \evol\ is due to uncertainties in the evolution
procedure.  The evolution of \asmt\ to \asmz\ takes the crossing of
the charm and bottom quark threshold into account by setting the charm
quark mass to $m_c=1.778$~GeV.  The uncertainty due to the evolution
procedure is evaluated following~\cite{rodrigo98} by setting the
matching thresholds to $m_c=4$~GeV or $m_b=20$~GeV and by changing the
absolute quark masses within their uncertainties~\cite{pdg04}.

The analogous determination of \asmt\ and \asmz\ using the calculation
of~\cite{neubert96} results in $\asmt=0.303\pm0.003\expt\pm0.002
\nonp\pm0.011\theo$ and $\asmz=0.1166\pm0.0004\expt
\pm0.0003\nonp\pm0.0015\theo\pm0.0003\evol$.  The difference between 
the result in equation~(\ref{equ_asmztau}) and this result is
$\Delta\asmz=0.0055$ which corresponds to less than three standard
deviations of the total errors.  However, since the calculation
of~\cite{neubert96} is rather different  compared to
e.g.~\cite{diberder92a} such discrepancies might be expected.
In~\cite{koerner00} the strong coupling constant is extracted using the
ALEPH data~\cite{aleph173} and a renormalisation group invariant formulation
of the perturbative QCD prediction.  Their result is $\asmz=0.1184\pm0.0007
\expt\pm0.0019\theo\pm0.0006\evol$ which is in fair agreement 
with~(\ref{equ_asmztau}).  A similar result was found
in~\cite{raczka96a}.  The difference between using the calculation
of~\cite{braaten92} and the improved calculation of~\cite{diberder92a}
is $\Delta\asmz\simeq 0.003$~\cite{OPALPR246}.  Using the prediction
of~\cite{diberder92a} including the estimate of the \order{\as^4} term
as in~\cite{OPALPR246} we find $\Delta\asmz=-0.0006$.

This measurement of \asmz\ shown in equation~(\ref{equ_asmztau}) has
comparable uncertainties to and is consistent with the measurement
using the \znull\ line shape observables discussed in
section~\ref{sec_zpeak}.  Even though the difference between different
methods of extracting \asmz\ from the hadronic $\tau$ data appears
large it corresponds to about three times the total uncertainties.  We
conclude that the determination of \asmz\ from hadronic $\tau$ decays
is reliable within the reasonably well estimated total errors.

\subsection{\epem\ annihilation at low energies}
\label{sec_rhad}

In the \epem\ annihilation to hadrons at low energies the intermediate
gauge boson is predominantly a photon thus giving access to the
process $\gamma\rightarrow\qqbar$.  The quantity of interest is
$\rhad=\sigma(\epem\rightarrow\mathrm{hadrons})/\sigma(\epem\rightarrow\mpmm)$,
because many experimental and theoretical uncertainties introduced by
selection efficiency corrections or radiative corrections are reduced
in the ratio.  Many measurements exist and have been summarised in
other reviews, see e.g.~\cite{bethke00a,hinchliffe00,pdg04}.
Theoretical predictions for this observable exist in
\oaaa~\cite{gorishnii91,surguladze91}.

In~\cite{bethke00a} a consistent analysis of \rhad\ using TRISTAN and
PETRA data~\cite{haidt95} is used to extract the strong coupling
constant in \oaaa\ with the result
$\as(42.4~\mathrm{GeV})=0.175\pm0.028$ or $\asmz=0.126\pm0.022$.  The
measurements of \rhad\ take account of the \znull\ mass measured at LEP
and also include improved electroweak radiative corrections.

The CLEO collaboration has measured \rhad\ at $\roots=10.52$~GeV, i.e.\
just below the Y(4S) resonance.  A QCD prediction including heavy
quark masses and corrected for QED radiation is used to extract
$\as(10.52~\mathrm{GeV})=0.20\pm0.01\stat\pm0.06\syst$.  For the
evolution to the \znull\ mass scale we use a b quark threshold of
10.53~GeV such that the value $\as(10.52~\mathrm{GeV})=0.20$ is
interpreted as the strong coupling in a regime with four active quark
flavours.  The result is
\begin{equation}
  \asmz = 0.130\pm0.004\stat\pm0.025\syst
\end{equation}
in good agreement with previous results.  The dependence of this
result on the actual value of the b quark threshold and the b quark
mass $m_b=4.25\pm0.15$~GeV~\cite{pdg04} in the \msbar\ scheme is
negligible.

The behaviour of $\rhad(s)$ at low values of $s$ in the region of
hadronic resonances may be described by perturbative QCD in the same
way as the spectral functions from hadronic $\tau$
decays~\cite{davier98}.  The perturbative prediction for \rhad\ is
written in terms of a the so-called Adler D-function:
\begin{equation}
  \rhad(s_0) = \frac{1}{2\pi i} \oint_{s=s_o} \frac{D(s)}{s} ds
\end{equation}
where the D-function $D(s)$ is known in
\oaaa~\cite{gorishnii91,surguladze91}.  Non-perturbative effects are
treated by applying the OPE to moments of $\rhad(s)$:
\begin{equation}
  \rhad^{kl}(s_0) = \int_{4m_{\pi}^2}^{s_0} \left( 1-\frac{s}{s_0} \right)^k
                           \left( \frac{s}{s_0} \right)^l 
                           \frac{\rhad(s)}{s_0} ds\;\;\;.
\end{equation}
Similar to the case of hadronic $\tau$ decays the complete prediction
for $D(s)$ is a sum of perturbative and non-perturbative
terms\footnote{ Corrections for finite quark masses are viewed as part
of the perturbative prediction.}.  The non-perturbative part of $D(s)$
in the OPE is a power series in terms of $1/s^n$ starting at $n=3$.

In~\cite{menke01} the same data as in~\cite{davier98} are used to
determine the moments $\rhad^{kl}(4~\mathrm{GeV}^2), kl=20, 30, 31,
32, 33$.  A simultaneous fit of the QCD prediction for these moments
with $\as(2~\mathrm{GeV})$, the strange quark mass and the first two
non-perturbative terms proportional to $1/s^3$ and $1/s^4$ gives a
good description of the data.  The result for the strong coupling
constant is $\as(2~\mathrm{GeV})=0.286\pm0.031\expt\pm0.015\theo$ 
corresponding to
\begin{equation}
  \asmz= 0.117 \pm 0.005 \expt \pm 0.002 \theo\;\;\;.
\end{equation}
Compared to the result from hadronic $\tau$ decays shown in
equation~(\ref{equ_asmztau}) the theoretical uncertainties are similar
while the experimental errors are larger.  Within the uncertainties
the measurements from hadronic $\tau$ and \znull\ decays and from
hadron production in \epem\ annihilation are consistent.  New
measurements of hadron production in \epem\ annihilation at low
\roots\ promise to allow reduced experimental uncertainties in the
future~\cite{kloesigmahad} such that this measurement should have
competitive uncertainties.

\section{Summary of \as\ determinations}
\label{sec_as}

In this section we collect and systematically compare all significant
measurements of the strong coupling constant \asmz\ discussed in this
report.  We concentrate on reliable measurements with comparatively
small uncertainties in order to probe the consistency of the theory 
in the best possible way.

Table~\ref{tab_as} presents the results selected from various sections
of this report.  The first two rows of the table contain the results
based on \oaa+NLLA QCD with hadronisation correction performed using
Monte Carlo simulation.  As discussed in section~\ref{sec_lepjadeas}
the analyses from JADE and LEP use the same set of 3-jet event shape
observables sensitive to \as\ in \oa.  The ALEPH, DELPHI and OPAL
analysis (see section~\ref{sec_r4as}) based on the 4-jet fraction
\rfour\ are sensitive to \as\ in \oaa.  ALEPH and OPAL use NLO+NLLA
calculations and the Durham algorithm while DELPHI uses NLO
calculations with experimentally optimised renormalisation scale and
the Cambridge algorithm.

A direct comparison of the results from JADE and LEP shows very good
agreement.  As the set of observables is identical and the analysis
procedures are rather similar the good agreement shows that a
consistent analysis of 3-jet event shape observables is possible from
the lowest PETRA/JADE cms energies of 14~GeV to the highest LEP~2 cms
energies of about 200~GeV.  The results from analysis of \rfour\ are
consistent with the other NLO values within the uncertainties.  This
confirms that the calculation of radiative corrections within
perturbative QCD is consistent within this set of rather different
observables.  

\begin{table}[htb!]
\caption[ bla ]{ Summary of determinations of \asmz\ based on
observables, theoretical calculations and experiments as indicated.
The experimental errors (exp.) also include statistical errors.
Numbers covering two columns represent combined errors.  The observables
\rfourd\ and \rfourc\ refer to \rfour\ using the Durham or Cambridge
jet finding algorithm.  }
\label{tab_as}
\begin{tabular}{lllcccc} \hline\hline
 & & Experiments  & & & & \\ 
Observable & Theory & (Q [GeV]) & \asmz & $\pm$exp. & $\pm$had. & $\pm$theo. \\
\hline
3-jet shp. & NLO+NLLA & JADE (14-44) & 0.1203 & 0.0018 & 0.0053 & 0.0050 \\
3-jet shp. & NLO+NLLA & LEP (91-207) & 0.1201 & 0.0009 & 0.0019 & 0.0049 \\
\rfourd & NLO+NLLA & ALEPH (91) & 0.1170 & 0.0009 & 0.0010 & 0.0017 \\
\rfourd & NLO+NLLA & OPAL (91-207) & 0.1182 & 0.0015 & 0.0011 & 0.0018 \\
\rfourc & NLO & DELPHI (91) & 0.1175 & 0.0011 & 0.0027 & 0.0007 \\
sc. viol. & NLO & LEP (14-207) & 0.1192 & 0.0056 & \multicolumn{2}{c}{0.0070} \\
$\sigl/\sigtot$ & NLO & LEP (91) & 0.1169 & 0.0035 & 0.0018 & 0.0072 \\
\gammah, \rz, \sigmah, \sigmal & NNLO & LEP/SLD (91)
 & 0.1189 & 0.0027 & - & 0.0015 \\
\rtau & NNLO & \epem\ (1.777) & 0.1221 & 0.0006 & 0.0004 & 0.0019 \\
\rhad & NNLO & \epem\ (2) & 0.117 & 0.005 & \multicolumn{2}{c}{0.002} \\
\hline\hline
\end{tabular}
\end{table}

Having observed good consistency between the 3-jet results we
construct an average value of the results from JADE and LEP assuming
that statistical errors are uncorrelated and experimental and
hadronisation errors are partially correlated.  Theoretical errors are
treated as in the combinations of \as\ in section~\ref{sec_lepjadeas}.
The result for \asmz\ from 3-jet event shapes using NLO+NLLA QCD
predictions is
\begin{eqnarray}
 & \asmz = 0.1202 & \pm0.0004\stat \pm0.0009\expt \\ \nonumber
 &                & \pm0.0025\had \pm0.0049\theo \;\;.
\end{eqnarray}
The weights are 0.32 for the
JADE result and 0.68 for the LEP result.  The determination of \asmz\
in \oaa+NLLA from 3-jet event shape observables has a fairly large
theoretical uncertainty compared with some other measurements.  This
uncertainty might be reduced substantially when complete NNLO
calculations for 3-jet production in \epem\ annihilation will become
available~\cite{gehrmann-deridder04,gehrmann-deridder05}.

The measurements of \asmz\ from \rfour\ discussed in
section~\ref{sec_r4as} have small uncertainties.  However, since no
other comparable measurements based on different 4-jet observables
exist consistency checks were not possible.  For the purpose of
presentation we combine the ALEPH and OPAL measurements based on
\rfourd\ using the Durham algorithm and NLO+NLLA theory and the DELPHI
measurement based on \rfourc\ using the Cambridge algorithm and NLO
theory.  Experimental errors are assumed as partially correlated while
hadronisation and theoretical uncertainties are treated by
simultaneous variation of the input values within their individual
errors.  The result from analyses of 4-jet fractions \rfour\ becomes
$\asmz= 0.1175 \pm0.0002\stat \pm0.0010\expt \pm0.0014\had
\pm0.0015\theo$.

The result from the NLO analyses of scaling violation (sc.~viol.) in
\epem\ annihilation to hadrons is discussed in section~\ref{sec_sv}.
The method is inclusive and thus has negligible hadronisation
uncertainties.  There is good agreement with the results from other
observables.  However, since the QCD calculations are limited to NLO a
direct comparison and combination with other inclusive measurements of
\asmz\ based on NNLO QCD discussed below is not attempted.

The seventh row of table~\ref{tab_as} gives the result of the NLO
analysis of the ratio of longitudinal to total cross section
$\sigl/\sigtot$ from section~\ref{sec_flft}.  Combining all available
measurements of this observable and using the hadronisation
corrections derived from~\cite{OPALPR133} yields a consistent
measurement of \asmz\ with a significantly reduced experimental error
compared with~\cite{delphi124}.  This measurement is combined with the
determination of \asmz\ from scaling violations assuming experimental
errors as partially correlated while hadronisation and theory errors
are treated by simultaneous variation of the input values within their
hadronisation or theoretical uncertainties.  We find for \asmz\ from
analysis of fragmentation in \epem\ annihilation to hadrons:
\begin{equation}
  \asmz = 0.1179 \pm 0.0040\expt \pm 0.0010\had \pm 0.0071\theo\;\;.
\end{equation}

The lower three rows of table~\ref{tab_as} present results from the
inclusive observables discussed in section~\ref{sec_inclobs} derived
using \oaaa\ (NNLO) QCD calculations.  The inclusive nature of the
observables allows the calculation in \oaaa\ containing two-loop
radiative corrections and reduces uncertainties from non-perturbative
effects.  The results for \asmz\ from the \znull\ line shape
observables \gammah, \rz, \sigmah\ and \sigmal\ and from analysis of
\rtau\ are in agreement within less than two standard deviations of
their total uncertainties.  The good agreement of the precise values
of \asmz\ from the \znull\ line shape observables and \rtau\ at the
level of $\Delta\asmz=0.0032\pm0.0032$ may be viewed as a strong test
of asymptotic freedom, since two measurements obtained at energy
scales differing by a factor of \order{100} are compared.  In the
calculation of the error of $\Delta\asmz$ only the smaller of the
theoretical uncertainties was considered.

We derive an average value of the NNLO results assuming statistical
errors as uncorrelated and experimental errors to be partially
correlated.  Hadronisation and theoretical uncertainties are treated
by simultaneous variation of the input values within their uncertainties.
The result for \asmz\ from inclusive observables in NNLO QCD is
\begin{equation}
  \asmz = 0.1211 \pm 0.0010 \expt \pm 0.0018 \theo
\label{equ_asnnlo}
\end{equation}
where statistical errors have been included in the experimental error
and hadronisation uncertainties are included in the theory error. 
The weights are 0.27 for the \znull\ line shape observables, 0.70 for
\rtau\ and 0.03 for \rhad\ with $\chisqd=1.05/2$.  The consistency between
the three results is impressive as indicated by the \chisqd\ value.
This result is considered to be the most reliable determination of the
strong coupling constant \asmz\ discussed in this report, because it
is based on NNLO QCD calculations and has only small uncertainties
from modelling of non-perturbative effects.  Further improvements of
this determination of \asmz\ may be expected from new measurements of
\rhad\ in the low energy region leading to a reduction of the
experimental uncertainties~\cite{kloesigmahad}.

\begin{figure}[htb!]
\includegraphics[width=0.75\textwidth]{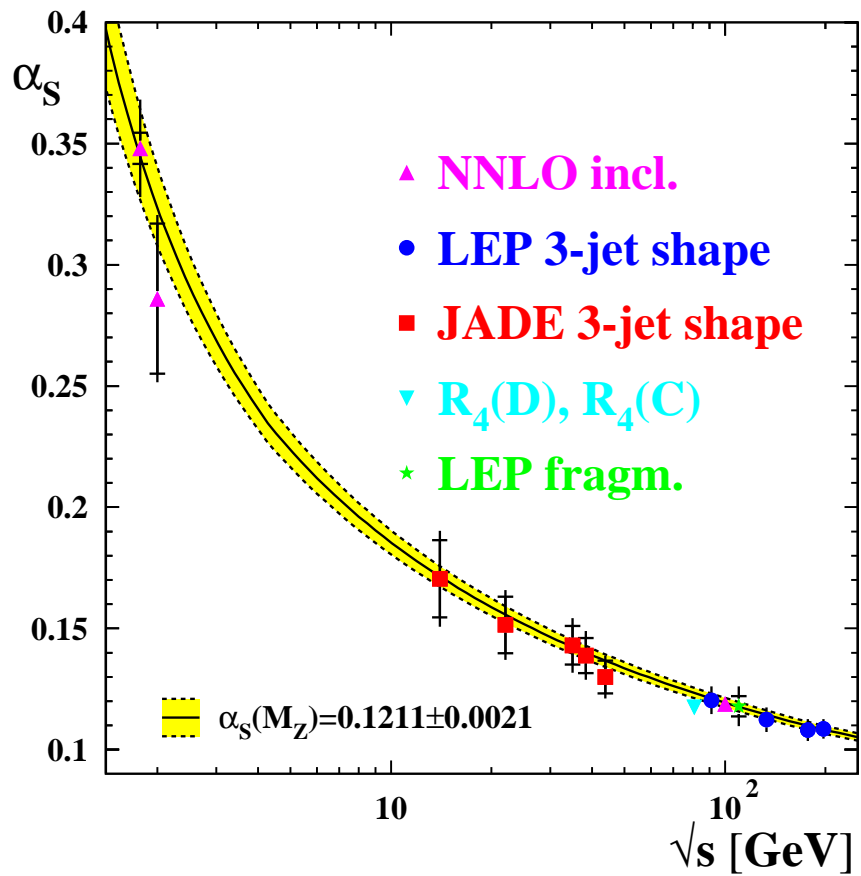} \\
\caption[ bla ]{ Values of \as\ determined at cms energy \roots\ from
various processes as indicated.  The NNLO incl.\ data are from \rtau(\mtau), 
\rhad(2 GeV) and the \znull\ line shape observables.  The error bars give 
experimental and total uncertainties.  Invisible error bars are smaller
than the symbols.  The measurements at $\roots\simeq\mz$
have been separated horizontally for clarity.  The lines and shaded band
show the running of the strong coupling based on the average value
shown in equation~(\ref{equ_asnnlo}). }
\label{fig_assummplot}
\end{figure}

Figure~\ref{fig_assummplot} summarises the high precision measurements
of \as\ discussed in this section.  The data are compared with our
average of NNLO QCD based determinations of \asmz\ given in
equation~(\ref{equ_asnnlo}) shown as lines and grey band.  The value
of \asmz\ is evolved in NNLO to other cms energies \roots\ as
explained in section~\ref{sec_qcdrun}.  The consistency between the
predicted behaviour of $\as(\roots)$ using only the NNLO observables
with the other measurements is good; this shows that QCD successfully
describes many important features of hadron production in \epem\
annihilation\footnote{More generally speaking, the analyses cover
hadronic decays of electroweak gauge bosons.} at scales covering two
orders of magnitude from \mtau\ to the highest LEP~2 energies of about
200~GeV.  The most significant difference is $\Delta\asmz=
0.0035\pm0.0021$ between the combined results from \rfour\ and from
inclusive observables in NNLO.  Both errors are combined taking
experimental and hadronisation uncertainties as uncorrelated and
theoretical errors as fully correlated.  Since \rfour\ is an exclusive
observable based on 4-jet final states in contrast to the fully
inclusive observables assuming experimental and hadronisation
uncertainties to be uncorrelated seems reasonable.

\section{Quark and gluon jets}
\label{sec_qg}

Differences between the properties of jets originating from quarks or
gluons are a basic consequence of QCD.  The quarks and gluons can be
viewed as carriers of colour charges given by their colour factors
which are $\cf=4/3$ for quarks and $\ca=3$ for gluons.  Predictions
for e.g. the ratio of parton multiplicities
\begin{equation}
\frac{\momone{n}_g}{\momone{n}_q} = \frac{\ca}{\cf}
\end{equation}
in jets originating from hard gluons or
quarks~\cite{brodsky76,konishi78} can be related to observed
quantities like charged hadron multiplicities measured in
experimentally identified quark or gluon jets by assuming local parton
hadron duality~\cite{azimov85a}.  One therefore expects the particle
multiplicities in gluon jets to be larger than in comparable quark
jets.  

With the relations $\momone{x}_p\momone{n}_p=1, p=q,g$, where
$x=p_{\mathrm{parton}}/p_{\mathrm{jet}}$ is the scaled momentum of a
parton produced in a quark or gluon jet and $\momone{n}_p$ is the
parton multiplicity in the jet, one gets
$\momone{x}_q/\momone{x}_g=\ca/\cf$.  Thus gluon jets can be expected
to have a softer scaled hadron momentum spectrum (fragmentation
function) compared to quark jets of the same energy~\cite{konishi78}.
The tails at large $x\simeq 1$ of the fragmentation functions in gluon
or quark jets can be predicted to have a ratio $\sim
(1-x)/\ln(1-x)$~\cite{konishi78}.  This implies that the gluon jet
fragmentation function is expected to be softer by a factor of
approximately $1-x$.

In experimental studies of differences between jets stemming from hard
quarks or gluons the main difficulty lies in achieving a clear and
unambiguous identification of the original parton.  Due to the
confinement property of QCD free partons are not observed but instead
they form more or less collimated showers of softer partons and in
turn of hadrons.  This may cause a loss of the close correspondence
between the hadronic final state and the underlying hard parton state.
In addition the correlated colour flow between the quarks and gluons
leads to effects of colour coherence with the consequence that the
properties of gluon jets depend on the topology of the state in which
they were produced, see e.g.~\cite{basicspqcd}.

Most theoretical predictions of differences between e.g.\ the average
multiplicity of charged hadrons \nch\ in quark or gluon jets don't
take experimental definitions of jets into account~\cite{gary94}.  The
jets are simply defined as \qqbar\ or gg systems produced from a
colour singlet point source.  Hadronic final states in \epem\
annihilation are a source of \qqbar\ systems but gg systems from a
colour singlet point source rarely occur in nature, the decay
Y(3S)$\rightarrow\gamma\chi(2P)_{b,J=0,2}$,
$\chi(2P)_{b,J=0,2}\rightarrow$gg~\cite{alam92} or production via
2-photon interactions~\cite{khoze91} being examples.  This prevents a
clean interpretation of the experimental results based on jet finding
algorithms to define jets and thus experimental techniques matching
the theory more closely had to be found.

The next sections describe results obtained using exclusively defined
jets employing jet finding algorithms and results based on inclusive
jet definitions without involvement of jet finding algorithms.

\subsection{Exclusive jets in 3-jet events}
\label{sec_excljets}

The first analysis showing significant differences between quark and
gluon jets used LEP data from the \znull\ peak and a novel technique
to obtain clean samples of quark and gluon jets of the same
energies~\cite{OPALPR038}.  The method~\cite{nilles81} employs lepton
tagging of quark jets in one-fold symmetric 3-jet events, so-called
Y-events.  The selection demands the angles between the highest energy
jet and the other two jets to be within $150\degr\pm10\degr$ yielding
jet energies of about 24~GeV for the lower energy jets.  The
presence of a high momentum lepton (electron or muon) is required in
one of the two lower energy jets.  Since most such leptons stem from
weak heavy quark decays the corresponding jet is identified as a quark
jet.  The production of heavy quark pairs from gluons of the
perturbative cascade contributes ($3.1\pm0.4$)\% to c and
($0.27\pm0.05$)\% to b production and is thus a small
effect (see section~\ref{sec_gqq}).  The most energetic jet is also a
quark jet with high probability due to the bremsstrahlung nature of
hard gluon radiation and thus the remaining jet is likely to be a
gluon jet.  The selection of a second event sample requiring the same
selection criteria except the lepton tagging allows to correct
measurements of quark jet properties for a possible lepton tagging
bias.  The gluon jets are assumed to be unbiased.

\begin{figure}[htb!]
\begin{tabular}{cc}
\includegraphics[width=0.475\textwidth,clip]{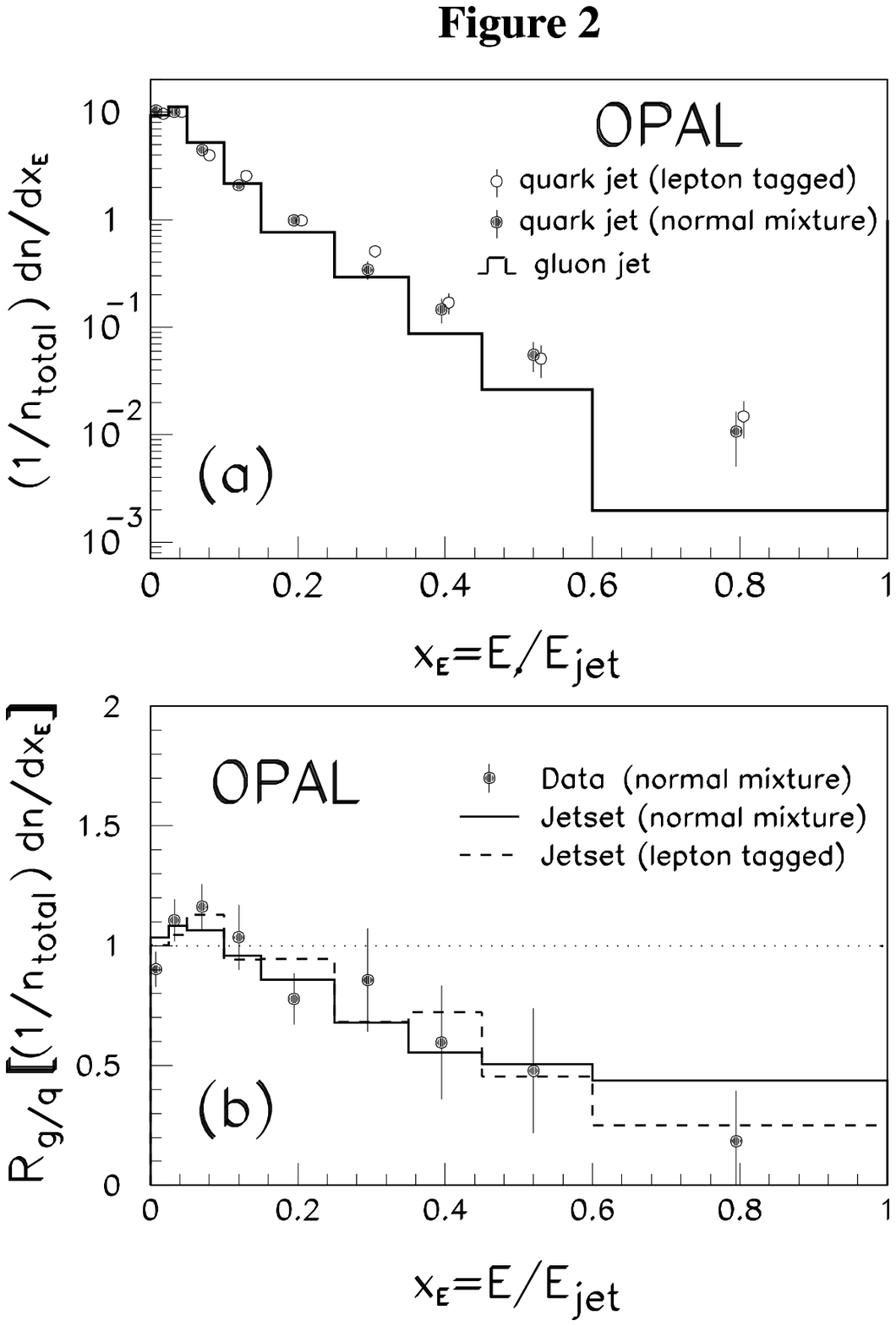} &
\includegraphics[width=0.475\textwidth,clip]{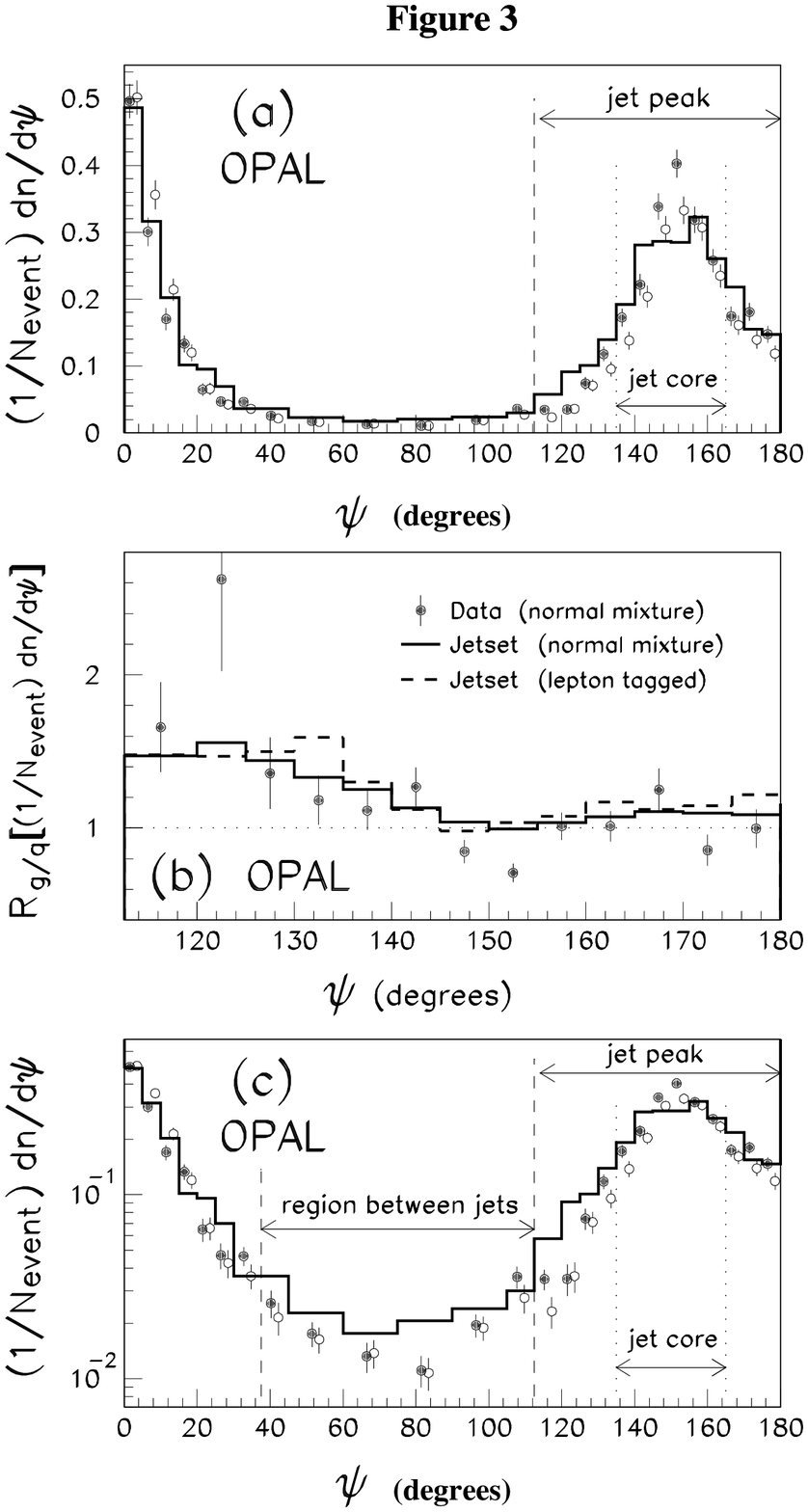} \\
\end{tabular}
\caption[ bla ]{ (left) (a) Scaled energy spectrum of particles from
quark (points) or gluon (line) jet cores.  The solid points are
corrected for the tagging bias.  (b) Ratio of gluon to corrected quark
jet data of (a).  The lines in (b) are predictions from JETSET~7.2.
(right) (a) and (c) Multiplicity spectrum as a function of the angle
in the event plane $\psi$ for quark (points) and gluon (line) jets on
a linear or logarithmic scale.  The solid points are corrected for
the tagging bias.  The angle $\psi$ runs for points on the quark jet
side and for the line on the gluon jet side.  (b) Gluon/quark jet
ratio of the data in (a) and (c).  All figures from~\cite{OPALPR038}.
}
\label{fig_opalpr038}
\end{figure}

The main results of the analysis are summarised in
figure~\ref{fig_opalpr038}.  On the left in (a) the scaled energy
spectrum of particles in the cores of the quark or gluon jets is shown
where the open (solid) points represent the data after lepton tagging (after
lepton tagging and bias correction).  The line displays the result for gluon
jets.  The jet cores are defined as the angular region $135\degr<\psi<
165\degr$ where $\psi$ is the angle in the event plane w.r.t. the
highest energy jet.  In left (b) the ratio of the bias corrected quark
to the gluon jet data is given.  The plots show convincing evidence
for a softer scaled energy spectrum in gluon jets, as predicted.  On
the right in (a) and (c) the multiplicity spectrum obtained in quark
(points) and gluon (line) jets is presented as a function of the angle
$\psi$.  The open and solid points are defined as above.  The angle
$\psi$ runs on the quark jet side of the event plane for points and on
the gluon jet side for the line.  Figure (b) on the right shows the
gluon/quark jet ratio for $112.5\degr<\psi< 180\degr$.  The figures
show evidence for a broader particle distribution in gluon jets
compared to quark jets.  The discrepancy visible in (c) in the {\em
region between jets} is a manifestation of the string effect, see
e.g.~\cite{OPALPR030}.

In a similar analysis using the same technique of lepton tagging of
quark jets in 3-jet events the OPAL collaboration studied
distributions of electric charge in quark and gluon
jets~\cite{OPALPR067}.  The electric charge of a jet is defined by the
sum $Q$ of charges of all charged particles assigned to the jet.  The
distributions of $Q$ filled separately for events with a positive or
negative lepton tag are observed to have mean values significantly
different from zero for the tagged and the highest energy jet.  The
$Q$ distribution for gluon jets has a mean value consistent with zero.
The signs of the quark jet $Q$ distribution mean values correspond to
the expected charge of the primary quark or antiquark.  This
observation constitutes experimental evidence based on \epem\ data
that gluons don't have electric charge.

These analyses have been refined in several ways.  The use of vertex
instead of lepton tagging of heavy quarks employs that the decay
vertex of heavy hadrons can be reconstructed using micro-vertex
detectors.  The technique uses a larger fraction of heavy hadron
decays and allows better efficiencies and purities.  The use of
alternative jet finding algorithms like the cone~\cite{OPALPR097} or
the JADE algorithm quantified the influence of the jet reconstruction
method on the results~\cite{OPALPR136,delphi115}, in particular the
ratio of average charged particle multiplicities in quark and gluon
jets at 24~GeV jet energy was found to be biased by about 10\%.  A
different correction method (\qqbar g/$\gamma$) uses hadronic events
consisting of two jets and one isolated and energetic photon as a pure
sample of quark jets~\cite{delphi115}.  This method allows analysis of
quark and gluon jets at varying energies finding e.g.\ that the ratio
of charged particle multiplicities in gluon and quark jets depends on
the jet energy.

A direct comparison of quark jets originating from light (u, d or s)
or b quarks with gluon jets was performed by tagging the high energy
jets of the one-fold symmetric samples~\cite{OPALPR141,aleph113}.
Vertex tagging is used to select b jets while the absence of charged
particle tracks with large positive impact parameter indicates light
quarks.  Comparing the gluon jet tagged sample (vertex tag in a lower
energy jet) with the light or b tagged sample (impact parameter or
vertex tag in the high energy jet) allows to extract the properties of
pure gluon, light or b jets at jet energies of about 24~GeV.  Some
results for the ratios of charged particle multiplicities are shown in
table~\ref{tab_rnch}.

\begin{table}[htb!]
\caption[ bla ]{ Ratios of charged particle multiplicities of gluon 
and inclusive quarks (\rgq), gluon and light quarks (\rguds) or gluon
and b quark jets (\rgb) at $\sim 24$~GeV jet energy based on the
Durham, JADE or the Cone algorithm.  The methods ``Y'' and
``\qqbar g/$\gamma$'' refer to the analysis methods based on one-fold
symmetric Y-events or \qqbar g and \qqbar$\gamma$ events to obtain
results for pure quark and gluon jet samples from the data. }
\label{tab_rnch}
\begin{indented}\item[]
\begin{tabular}{lcccccc} 
\hline\hline
Jet alg. & \ycut & method & refs. & \rgq & \rguds & \rgb \\ 
\hline
Durham & 0.01 & Y & \cite{aleph113} & $1.19\pm0.03$ & $1.25\pm0.09$ &
$1.06\pm0.05$ \\ 
Durham & 0.02 & Y & \cite{OPALPR136,OPALPR141} & $1.25\pm0.04$ & 
$1.39\pm0.05$ & $1.09\pm0.03$ \\ 
Durham & 0.015 & Y & \cite{delphi115} & $1.25\pm0.04$ 
& -- & -- \\
Durham & 0.01 & \qqbar g/$\gamma$ & \cite{delphi115} & $1.24\pm0.06$ 
& -- & -- \\
JADE & 0.04 & \qqbar g/$\gamma$ & \cite{delphi115} & $1.35\pm0.08$ 
& -- & -- \\
Cone & -- & Y & \cite{OPALPR136,OPALPR141} & $1.10\pm0.03$ & $1.14\pm0.04$ & 
$0.92\pm0.04$ \\
\hline\hline
\end{tabular}
\end{indented}
\end{table}

The multiplicity ratios \rgq\ and \rguds\ are significantly larger
than one in all cases proving that the charged particle multiplicity
is greater in gluon than in quark jets.  However, the difference
between the ratios \rgq\ and \rguds\ based on the different jet
finding algorithms (Durham, JADE or Cone) implies that the results
cannot be compared directly to QCD
predictions~\cite{gary94,OPALPR141}.  The values of
\rgb\ show that at jet energies of $\sim 24$~GeV the multiplicities of
gluon and b jets are comparable.

\begin{figure}[htb!]
\begin{tabular}{cc}
\includegraphics[width=0.475\textwidth,clip]{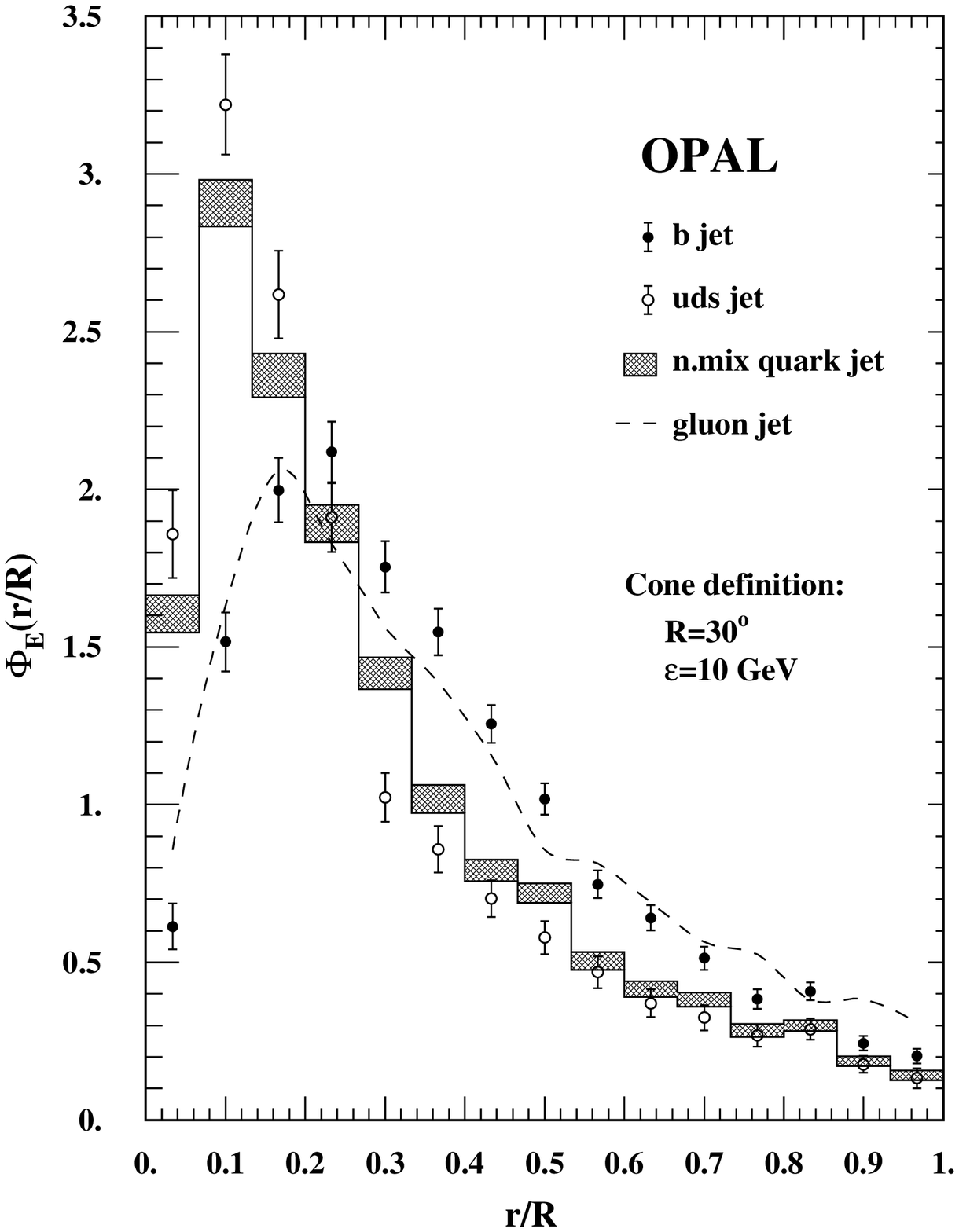} &
\includegraphics[width=0.475\textwidth,clip]{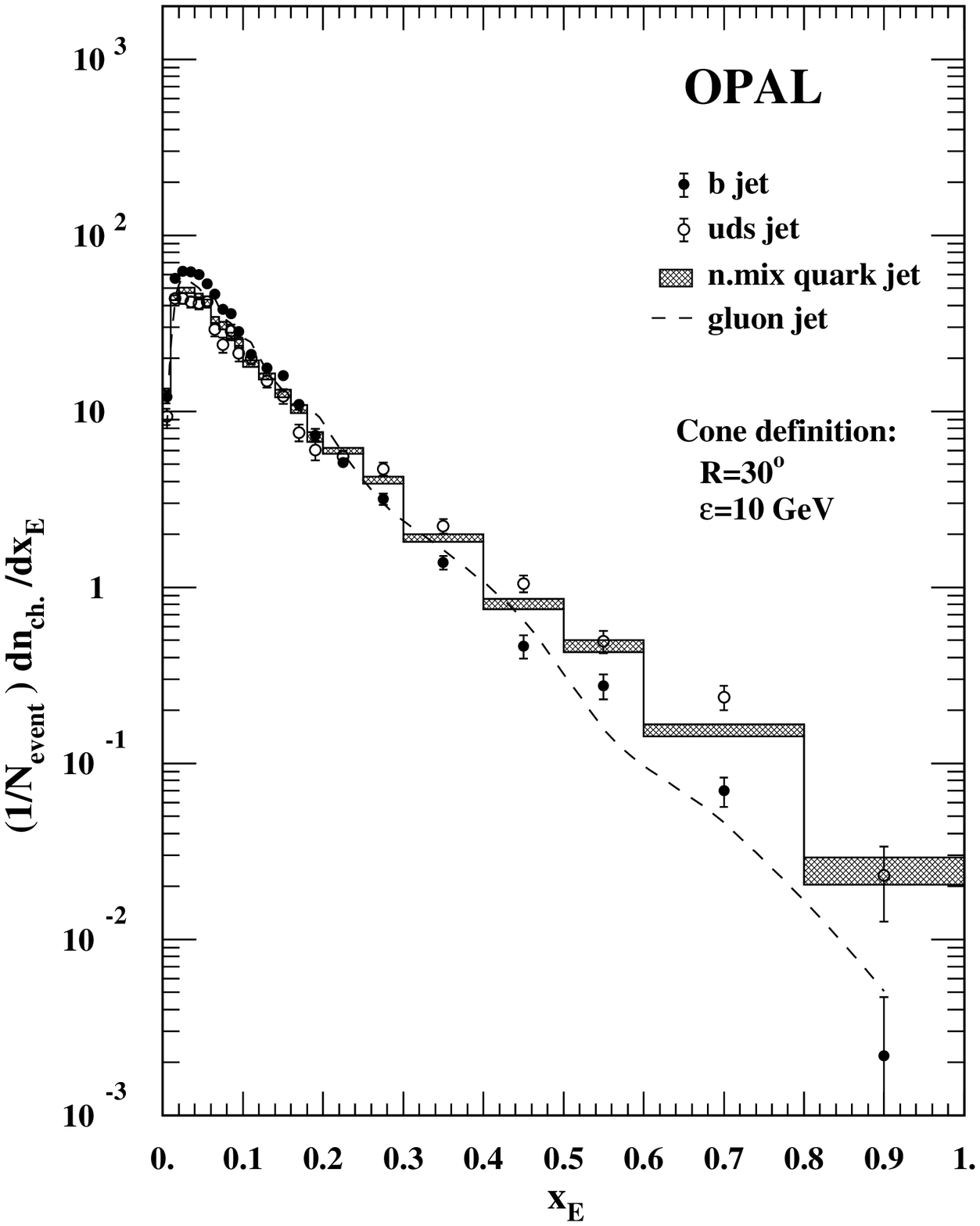} \\
\end{tabular}
\caption[ bla ]{ (left) Comparison of b, uds and inclusive quark jets
and gluon jets using the cone jet algorithm for the differential
energy profile $\phi_E(r/R$, see text.  (right) Comparison of the same
b, uds and inclusive quark jets and gluon jets for the charged
particle fragmentation function.  All error bars show statistical
uncertainties.  Both figures from~\cite{OPALPR141}.  }
\label{fig_gudsb}
\end{figure}

Figure~\ref{fig_gudsb} (left)~\cite{OPALPR141} shows the differential
energy profile $\phi_E(r/R)=\Delta E/\Delta(r/R)$ where $\Delta E$ is
the energy contained in an annulus of width $\Delta(r/R)$ at half
angle $r$ aligned with the axis of the Cone jet with half angle $R$.
A broader energy profile of gluon compared to light quark jets is
clearly visible.  The energy profile of b jets at $\sim 24$~GeV jet
energy turns out be comparable to the gluon jet profile.
Figure~\ref{fig_gudsb} (right)~\cite{OPALPR141} presents the
fragmentation function (FF) of charged particles.  The gluon FF is
clearly softer than the light quark (uds) FF, but the b quark FF is
again observed to be similar to the gluon jet FF at jet energies of
about 24~GeV.

All studies discussed so far did not take into account the topology
dependence of jet properties, see e.g.~\cite{dokshitzer91}, because
with symmetric 3-jet events the effects of topology dependence are
suppressed.  The topology dependence of particle multiplicity in jets
in 3-jet configurations has been studied theoretically in detail, see
e.g.~\cite{eden99} and references therein.  

In the MLLA the total particle multiplicity of 3-jet events
$N_{\qqbarg}(s,p_{\perp})$ produced at cms energies \roots\ also
depends on the transverse momentum $p_{\perp}$ of the gluon jet, due
to colour coherence~\cite{dokshitzer88,dokshitzer91}.  The quantity
$N_{\qqbarg}$ is expressed as a sum of particle multiplicities in
biased \qqbar\ events $\hat{N}_{\qqbar}(s,p_{\perp})$ and in ideal and
unbiased gg events $N_{\mathrm{gg}}(p_{\perp})$ taken at appropriate
scales~\cite{eden98,eden99}:
\begin{equation}
  N_{\qqbarg}(s) = \hat{N}_{\qqbar}(L,p_{\perp,Lu})
    +\frac{1}{2}N_{gg}(p_{\perp,Lu})
\label{equ_nch3jet}
\end{equation}
with $L=\ln(s/\lmqcd^2)$.  The contribution
$\hat{N}_{\qqbar}(L,p_{\perp})$ carries an explicit dependence on the
gluon jet scale $p_{\perp}$ which relates it to the unbiased particle
multiplicity $N_{\qqbar}$ measured without jet
requirements~\cite{eden98}: 
\begin{equation}
  \hat{N}_{\qqbar}(L,p_{\perp}) = N_{\qqbar}(L')
  +(L-L')\frac{\ddel N_{\qqbar}(L')}{\ddel L}\;\;,
\label{equ_nhat}
\end{equation}
with $L'=2\ln(p_{\perp}/\lmqcd)+3/2)$.  The transverse momentum of the
gluon jet is defined by $p_{\perp,Lu}=\sqrt{s_{\mathrm{qg}}s_{\qbar
g}/s}$ with $s_{\mathrm{qg}}$ and $s_{\qbar\mathrm{g}}$ the invariant
masses in the qg and \qbar g systems.  In an alternative expression
for $N_{\qqbarg}$ the variable $L$ is replaced by
$L_{\qqbar}=\ln(s_{\qqbar}/\lmqcd^2)$ and $p_{\perp,Lu}$ is replaced
by $p_{\perp,Le}=\sqrt{s_{\mathrm{qg}}s_{\qbar
g}/s_{\qqbar}}$~\cite{dokshitzer88,dokshitzer91}.  

These predictions are only valid for 3-jet events selected without a
cut on a fixed value of the jet resolution parameter \ycut, since such
a cut introduces an upper limit on the transverse momenta of any
subjets.  In contrast, when three jets are reconstructed in all events
and the events are classified by the $p_{\perp}$ of the third jet the
only remaining bias is given by the limitation of subjet transverse
momenta.  This bias is taken into account by the quantity
$\hat{N}_{\qqbar}$ discussed above.  Also, the jet reconstruction
should be done using a jet resolution definition based on transverse
momentum, e.g. the Durham, Cambridge or LUCLUS/PYCLUS algorithms.

For 3-fold symmetric 3-jet events with all three angles between the
jets $\simeq 120$\degr, so called Mercedes\footnote{Trademark of
DaimlerChrysler AG acknowledged} events, the alternative expression
discussed above for the transverse momentum of the gluon jet
simplifies~\cite{dokshitzer88,dokshitzer91}:
\begin{equation}
p_{\perp,Le} = \sqrt{\frac{ s_{\mathrm{qg}}s_{\qbar\mathrm{g}}}{s_{\qqbar}}}
  \sim \frac{\roots}{3}\sin\frac{\Theta}{2}\;\;.
\end{equation}
This has been used as a motivation in experimental studies to use
gluon and quark jet scales $\qjet=\ejet\sin(\Theta/2)$ where
$\Theta$ is the angle to the closest jet and thus takes the event
topology into account~\cite{aleph142,delphi166,delphi207,OPALPR312}.
The study~\cite{aleph142} showed that using \qjet\ for quark jets
and $\qbarjet=\sqrt{Q_{\mathrm{qg}}Q_{\qbar\mathrm{g}}}$
removes biases in the jet multiplicities which appear when jets with
similar energies \ejet\ but at different angles $\Theta$ are
compared.  

\begin{figure}[htb!]
\includegraphics[width=0.75\textwidth,clip]{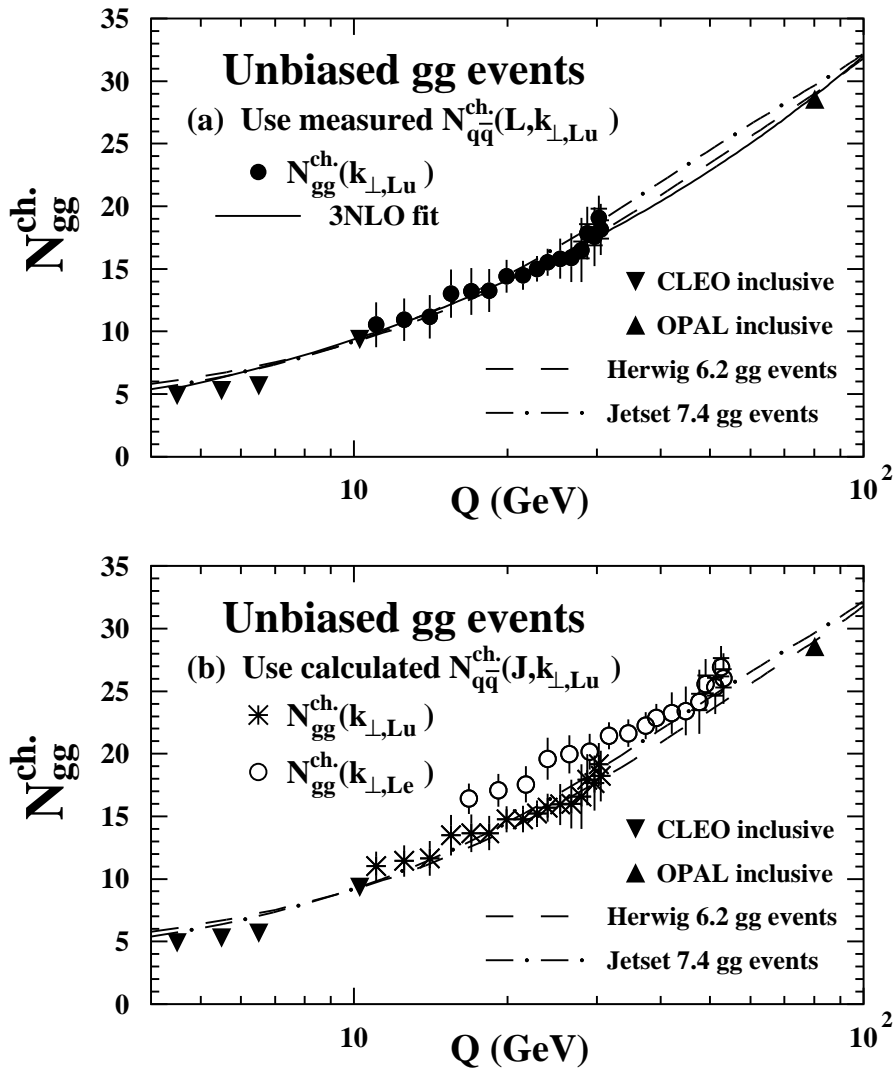}
\caption[ bla ]{ (a) Results for $N_{\mathrm{gg}}$ shown by
solid points as function of the jet energy scale $Q$ based on measured
$\hat{N}_{\qqbar}(L',p_{\perp})$~\cite{OPALPR347}.  The error bars
indicate total uncertainties.  (b) Results for $N_{\mathrm{gg}}$
based on calculated $\hat{N}_{\qqbar}(L',p_{\perp})$ using
$p_{\perp,Lu}$ (stars) or $p_{\perp,Le}$ (open points).  Both figures
show by the solid and dashed or dash-dotted lines the same predictions
from pQCD and Monte Carlo simulations and by the solid triangles other
measurements based on inclusive gluon jets~\cite{OPALPR347}.  }
\label{fig_ngg}
\end{figure}

The theoretical description of multiplicity in 3-jet events of
equations~(\ref{equ_nch3jet}) and~(\ref{equ_nhat}) has been tested.
In~\cite{OPALPR347} the charged particle multiplicity in 2-jet events
selected with the Durham jet algorithm with fixed vales of \ycut\ has
been measured.  The measurements have been compared with
equation~(\ref{equ_nhat}) using the relation
$p_{\perp}=E_{\mathrm{vis}}\sqrt{\ycut}$ and reasonable agreement
within the total uncertainties of the data has been observed.  Based
on this result the charged particle multiplicity in unbiased gg events
has been extracted as a function of the jet energy scale $Q$.  The
results are shown in figures~\ref{fig_ngg} (a) and (b).  In
figure~\ref{fig_ngg} (a) the measured $\hat{N}_{\qqbar}(L,p_{\perp})$
have been used to extract $N_{\mathrm{gg}}$ shown by the solid points
at $Q=p_{\perp}$.  The measurement of $N_{\mathrm{gg}}$ is compared
with a pQCD prediction~\cite{capella00} which has been fitted to the
data (solid line) and predictions by Monte Carlo simulations (dashed
and dash-dotted lines) of artificial gg events.  The predictions are
found to agree well with the data including other measurements from
inclusively defined gluon jets shown by solid triangles (see below).
In figure~\ref{fig_ngg} (b) the $\hat{N}_{\qqbar}(J,p_{\perp})$ have
been calculated using equation~(\ref{equ_nhat}) using both definitions
of $p_{\perp}$; stars correspond to $p_{\perp,Lu}$ and $J=L$ and open
points to $p_{\perp,Le}$ and $J=L_{\qqbar}$.  The Monte Carlo
predictions and inclusive gluon jet data are the same as in
figure~\ref{fig_ngg} (a).  The results based on $p_{\perp,Lu}$ are in
better agreement with the inclusive gluon jet data at lower and higher
$Q$ than the results employing $p_{\perp,Le}$ as the jet energy scale.

The quantities $N_{\mathrm{gg}}$ for gluon jets and $N_{\qqbar}$ for
quark jets in two-parton systems are related by $\ddel
N_{\mathrm{gg}}(L')/\ddel L = \ca/\cf(1-\alpha_0c_r/L)\ddel
N_{\qqbar}(L)/\ddel L$ with $\alpha_0=6\ca/(11\ca-2\nf)$ and
$c_r=10\pi^2/27-3/2$~\cite{eden98}.  The measurements of
$N_{\mathrm{gg}}$ together with measured $N_{\mathrm{\qqbar}}$ thus
allow a precise determination of ratio of colour factors
$\ca/\cf$~\cite{OPALPR347}:
\begin{equation}
\ca/\cf=2.23\pm0.01\stat\pm0.14\syst
\end{equation}
The data in figure~\ref{fig_ngg} (a) above $Q=7$~GeV have been used in
the fit and the fit result corresponds to the solid line.
In\cite{delphi207} a similar measurement has been performed for the
first time using slightly different predictions~\cite{khoze97} and
involving an additional non-perturbative correction.  The result was
$\ca/\cf=2.25\pm0.06\stat\pm0.12\syst$.  Both results for $\ca/\cf$
agree well with the QCD expectation of $\ca/\cf=2.25$ and have small
systematic uncertainties compared to other measurements of this
quantity.  A compilation of other measurements of $\ca/\cf$ is shown
in section~\ref{sec_gauge}.

The FFs of quark and gluon jets in mirror symmetric Y-events and
completely symmetric Mercedes events were studied as function of the
scaled jet energy \qjet~\cite{delphi166}.  The comparison of the
gluon FF in Y and Mercedes events found evidence for scaling
violations in the gluon FF.  The scaling violation of the gluon jet FF
was seen to be stronger than the comparable quark jet FF consistent
with the QCD expectation motivated by the larger colour charge of
gluons compared to quarks.  This observation was confirmed by a study
of subjets in quark and gluon jets.  To find subjets the jet finding
algorithm is run only on the particles associated with the gluon or
quark jet~\cite{delphi166}.  The fraction of events without subjet
production was found to be smaller in the quark jet sample in
agreement with the expectation and an NLLA QCD calculation.

Other subjects of study have been the production rates of identified
particles in quark and gluon jets.  In several analyses the production
of \pizero\ and $\eta$~\cite{l3089,OPALPR312}, of
\kshort~\cite{OPALPR235,delphi136,l3118,OPALPR312},
\kplus\ and p~\cite{delphi136}, and
$\Lambda^0$~\cite{l3118,delphi136,OPALPR235} was investigated.
In~\cite{l3089} the production of \pizero\ and $\eta$ was measured in
2- and 3-jet events and $\eta$ production was found to increase
stronger than \pizero\ production in 3-jet events.  This observation
was not confirmed in~\cite{OPALPR312} where production rates of
\pizero, $\eta$ and \kshort\ in gluon jets were found to be larger than
in quark jets by the same factor as for the increase of the inclusive
charged particle multiplicity.  In~\cite{delphi136} it was observed
that the momentum spectra of \kshort, \kplus, p and $\Lambda^0$ were
softer in gluon compared to quark jets with a higher total particle
multiplicity.

\subsection{Inclusive jets}

In the definition of inclusive gluon and quark jets no jet finding
algorithm is directly involved.  This removes systematic uncertainties
connected with the assignment of particles to jets, as discussed in
the previous section.  Studies using inclusive gluon jets are based on
i) particle decays to hadronic final states via gluons
only~\cite{alam92,alam97} or ii) $\epem\rightarrow\qqbarg$ events at
high energy in the rare kinematic configuration where a hard gluon g
recoils against the \qqbar\ system~\cite{gary94}.

The first study using inclusive gluon jets was~\cite{alam92} using the
decays Y(3S)$\rightarrow\gamma\chi_b(2P)$, Y(1S)$\rightarrow$ggg, and
\qqbar\ events produced in the continuum at $\roots=10.55$~GeV.  The
$\chi_b(2P)$ states with $J=0,2$ can decay to gg while the
$\chi_b(2P), J=1$ state decays to \qqbarg.  The three different
$\chi_b(2P), J=0,1,2$ states are disentangled employing their slightly
differing invariant masses.  The distributions of the event shape
observable $R_2$\footnote{ratio of 2nd to 0th Fox-Wolfram moment using
charged and neutral particles~\cite{fox78}} derived from
$\chi_b(2P), J=0,2$ decays or continuum events evolved to the
$\chi_b(2P), J=1$ scale show clear differences.

In a similar study~\cite{alam97} events of the type
$\mathrm{Y(1S)}\rightarrow\mathrm{gg}\gamma$ are compared with
$\epem\rightarrow\qqbar\gamma$ events produced in the continuum.  The
mean charged particle multiplicity is measured in both samples as a
function of the recoil mass of the photon.  The ratio of
multiplicities in gg and \qqbar\ systems for recoil masses below 7~GeV
is found to be compatible with unity.  The gluon multiplicities are
also shown in figures~\ref{fig_ngg} (a) and (b).  

A method to obtain inclusive gluon jets at high energy has been
proposed in~\cite{gary94}.  The inclusive gluon jet definition is
achieved experimentally by selecting hadronic \znull\ decays were both
q and \qbar\ are detected in the same hemisphere of the event.  The
quarks are identified via tagging of heavy quark decays and the gluon
jet is identified with the other hemisphere of the event.  In two
studies the mean charged particle multiplicity~\cite{OPALPR172} and
the charged particle multiplicity distributions~\cite{OPALPR214} were
measured in inclusive gluon and quark jets.  The multiplicity
distribution was analysed in terms of higher moments and agreement
with a NNLO QCD calculation was found for gluon and quark jets.  The
latest result for the ratio of mean charged particle multiplicities in
inclusive quark and gluon jets of $\ejet=40.1$~GeV is
$r_{ch}=1.51\pm0.02\stat\pm0.03\syst$~\cite{OPALPR273}.  This is only
in qualitative agreement with results from exclusive jets discussed in
the previous section.

\begin{figure}[htb!]
\includegraphics[width=0.75\textwidth]{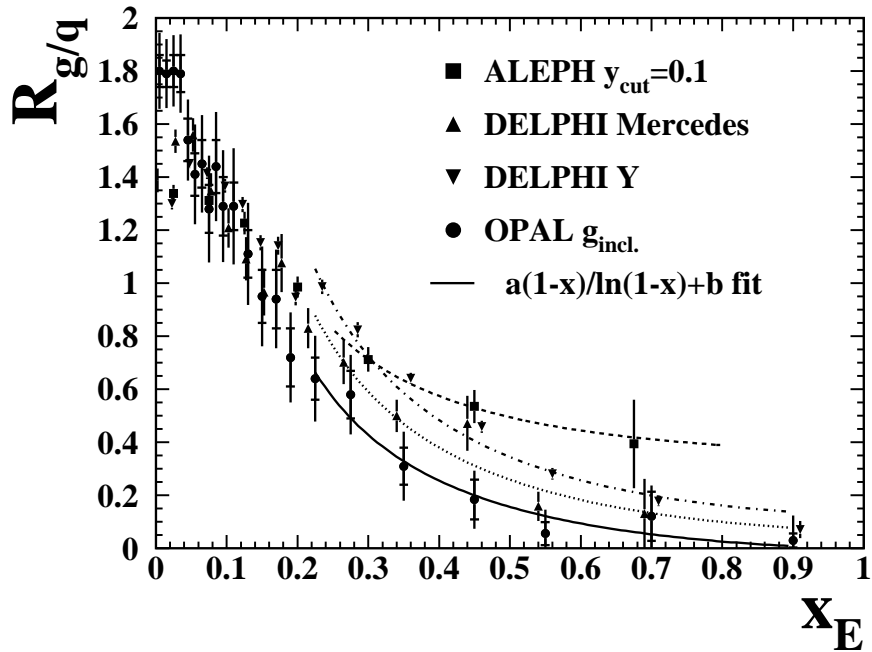}
\caption[ bla ]{ Ratio of scaled momentum distributions of charged
particles measured in inclusive gluon and quark jets at
$\ejet=40.1$~GeV.  Superimposed are fits of the form $(1-x)/\ln(1-x)$
as predicted in~\cite{konishi78}.  The DELPHI data points are
horizontally displaced for clarity. Data
from~\cite{delphi166,OPALPR273,aleph174}.  }
\label{fig_rxe}
\end{figure}

In~\cite{OPALPR273} many properties of inclusive gluon and quark jets
at high jet energies were studied.  The measurement of the
distributions of charged particle energy fractions $x_E$ allows to
test some of the predictions discussed above.  From
the data for $x_E$ in table~2 of~\cite{OPALPR273} one can extract
$\momone{x_E}_g=0.044\pm0.002$, $\momone{x_E}_q=0.0635\pm0.0003$.  The
ratio is found as
\begin{equation}
  r_{\momone{x}} = \momone{x_E}_q/\momone{x_E}_g = 1.44\pm0.07\;\;,
\end{equation}
where the error contains an additional contribution of 3\% to cover
the energy difference between the inclusive gluon jets at 40.1~GeV and
the quark jets at 45.6~GeV.  The result for $r_{\momone{x}}$ is
consistent with the result for $r_{ch}$ shown above.  The prediction
of~\cite{konishi78} discussed above for the behaviour of the ratio of
$x_E$ distributions can be tested by fitting a function
$a(1-x)/\ln(1-x)+b$, where $a$ and $b$ free parameters, to the data
for $x_E>0.2$.  Figure~\ref{fig_rxe} presents the results of such fits
using the data of~\cite{delphi166,OPALPR273,aleph174}.  The fits based
on the total errors agree with the data; the results are shown in
table~\ref{tab_rxe}.  The fit to OPAL data was performed with $b=0.0$
fixed since the fitted value of $b$ was consistent with zero.

\begin{table}[htb!]
\caption[ bla ]{ Results of fits of $a(1-x)/\ln(1-x)+b$ as explained
in the text to data for ratios of FFs from gluon and quark jets. }
\label{tab_rxe}
\begin{indented}\item[]
\begin{tabular}{lccc} 
\hline\hline
 & a & b & \chisqd \\
\hline
ALEPH $\ycut=0.1$ & $0.176\pm0.064$ & $0.36\pm0.11$ & 0.03/1 \\
DELPHI Merc. & $0.267\pm0.033$ & $0.067\pm0.057$ & 4.3/4 \\
DELPHI Y  & $0.306\pm0.010$ & $0.124\pm0.014$ & 25/5 \\
OPAL $g_{incl}$ & $0.218\pm0.035$ & 0.0 (fixed) & 1.3/6 \\ 
\hline\hline
\end{tabular}
\end{indented}
\end{table}

The results for $a$ can be averaged yielding $\bar{a}= 0.295\pm0.009$
with a $\chisqd=10/3$.  The data qualitatively support the prediction
of~\cite{konishi78}.  However, there are significant differences
between the fit results indicating substantial systematic differences
between the different data sets, which have been obtained using
different experimental techniques.

\section{QCD gauge structure}
\label{sec_gauge}

QCD as the gauge theory of strong interactions assumes that quarks
carry one out of three strong charges, referred to as colour.  The
requirement of local gauge symmetry under SU(3) transformations in the
colour space generates the gauge bosons of QCD: an octet of gluons
each carrying colour charge and anti-charge, see e.g.~\cite{ellis96}.
The gluons can thus interact with themselves; it is due to this
property of the QCD gauge bosons that the theory explains confinement
and asymptotic freedom via the running of the strong coupling \as.  

In perturbative QCD at NLO three fundamental vertices contribute: i)
the quark-gluon vertex with colour factor \cf, ii) the gluon-gluon
vertex with colour factor \ca\ and iii) the \qqbar\ production from a
gluon with colour factor $\tf\nf$, see e.g.~\cite{magnoli90}.  The
colour factors are $\cf=4/3$, $\ca=3$ and $\tf\nf=1/2\cdot 5$ in QCD
with SU(3) gauge symmetry and specify the relative contribution of the
corresponding vertex to observables.  In NLO QCD the prediction for an
observable $R$ is $R= A\as + (B_{\cf}\cf + B_{\ca}\ca +
B_{\tf}\tf\nf)\cf\as^2$; NLLA predictions e.g.\ for event shapes or
jet rates decompose in a similar way.  For NLO predictions for 4-jet
observables an analogous decomposition in terms of the six possible
products of two out of the three colour factors holds~\cite{nagy98a};
a seventh term is negligible and left out.

Experimental investigations of the gauge structure of QCD are possible
because of the different angular momenta in the initial and final
states of the fundamental vertices.  It is an important test of QCD to
probe the gauge structure in experiments.  Several techniques with
rather different experimental and theoretical uncertainties have been
developed; we will discuss here some recent results.

\subsection{Four-jet events}

The LEP experiments ALEPH~\cite{aleph249} and OPAL~\cite{OPALPR330}
have analysed 4-jet final states from hadronic \znull\ decays using
the recent QCD NLO predictions (see e.g.~\cite{nagy98a} and references
therein).  The 4-jet final states are selected by clustering events
using the Durham algorithm~\cite{durham} with $\ycut=0.008$ and
demanding four jets.  At this value of \ycut\ the 4-jet fraction is
relatively large ($\rfour\simeq 7$\%) and the four jets are well
separated.  

The energy-ordered 4-momenta $p_i, i=1,\ldots, 4$ of the jets are used
to calculate the angular correlation observables
(see~\cite{aleph249,OPALPR330} for details).  As an example, the
Bengtsson-Zerwas angle \chibz\ is defined by~\cite{bengtsson88}
$\chibz=\angle([\vec{p}_1\times\vec{p}_2],[\vec{p}_3\times\vec{p}_4])$,
i.e. the angle between the two planes spanned by the momentum vector
pairs $(\vec{p}_1,\vec{p}_2)$ and $(\vec{p}_3,\vec{p}_4)$.  Assuming
that energy ordering selected the primary quarks from the \znull\
decay as $(p_1,p_2)$ the observable \chibz\ is sensitive to the decay
of an intermediate gluon to gluons (vertex ii)) or quarks (vertex
iii)) and the competing process of radiation of a second gluon from a
primary quark (vertex i)).

\begin{figure}[!htb]
\begin{tabular}{cc}
\includegraphics*[width=0.45\textwidth]{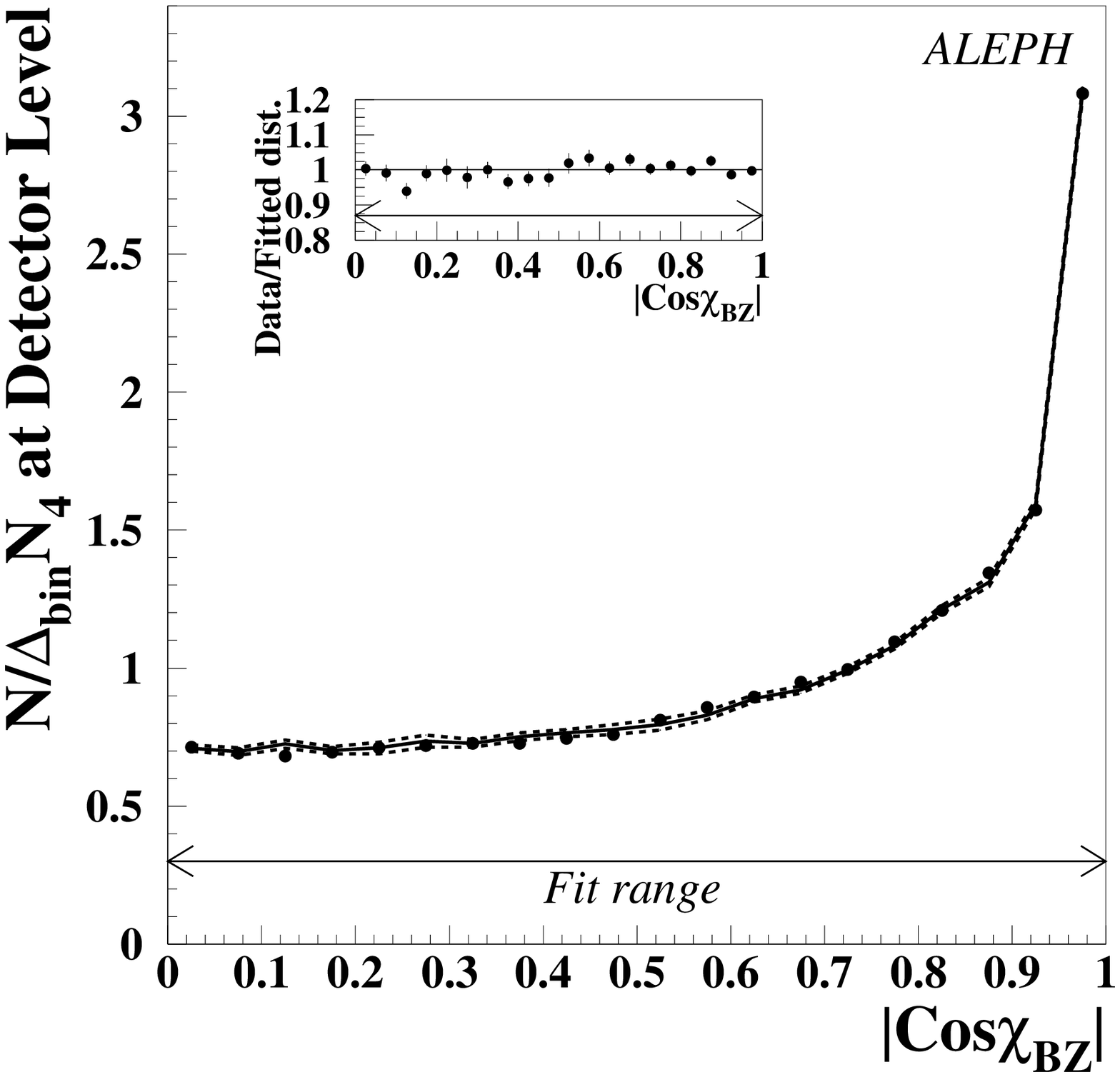} &
\includegraphics*[width=0.5\textwidth]{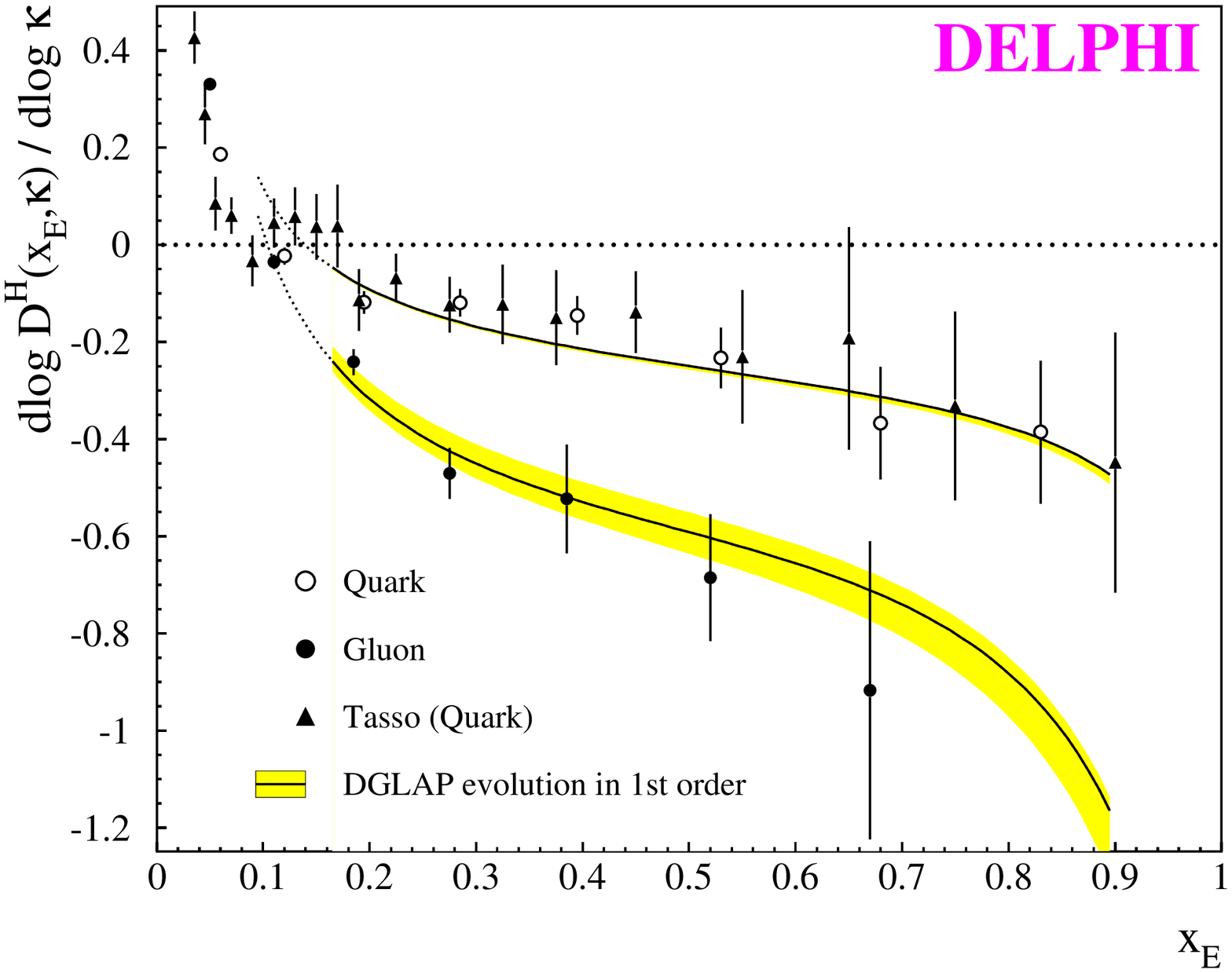} \\
\end{tabular}
\caption{ (left) Distribution of $|\cos\chibz|$ before corrections 
(solid points).  Superimposed is a fit of NLO QCD (solid
lines)~\cite{aleph249}. (right) Measurements of $\ddel
D^H(\xe,\kaph)/\ddel\log(\kaph)$ for gluon (solid points)
and quark jets (open circles and solid triangles).  The lines
represent a fit of the LO QCD prediction~\cite{delphi236}.  }
\label{fig_4jqgff}
\end{figure}

Figure~\ref{fig_4jqgff} (left) presents the uncorrected distribution
of $\cos(\chibz)$ measured by ALEPH~\cite{aleph249}.  Superimposed on
the data points is the result of a simultaneous fit of the NLO QCD
predictions to four angular correlations including \chibz\ and the
distribution of the 4-jet rate \rfour.  From the fit values for \asmz\
and the colour factors \ca\ and \cf\ are extracted.  The results from
ALEPH are $\asmz=0.119\pm0.027$, $\ca=2.93\pm0.60$ and
$\cf=1.35\pm0.27$; from OPAL we have $\asmz=0.120\pm0.023$,
$\ca=3.02\pm0.55$ and $\cf=1.34\pm0.26$.  The systematic errors are
dominated by uncertainties from the hadronisation corrections and
theoretical errors from estimating missing higher order contributions.
The hadronisation corrections are implemented using Monte Carlo models
with standard values for the colour factors.  Both analyses use an
unconventional method of estimating systematic uncertainties.  A more
conservative evaluation based on the information available
in~\cite{aleph249,OPALPR330} leads to total errors for \asmz, \ca\ and
\cf\ of $\pm0.048$, $\pm1.06$ and $\pm0.46$ for ALEPH and $\pm0.049$,
$\pm1.07$ and $\pm0.47$ for OPAL; i.e.\ the errors turn out to be
approximately twice as large.

\subsection{Scaling violation in gluon and quark jets}

Jets originating from quarks or gluons should have different
properties due to the different colour charge carried by quarks or
gluons, see section~\ref{sec_qg}.  In an analysis by
DELPHI~\cite{delphi236} the scaling violation of the fragmentation
function (FF) in gluon and quark jets at different energies is
compared.  From a sample of $3.7\cdot 10^6$ hadronic \znull\ decays
planar 3-jet events with well reconstructed jets are selected using
the Durham or Cambridge algorithms, but without imposing a fixed value
of \ycut\ in order to reduce biases~\cite{eden98} (see
section~\ref{sec_excljets}) .  In events with the two angles between
the most energetic jet and the other jets in the range between
100\degr\ and 170\degr\ a b-tagging procedure is applied to the jets.
Gluon jets are identified indirectly as the jets without a successful
b-tag.  Jets originating from light (udsc) quarks are taken from
events which failed the b-tagging.  A correction procedure based on
efficiencies and purities determined by Monte Carlo simulation yields
results for pure gluon or quark jets.

The jet energy scale is calculated according to
$\kaph=\ejet\sin(\theta/2)$, where $\theta$ is the angle w.r.t.\ to the
closest jet.  This definition takes colour coherence effects into
account~\cite{basicspqcd}.  The FF in a jet $D^H(\xe,\kaph)$ is given
by the distribution of $\xe=E_{\mathrm{hadron}}/\kaph$ for the hadrons
assigned to the jet.  The evolution of the FFs of gluon and quark jets
with jet energy scale \kaph\ is studied in intervals of \xe.
Figure~\ref{fig_4jqgff} (right) presents the quantity $\ddel
D^H(\xe,\kaph)/\ddel\log(\kaph)$, i.e.\ the slopes of the scaling
violation for a given interval of \xe, for quark and gluon jets.  The
steeper slopes corresponding to stronger scaling violations of gluon
compared to quark jets are clearly visible.  A fit of the scaling
violations based on the LO QCD prediction (DGLAP equation) allows to
extract the ratio $\ca/\cf$ resulting in $\ca/\cf=2.26\pm0.16$.

\subsection{Event shape fits}

In this analysis~\cite{colrun} the decomposition of the \oaa+NLLA QCD
predictions for event shape observables into terms proportional to the
colour factors is used.  Since the sensitivity of event shape
distributions measured at LEP~1 alone is not
sufficient~\cite{OPALPR134} data from $\roots=14$ to 189 GeV are used.
In this way the colour structure of the running of the strong coupling
contributes as well.  Hadronisation corrections are implemented using
power corrections, see section~\ref{sec_powcor}.  The advantage of
using power corrections instead of Monte Carlo model based
hadronisation corrections is that the colour structure of the power
corrections is known and can be varied in the fit.

In the analysis simultaneous fits of \asmz, \ca\ and \cf\ to data for
the event shape observables \thr\ at $\roots=14$ to 189~GeV and \cp\
at $\roots=35$ to 189~GeV are performed.  The data for \thr\ and \cp\
are analysed separately and the results are combined.  The results are
$\asmz=0.119\pm0.010$, $\ca=2.84\pm0.24$ and $\cf=1.29\pm0.18$ and are
shown on figure~\ref{fig_cacf} below.  The errors are dominated by
uncertainties from the hadronisation correction and from experimental
effects.

\subsection{Colour factor averages}

The measurements of the colour factors \ca\ and \cf\ or of $x=\ca/\cf$
discussed above or in section~\ref{sec_qg} can be combined into
average values of \ca\ and \cf\ taking into account correlations
between \ca\ and \cf\ as well as between different experiments.  The
variables $x$ and $y=\tf/\cf$ are used to define the \chisq\ function
\begin{eqnarray}
\label{equ_covxy}
 \chisq & = & \sum_i (x_i-\bar{x})v^{(xy)}_{i}(y_i-\bar{y}) + \\ \nonumber
 & &        \sum_{ij} (x_i-\bar{x})v^{(x)}_{ij}(x_j-\bar{x}) + \\ \nonumber
 & &        \sum_{ij} (y_i-\bar{y})v^{(y)}_{ij}(y_j-\bar{y})\;\;, 
\end{eqnarray}
where $\bar{x}$ and $\bar{y}$ are the averages, the indices $i$ and $j$
count experiments, $v^{(xy)}_{i}$, $v^{(x)}_{ij}$ and $v^{(y)}_{ij}$
are elements of the inverses of the corresponding covariance matrices
for the $x_i$ and $y_i$ within experiment $i$ and for the $x_i$ and
$y_i$ between experiments.  The averages $\bar{x}$ and $\bar{y}$ are
converted to the average values \bca\ and \bcf\ after the fit is
performed.  The input data for $x$ and $y$ are directly taken
from~\cite{aleph249,OPALPR330,delphi236,OPALPR347} while the results
from~\cite{colrun} have to be converted\footnote{The results are
$x=2.20\pm0.21\stat\pm0.25\syst$, $y=0.388\pm0.021\stat\pm0.051\syst$,
$\rho_{\mathrm{stat.}}=0.98$ and $\rho_{\mathrm{syst.}}=0.86$.}.

The covariance matrices between experiments are constructed as
follows: the ALEPH and OPAL 4-jet analyses have experimental and
hadronisation errors partially and theory errors fully correlated, the
4-jet analyses and the event shape analysis have hadronisation errors
partially and theory errors fully correlated, and the 4-jet and event
shape analyses and the DELPHI FF and the OPAL \nchgg\ analyses have
their theory errors partially correlated.

The averaging fit is done using only the first term of
equation~(\ref{equ_covxy}) to determine the averages \bca\ and \bcf\
avoiding possible biases from our assumptions on correlations of
systematic uncertainties between different sets of measurements.  The
fit is repeated using only statistical correlations in the first term
to determine the statistical errors and using the full covariance
matrix to determine the total errors.  The final results are
\begin{eqnarray}
  \bca & = & 2.89\pm0.03\stat\pm0.21\syst \\ \nonumber
  \bcf & = & 1.30\pm0.01\stat\pm0.09\syst
\end{eqnarray}
with correlation coefficient $\rho=0.82$.  The relative total
uncertainties are about 8\% for both \bca\ and \bcf.  The fit using
the full covariance matrix yields $\bca=2.83$ and $\bcf=1.29$ in good
agreement with our main results.

\begin{figure}[!htb]
\includegraphics*[width=0.75\textwidth]{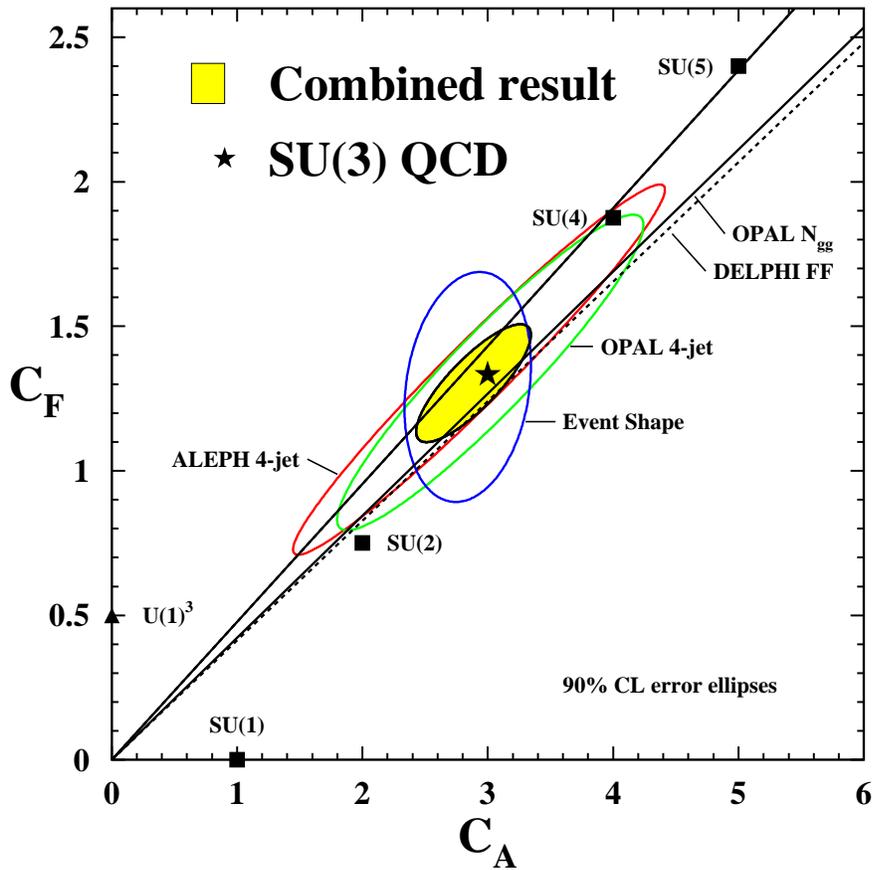}
\caption{ Measurements of the colour factors \ca\ and \cf\ discussed 
in this report. The ellipses show the correlated measurements using
4-jet events~\cite{aleph249,OPALPR330} or event shape
distributions~\cite{colrun} while the lines represent the results of
determinations of $\ca/\cf$ from DELPHI~\cite{delphi236} (dashed) and
OPAL~\cite{OPALPR347} (solid).  The upper solid and dashed lines
overlap.  The grey filled ellipsis displays the combined result for
\bca\ and \bcf\ (see text).  The solid triangle and squares show the
expectations for various assumptions for the gauge symmetry of QCD as
indicated on the figure. }
\label{fig_cacf}
\end{figure}

Figure~\ref{fig_cacf} presents the results of the individual analysis
in a \cf\ vs.\ \ca\ plane together with the combined result and the
expectations of QCD based on the SU(3) gauge symmetry and various
other gauge symmetries.  The correlation coefficients
for~\cite{aleph249,OPALPR330} were calculated from the references with
the results $\rho=0.97$ (ALEPH) and $\rho=0.93$ (OPAL).  The error
ellipses refer to 90\% CL.  The combined result is in good agreement
with the individual analyses and with standard SU(3) QCD while the
total uncertainties are substantially reduced.  The other
possibilities for gauge symmetries shown on the figure are clearly
ruled out.

\section{Conclusions and outlook}
\label{sec_conc}

We have shown in this report that hadron production in \epem\
annihilation is a useful and fruitful environment to test the theory
of strong interactions, QCD.  The absence of interference between
initial and final states and the large range of cms energies probed by
the experiments make many stringent tests of the theory possible.  The
early results from hadron production in \epem\ annihilation at low
energies were crucial to establish QCD as the theory of strong
interactions and thus as an integral part of the standard model
of high energy physics.

Studies of differences between quark and gluon jets reveal many
properties of the gauge bosons of QCD, the gluons, which are correctly
predicted by the theory.  The measurements of jet production rates and
event shape observables using theoretically and experimentally well
behaved observables allow direct tests of advanced perturbative QCD
predictions and precise determinations of the value of the strong
coupling constant \as.  A fundamental prediction by QCD is asymptotic
freedom of the coupling at high energies and this has been verified
directly using data over a large range of cms energies.  More indirect
tests of asymptotic freedom stem from successful comparison of
precision determinations of \as\ at different energy scales.

The most reliable and precise determinations of \as\ in \epem\
annihilation to hadrons employ inclusive observables such as the
hadronic branching ratios of the gauge bosons of the electroweak
theory, $\gamma$, \znull\ and \wpm.  The inclusive observables are
mostly unaffected by hadronisation corrections and they are accessible
to NNLO perturbative QCD calculations with 2-loop radiative
corrections.  The final result for \asmz\ from inclusive observables
calculated in NNLO perturbative QCD is
\begin{displaymath}
  \asmz = 0.1211 \pm0.0010\expt \pm0.0018\theo\;\;.
\end{displaymath}
This result is in good agreement with recent world averages of
measurements of the strong coupling constant
$\asmz=0.1184\pm0.0027$~\cite{bethke04} or
$\asmz=0.1187\pm0.0020$~\cite{pdg04} which are based on all available
analyses.  The best measurements of \asmz\ including ours shown above
yield errors of about 2\%.  The other determinations of \as\ discussed
in this report are found to be in good agreement with our main result
and we note that the results from analyses of the 4-jet rate using the
Durham or Cambridge algorithms are of similar precision.  Further
improvements in the near future may be expected from NNLO calculations
for 3-jet observables which are expected to reduce the otherwise
dominating theoretical uncertainties and might allow total
uncertainties of 1\%~\cite{burrows96b}.

Studies of the gauge structure of QCD using angular correlations in
4-jet final states, 3-jet observables at different energy scales and
quark and gluon jet differences were discussed.  The results for the
colour factors \ca\ and \cf\ were observed to be in good agreement and
combined values were calculated.  The results are
\begin{eqnarray} \nonumber
  \ca & = & 2.89\pm0.03\stat\pm0.21\syst \\ \nonumber
  \cf & = & 1.30\pm0.01\stat\pm0.09\syst \nonumber
\end{eqnarray}
with correlation $\rho=0.82$.  These results are in good agreement
with expectations from QCD with the SU(3) gauge symmetry $\ca=3$ and
$\cf=4/3$.  

In summary, \epem\ annihilation to hadrons has been and still is
indispensable to stringently probe the theory of strong interactions.
QCD has been established as a part of the standard model of high
energy physics and has been tested to a precision of 2\% with some
well suited observables.  This impressive result has been achieved by
the LEP experiments, SLD, the re-analysis of JADE data and the
parallel theoretical work stimulated by the availability of precise
data.  Future high energy physics programmes such as the LHC or the
ILC will benefit greatly from this achievement, because precise and
reliable QCD predictions will be essential for disentangling possible
new physics signals from standard processes.

\newpage

\section{References}

\end{document}